%% file: main.tex
\documentclass{article}
\usepackage[margin=0.75in]{geometry} % Narrow margins similar to IEEE
\usepackage{authblk} % For affiliations
\usepackage{cite}
\usepackage{amsmath,amssymb,amsfonts}
\usepackage{algorithmic}
\usepackage{graphicx,color}
\usepackage{textcomp}
\usepackage{multirow}
\usepackage{multicol}
\usepackage{lipsum}
\usepackage{float}
\usepackage{subcaption}
\usepackage{array}
\usepackage{pifont}
\usepackage{longtable}
\usepackage{pdfpages}
\usepackage{tablefootnote}

\usepackage[
  pdftitle={Measurement Campaigns, Datasets, and Curve Fitting Used by 3GPP in the Rel-19 Channel Model Validation of TR 38.901 for 7--24 GHz},
  pdfauthor={Hitesh Poddar; Ximan Lu},
  pdfsubject={3GPP channel modeling, TR 38.901, Release 19, 7--24 GHz},
  pdfkeywords={3GPP, TR 38.901, Release 19, channel model validation, 7--24 GHz, NR channel modeling, measurement campaigns, datasets, curve fitting, 5G-Advanced, 6G}
]{hyperref}

\title{Measurement Campaigns, Datasets, and Curve Fitting Used by 3GPP in the Rel-19 Channel Model Validation of TR 38.901 for 7--24 GHz}

\author[1]{Hitesh Poddar}
\author[2]{Jianhua Zhang}
\author[2]{Ximan Liu}
\author[3]{Mansoor Shafi}

\affil[1]{Sharp Laboratories of America, USA}
\affil[2]{State Key Laboratory of Networking and Switching Technology, Beijing University of Posts and Telecommunications, China}
\affil[2]{State Key Laboratory of Networking and Switching Technology, Beijing University of Posts and Telecommunications, China}
\affil[3]{Spark NZ Ltd, NZ}
\date{}

\begin{document}

\maketitle
\tableofcontents
\newpage
\section{Introduction}
\label{sec:intro}
\input{intro.tex}

\section{Measurement Campaigns}
\label{sec:meas_camp}
\input{meas_camp.tex}

\section{Measurement Datasets}
\label{sec:meas_dataset}

\input{meas_dataset.tex}

\section{Curve Fitting Results}
\label{sec:curve_fit_results}
\input{curve_fit_results.tex}

\section{Conclusion}
\label{sec:conclusion}
\input{conclusion.tex}

\bibliographystyle{IEEEtran}

\end{document}

%% file: intro.tex
The Third Generation Partnership Project (3GPP) Technical Report (TR) 38.901~\cite{tr38901v18} provides a comprehensive channel model for the frequency range of 0.5–100 GHz for various scenarios, such as Urban Microcell (UMi), Urban Macrocell (UMa), Rural Macrocell (RMa), Indoor Office (InH), and Indoor Factory (InF) scenarios\footnote{In the updated 3GPP TR 38.901~\cite{tr38901v19}, a new scenario called the Suburban Macro (SMa) was introduced.}. However, more than 80\% of the data used for developing 3GPP TR 38.901~\cite{tr38901v18} was based on sub-6 GHz and above-24 GHz frequency bands, leaving the 7–24 GHz band significantly underrepresented. With the commencement of 3GPP Release (Rel)-20, focusing on 5G-Advanced and initial 6G studies in late 2025, and with the majority of 6G studies starting in the Radio Access Network (RAN) Working Groups (WGs) in early to mid 2026, the need for an accurate channel model for the 7–24 GHz band has become essential. As the 3GPP TR 38.901~\cite{tr38901v18} channel models serve as industry-standard benchmarks for system design and evaluation, timely updates are critical for enabling 6G use cases.
In this context, several studies related to channel modeling were introduced by 3GPP in Rel-19, which focused on 5G-Advanced and commenced in early 2024 and concluded in mid 2025. One of these studies focused on channel modeling enhancements for the 7–24 GHz frequency range for NR~\cite{rp234018}. This particular study comprised of two objectives:
\begin{itemize}
\item Validate using measurements the channel model of 3GPP TR 38.901~\cite{tr38901v18} at least for 7-24 GHz
\begin{itemize}
    \item Only stochastic channel model is considered for the validation.
    \item The validation may consider all existing scenarios such as UMi, UMa, InH, RMa and InF.
\end{itemize}
\item Adapt/extend as necessary the channel model of 3GPP TR 38.901~\cite{tr38901v18} at least for 7-24 GHz, including at least the following aspects for applicable scenarios:
\begin{itemize}
    \item Near-field propagation (with consideration being given to consistency between near-field and far-field)
    \item Spatial non-stationarity
\end{itemize}
\textit{Note:} The study emphasized that the continuity of the channel model in the frequency domain below 7 GHz and above 24 GHz shall be ensured.
\end{itemize}
This document specifically focuses on the Rel-19 study on validate using measurements the channel model of 3GPP TR 38.901~\cite{tr38901v18} at least for 7-24 GHz based on extensive measurement campaigns, datasets, and curve-fitting analyses officially used within 3GPP. The document is organized as follows. Section \ref{sec:meas_camp} provides a summary of measurement campaigns officially used by 3GPP for the Rel-19 study. Section \ref{sec:meas_dataset} provides a link to an Excel file containing the datasets used in 3GPP, and explains the organization of the Excel file. Section \ref{sec:curve_fit_results} presents the curve-fitting results adopted in 3GPP. Finally, concluding remarks are provided in Section \ref{sec:conclusion}.

\vspace{1em} \noindent\textbf{Index Terms—}
3GPP, TR 38.901, Release 19, channel model validation, 7--24 GHz, NR channel modeling, measurement campaigns, datasets, curve fitting, 5G-Advanced, 6G.

\clearpage

%% file: meas_camp.tex
This section provides a comprehensive summary of the various sources that contributed measurement (Meas.) and ray-tracing (RT) data at different frequencies (Freq.) and bandwidths (Band.) for the 3GPP Rel-19 study on validate using measurements the channel model of 3GPP
TR 38.901~\cite{tr38901v18} at least for 7-24 GHz~\cite{rp234018}. Specifically, the parameters documented in this section include path loss (PL), excess PL, shadow fading (SF), delay spread (DS), angular spread of departure (ASD), angular spread of arrival (ASA), elevation/zenith spread of departure (ZSD), elevation/zenith spread of arrival (ZSA), K-factor, the number of clusters and the absolute time of arrival (ToA) across scenarios such as UMi, UMa, SMa, RMa, InH, and InF. In addition, material penetration loss measurements for commonly encountered materials are included. 

% \textcolor{red}{@Ximan -- Missing parameters: Absolute time of arrival for UMi, UMa, RMa and InH. I think only Ericsson and ZTE provided the data and the final values from ZTE were used. In each scenario subsection you can add a paragraph and state the configuration and the reference for who provided the data for absolute time of arrival. In Section 2.2 UMa O2I ASD is missing, UMa cluster ASD LOS, NLOS, O2I is also missing. This was only provided by Ericsson and BUPT. You can add footnote at the sources in the UMa tables and mention the references where this data was provided and if the same config in the tables were used to get this data. If configs are different you can mention the details of the config and the sources. Finally, can you check if any sources provided the LOS probability data and K factor data for any scenario other than SMa. If you find any just mention in the main section that is section 2 as a paragraph at the end for me to check and handle it?}

\subsection{UMi-Street Canyon}
\begin{center}
\begin{longtable}{|p{5.5em}|p{2.2em}|p{2.7em}|p{3em}|p{3.5em}|p{1.2em}|p{1.8em}|p{1.8em}|p{1.8em}|p{1.8em}|p{4.5em}|p{5.1em}|}
  \caption{Meas. and RT data at various Freq. and Band. by various sources for the outdoor to outdoor (O2O) UMi-Street Canyon scenario. \ding{51} indicates that the parameter was reported whereas, \ding{55} indicates that the parameter was not reported. Some companies use continuous wave (CW) in measurement. `NA' indicates data was not provided.}
  \label{tab:umi_measurements}
    \\
\hline
\centering \textbf{Source} & 
\textbf{Type} & 
\textbf{Freq. (GHz)} &
\textbf{Band. (MHz)} & 
\textbf{\centering PL/\- Excess PL/SF} & 
\textbf{DS} & 
\textbf{ASD} &
\textbf{ASA} & 
\textbf{ZSA} & 
\textbf{ZSD} & 
\centering \textbf{Number of Clusters} & 
\textbf{References} \\
\hline
\endfirsthead

\hline
\textbf{Source} & 
\textbf{Type} & 
\textbf{Freq. (GHz)} &
\textbf{Band. (GHz)} &
\textbf{\centering PL/\- Excess PL/SF} & 
\textbf{DS} & 
\textbf{ASD} &
\textbf{ASA} & 
\textbf{ZSA} & 
\textbf{ZSD} & 
\textbf{Number of Clusters} & 
\textbf{References} \\
\hline
\endhead

\hline
\endfoot
        \multirow{2}{5.5em}{\centering Sharp, NYU, Nokia} & 
        \multirow{2}{3.6em}{Meas.} & 
        \centering 6.75 &
        \centering 1000 &
        \centering \ding{51} &
        \centering \ding{51} &
        \centering \ding{51} &
        \centering \ding{51} &
        \centering \ding{51} &
        \centering \ding{51} &
        \centering \ding{55} &
        \multirow{2}{5.1em}{\centering \cite{r1-2410319},\cite{shakya2024urban}} \\
        \cline{3-11}
        & 
        & 
        \centering 16.95 &
        \centering 1000 &
        \centering \ding{51} &
        \centering \ding{51} &
        \centering \ding{51} &
        \centering \ding{51} &
        \centering \ding{51} &
        \centering \ding{51} &
        \centering \ding{55} &
        \\
    \hline
        \multirow{3}{5.5em}{\centering Nokia, Anritsu} & 
        \multirow{3}{3.6em}{Meas.} & 
        \centering 3.5 &
        \centering 2000 &
        \centering \ding{51} &
        \centering \ding{51} &
        \centering \ding{55} &
        \centering \ding{55} &
        \centering \ding{55} &
        \centering \ding{55} &
        \centering \ding{55} &
        \multirow{3}{5.1em}{\centering \cite{r1-2403996}}\\
        \cline{3-11}
        & 
        & 
        \centering 11 &
        \centering 2000 &
        \centering \ding{51} &
        \centering \ding{51} &
        \centering \ding{55} &
        \centering \ding{55} &
        \centering \ding{55} &
        \centering \ding{55} &
        \centering \ding{55} &
        \\
        \cline{3-11}
        & 
        & 
        \centering 29 &
        \centering 2000 &
        \centering \ding{51} &
        \centering \ding{51} &
        \centering \ding{55} &
        \centering \ding{55} &
        \centering \ding{55} &
        \centering \ding{55} &
        \centering \ding{55} &
        \\
        \hline
        \centering Intel & 
        \centering Meas. & 
        \centering 10 &
        \centering 250 &
        \centering \ding{51} &
        \centering \ding{51} &
        \centering \ding{55} &
        \centering \ding{55} &
        \centering \ding{55} &
        \centering \ding{55} &
        \centering \ding{55} &
        \multirow{1}{5.1em}{\centering \cite{r1-2406007}}
        \\
        \hline
        \multirow{3}{5.5em}{\centering China Telecom} & 
        \multirow{3}{3.6em}{Meas.} & 
        \centering 7 &
        \centering NA &
        \centering \ding{51} &
        \centering \ding{51} &
        \centering \ding{55} &
        \centering \ding{51} &
        \centering \ding{51} &
        \centering \ding{55} &
        \centering \ding{55} &
        \multirow{3}{5.1em}{\centering \cite{r1-2400841}}\\
        \cline{3-11}
        & 
        & 
        \centering 10 &
        \centering NA &
        \centering \ding{51} &
        \centering \ding{51} &
        \centering \ding{55} &
        \centering \ding{51} &
        \centering \ding{51} &
        \centering \ding{55} &
        \centering \ding{55} &
        \\
        \cline{3-11}
        & 
        & 
        \centering 13 &
        \centering NA &
        \centering \ding{51} &
        \centering \ding{51} &
        \centering \ding{55} &
        \centering \ding{51} &
        \centering \ding{51} &
        \centering \ding{55} &
        \centering \ding{55} &
        \\
        \hline
        \multirow{2}{5.5em}{\centering Samsung} & 
        \multirow{2}{3.6em}{Meas.} & 
        \centering 6.5 &
        \centering 100 &
        \centering \ding{51} &
        \centering \ding{51} &
        \centering \ding{51} &
        \centering \ding{51} &
        \centering \ding{55} &
        \centering \ding{55} &
        \centering \ding{55} &
        \multirow{2}{5.1em}{\centering \cite{r1-2408659}}\\
        \cline{3-11}
        & 
        & 
        \centering 13.5 &
        \centering 100 &
        \centering \ding{51} &
        \centering \ding{51} &
        \centering \ding{51} &
        \centering \ding{51} &
        \centering \ding{55} &
        \centering \ding{55} &
        \centering \ding{55} &
        \\
        \hline
        \multirow{3}{5.5em}{\centering AT\&T} & 
        \multirow{3}{3.6em}{Meas.} & 
        \centering 7 &
        \centering 400 &
        \centering \ding{51} &
        \centering \ding{51} &
        \centering \ding{55} &
        \centering \ding{51} &
        \centering \ding{51} &
        \centering \ding{55} &
        \centering \ding{55} &
        \multirow{2}{5.1em}{\centering \cite{r1-2501371}}  \\
        \cline{3-11}
        & 
        & 
        \centering 8 &
        \centering 400 &
        \centering \ding{51} &
        \centering \ding{51} &
        \centering \ding{55} &
        \centering \ding{51} &
        \centering \ding{51} &
        \centering \ding{55} &
        \centering \ding{55} &
        \\
        \cline{3-11}
        & 
        & 
        \centering 15 &
        \centering 400 &
        \centering \ding{51} &
        \centering \ding{51} &
        \centering \ding{55} &
        \centering \ding{51} &
        \centering \ding{51} &
        \centering \ding{55} &
        \centering \ding{55} &
        \\
        \hline
        \multirow{2}{5.5em}{\centering Apple} & 
        \centering RT & 
        \centering 8 &
        \centering NA &
        \centering \ding{51} &
        \centering \ding{51} &
        \centering \ding{51} &
        \centering \ding{51} &
        \centering \ding{51} &
        \centering \ding{55} &
        \centering \ding{51} &
        \multirow{1}{5.1em}{\centering \cite{r1-2408484}} \\
        \cline{2-12}
        & 
        \centering  Meas. & 
        \centering  13 &
        \centering 550 &
        \centering  \ding{55} &
        \centering  \ding{51} &
        \centering \ding{51} &
        \centering  \ding{51} &
        \centering  \ding{51} &
        \centering  \ding{51} &
        \centering  \ding{55} &
        \multirow{1}{5.1em}{\centering \cite{r1-2502625}}\\
        \hline
        \centering BUPT, Spark & 
        \centering Meas. & 
        \centering 13 &
        \centering 400 &
        \centering \ding{55} &
        \centering \ding{51} &
        \centering \ding{51} &
        \centering \ding{51} &
        \centering \ding{51} &
        \centering \ding{51} &
        \centering \ding{51} &
        \multirow{1}{5.1em}{\centering \cite{r1-2408095}}\\
        \hline
        \centering Keysight & 
        \centering Meas. & 
        \centering 10.1 &
        \centering 500 &
        \centering \ding{51} &
        \centering \ding{51} &
        \centering \ding{51} &
        \centering \ding{51} &
        \centering \ding{51} &
        \centering \ding{51} &
        \centering \ding{55} &
        \multirow{1}{5.1em}{\centering \cite{r1-2406393,7918569}}\\
        \hline
        \multirow{6}{5.5em}{\centering Ericsson} & 
        \multirow{6}{3.6em}{Meas.} & 
        \centering 0.87 &
        \centering CW &
        \centering \ding{51} &
        \centering \ding{55} &
        \centering \ding{55} &
        \centering \ding{55} &
        \centering \ding{55} &
        \centering \ding{55} &
        \centering \ding{55} &
        \multirow{6}{5.1em}{\centering \cite{r1-2402613,10999643}}\\
        \cline{3-11}
        & 
        & 
        \centering 2.01 &
        \centering CW &
        \centering \ding{51} &
        \centering \ding{55} &
        \centering \ding{55} &
        \centering \ding{55} &
        \centering \ding{55} &
        \centering \ding{55} &
        \centering \ding{55} &
        \\
        \cline{3-11}
        & 
        & 
        \centering 5.02 &
        \centering CW &
        \centering \ding{51} &
        \centering \ding{55} &
        \centering \ding{55} &
        \centering \ding{55} &
        \centering \ding{55} &
        \centering \ding{55} &
        \centering \ding{55} &
        \\
        \cline{3-11}
        & 
        & 
        \centering 10.29 &
        \centering CW &
        \centering \ding{51} &
        \centering \ding{55} &
        \centering \ding{55} &
        \centering \ding{55} &
        \centering \ding{55} &
        \centering \ding{55} &
        \centering \ding{55} &
        \\
        \cline{3-11}
        & 
        & 
        \centering 22 &
        \centering CW &
        \centering \ding{51} &
        \centering \ding{55} &
        \centering \ding{55} &
        \centering \ding{55} &
        \centering \ding{55} &
        \centering \ding{55} &
        \centering \ding{55} &
        \\
        \cline{3-11}
        & 
        & 
        \centering 37 &
        \centering CW &
        \centering \ding{51} &
        \centering \ding{55} &
        \centering \ding{55} &
        \centering \ding{55} &
        \centering \ding{55} &
        \centering \ding{55} &
        \centering \ding{55} &
        \\
        \hline
        \centering Huawei & 
        \centering Meas. & 
        \centering 10 &
        \centering 500 &
        \centering \ding{55} &
        \centering \ding{51} &
        \centering \ding{51} &
        \centering \ding{51} &
        \centering \ding{55} &
        \centering \ding{51} &
        \centering \ding{51} &
        \multirow{1}{5.1em}{\centering \cite{r1-2403925}}\\
        \hline
        \centering Sony & 
        \centering Meas. & 
        \centering 15 &
        \centering 100 &
        \centering \ding{51} &
        \centering \ding{51} &
        \centering \ding{51} &
        \centering \ding{51} &
        \centering \ding{51} &
        \centering \ding{51} &
        \centering \ding{55} &
        \multirow{1}{5.1em}{\centering \cite{8012501,r1-2406485}}\\
\end{longtable}
\end{center}
\vspace{-2.5em}
Moreover, parameters for the absolute ToA were reported by ZTE and Sanechips using RT simulations \cite{R1-2406128}, and by Huawei and HiSilicon based on measurements \cite{r1-2409402}. Additionally, Keysight \cite{r1-2406393} and China Telecom \cite{r1-2400841} provided values for the K-factor based on measurements.
\clearpage

\subsection{UMa}
\begin{center}
\begin{longtable}{|p{5.5em}|p{2.2em}|p{2.7em}|p{3em}|p{3.5em}|p{1.2em}|p{2.5em}|p{1.8em}|p{1.8em}|p{1.8em}|p{4.5em}|p{5.1em}|}
\caption{Meas. and RT data at various Freq. and Band. by various sources for the outdoor to outdoor (O2O) UMa scenario. \ding{51} indicates that the parameter was reported whereas, \ding{55} indicates that the parameter was not reported. Some companies use continuous wave (CW) in measurement. `NA' indicates data was not provided.}
\label{tab:uma_measurements}
\\
\hline
\centering \textbf{Source} & 
\textbf{Type} & 
\textbf{Freq. (GHz)} &
\textbf{Band. (MHz)} &
\textbf{PL/\- Excess PL/SF} & 
\textbf{DS} & 
\textbf{ASD} &
\textbf{ASA} & 
\textbf{ZSA} & 
\textbf{ZSD} & 
\centering \textbf{Number of Clusters} & 
\textbf{References} \\
\hline
\endfirsthead

\hline
\textbf{Source} & 
\textbf{Type} & 
\textbf{Freq. (GHz)} &
\textbf{Band. (MHz)} &
\textbf{PL/\- Excess PL/SF} & 
\textbf{DS} & 
\textbf{ASD} \textsuperscript{a} &
\textbf{ASA} & 
\textbf{ZSA} & 
\textbf{ZSD} & 
\textbf{Number of Clusters} & 
\textbf{References} \\
\hline
\endhead
\hline
\endfoot

\hline
\endfoot
\multicolumn{12}{l}{} \\
\endlastfoot

\multirow{3}{5.5em}{\centering Samsung} & 
\multirow{3}{2em}{\centering Meas.} & 
\centering 6.5 &
\centering 100& 
\centering \ding{55}&
\centering \ding{55}&
\centering \ding{51}& 
\centering \ding{51}&
\centering \ding{55}&
\centering \ding{55}&
\centering \ding{55}&
\multirow{3}{5.1em}{\centering \cite{r1-2404129}}\\
\cline{3-11}
& & \centering 10.5 &
\centering 100& 
\centering \ding{55}&
\centering \ding{55}&
\centering \ding{51}& 
\centering \ding{51}&
\centering \ding{55}&
\centering \ding{55}&
\centering \ding{55}&
\\
\cline{3-11}
 & & \centering 13.5 & 
\centering 100& 
\centering \ding{55}&
\centering \ding{55}&
\centering \ding{51}& 
\centering \ding{51}&
\centering \ding{55}&
\centering \ding{55}&
\centering \ding{55}&
\\
\hline
\multirow{3}{5.5em}{\centering AT\&T} & \multirow{3}{2em}{\centering Meas.} & \centering 7 & \centering 400 & \centering \ding{51} & \centering \ding{51} & \centering \ding{55} & \centering \ding{51} & \centering \ding{51} & \centering \ding{55} & \centering \ding{55} & \multirow{3}{5.1em}{\centering \cite{r1-2501371}} \\
\cline{3-11}
 & & \centering 8 & \centering 400 & \centering \ding{51} & \centering \ding{51} & \centering \ding{55} & \centering \ding{51} & \centering \ding{51} & \centering \ding{55} & \centering \ding{55} &  \\
\cline{3-11}
 & & \centering 15 & \centering 400 & \centering \ding{51} & \centering \ding{51} & \centering \ding{55} & \centering \ding{51} & \centering \ding{51} & \centering \ding{55} & \centering \ding{55} &  \\
\hline
\multirow{2}{5.5em}{\centering Apple} & \centering RT & \centering 8 & \centering NA & \centering \ding{51} & \centering \ding{51} & \centering \ding{51} & \centering \ding{51} & \centering \ding{51} & \centering \ding{55} & \centering \ding{51} & \multirow{1}{5.1em}{\centering \cite{r1-2408484}} \\
\cline{2-12}
 & \centering Meas. & \centering 13 & \centering 100 & \centering \ding{51} & \centering \ding{51} & \centering \ding{55} & \centering \ding{55} & \centering \ding{55} & \centering \ding{55} & \centering \ding{55} & \multirow{1}{5.1em}{\centering \cite{r1-2404304}}  \\
\hline
\centering BUPT, Spark & \vspace{0.03em} \centering Meas. & \vspace{0.03em} \centering 13 & \vspace{0.03em} \centering 400 & \vspace{0.03em} \centering \ding{55} & \vspace{0.03em} \centering \ding{51} & \vspace{0.03em} \centering \ding{51} & \vspace{0.03em} \centering \ding{51} & \vspace{0.03em} \centering \ding{51} & \vspace{0.03em} \centering \ding{51} & \vspace{0.03em} \centering \ding{51} & \vspace{0.03em} \multirow{1}{5.1em}{\centering \cite{r1-2408095,11112530}}  \\
\hline
\centering Vodafone, Ericsson & \vspace{0.05em} \centering Meas. & \vspace{0.05em} \centering 3.4 & \vspace{0.05em} \centering 100 & \vspace{0.05em} \centering \ding{55} & \vspace{0.05em} \centering \ding{55} & \vspace{0.05em} \centering \ding{51} & \vspace{0.05em} \centering \ding{55} & \vspace{0.05em} \centering \ding{55} & \vspace{0.05em} \centering \ding{51} & \vspace{0.05em} \centering \ding{55} & \multirow{2}{5.1em}{\centering \cite{r1-2404521}}  \\
\hline
\multirow{10}{5.5em}{\centering Ericsson} & \multirow{10}{2em}{\centering Meas.} & \centering 3.5 & \centering 100 & \centering \ding{55} & \centering \ding{51} & \centering \ding{51} & \centering \ding{55} & \centering \ding{55} & \centering \ding{51} & \centering \ding{51} & \multirow{1}{5.1em}{\centering \cite{r1-2402613,11000032}} \\
\cline{3-12}
 & & \centering 13 & \centering 50 & \centering \ding{51} & \centering \ding{55} & \centering \ding{51} & \centering \ding{55} & \centering \ding{55} & \centering \ding{51} & \centering \ding{55} & \multirow{2}{5.1em}{\centering \cite{r1-2406717}} \\
\cline{3-11}
 & & \centering 28 & \centering 50 & \centering \ding{51} & \centering \ding{55} & \centering \ding{51} & \centering \ding{55} & \centering \ding{55} & \centering \ding{51} & \centering \ding{55} & \\
\cline{3-12}
 & & \centering 3.4 & \centering 100 & \centering \ding{55} & \centering \ding{55} & \centering \ding{51} & \centering \ding{55} & \centering \ding{55} & \centering \ding{51} & \centering \ding{55} & \multirow{1}{5.1em}{\centering \cite{r1-2404521}} \\
\cline{3-12}
 & & \centering 0.87 & \centering CW & \centering \ding{51} & \centering \ding{55} & \centering \ding{55} & \centering \ding{55} & \centering \ding{55} & \centering \ding{55} & \centering \ding{55} & \multirow{6}{5.1em}{\centering \cite{r1-2402613,10999643}} \\
\cline{3-11}
 & & \centering 2.01 & \centering CW & \centering \ding{51} & \centering \ding{55} & \centering \ding{55} & \centering \ding{55} & \centering \ding{55} & \centering \ding{55} & \centering \ding{55} & \\
\cline{3-11}
 & & \centering 5.02 & \centering CW & \centering \ding{51} & \centering \ding{55} & \centering \ding{55} & \centering \ding{55} & \centering \ding{55} & \centering \ding{55} & \centering \ding{55} & \\
\cline{3-11}
 & & \centering 10.29 & \centering CW & \centering \ding{51} & \centering \ding{55} & \centering \ding{55} & \centering \ding{55} & \centering \ding{55} & \centering \ding{55} & \centering \ding{55} & \\
\cline{3-11}
 & & \centering 22 & \centering CW & \centering \ding{51} & \centering \ding{55} & \centering \ding{55} & \centering \ding{55}  & \centering \ding{55} & \centering \ding{55} & \centering \ding{55} & \\
\cline{3-11}
 & & \centering 37 & \centering CW & \centering \ding{51} & \centering \ding{55} & \centering \ding{55} & \centering \ding{55} & \centering \ding{55} & \centering \ding{55} & \centering \ding{55} & \\
\hline
\multirow{4}{5.5em}{\centering Huawei} & \multirow{4}{2em}{\centering Meas.} & \centering 3.5 & \centering NA & \centering \ding{51} & \centering \ding{55} & \centering \ding{55} & \centering \ding{55} & \centering \ding{55} & \centering \ding{55} & \centering \ding{55} & \multirow{4}{5.1em}{\centering \cite{r1-2403925,r1-2407683,r1-2409402,r1-2405865,r1-2500087}} \\
\cline{3-11}
 & & \centering 6.5 & \centering 160 & \centering \ding{51} & \centering \ding{51} & \centering \ding{51} & \centering \ding{51} & \centering \ding{55} & \centering \ding{51} & \centering \ding{51} &  \\
\cline{3-11}
 & & \centering 13 & \centering 400 & \centering \ding{55} & \centering \ding{51} & \centering \ding{51} & \centering \ding{51} & \centering \ding{51} & \centering \ding{51} & \centering \ding{51} & \\
\cline{3-11}
 & & \centering 15 & \centering 250 & \centering \ding{51} & \centering \ding{51} & \centering \ding{51} & \centering \ding{51} & \centering \ding{55} & \centering \ding{51} & \centering \ding{51} & \\

        \hline
        \centering Sony & 
        \centering Meas. & 
        \centering 15 &
        \centering 100 &
        \centering \ding{51} &
        \centering \ding{51} &
        \centering \ding{51} &
        \centering \ding{51} &
        \centering \ding{51} &
        \centering \ding{51} &
        \centering \ding{55} &
        \multirow{1}{5.1em}{\centering \cite{8012501,r1-2406485}}\\
        \hline
        \multirow{2}{5.5em}{\centering BT, Ericsson} & \multirow{2}{2em}{\centering Meas.} & \centering 3.56 & \centering 40 & \centering \ding{55} & \centering \ding{55} & \centering \ding{51} & \centering \ding{55} & \centering \ding{55} & \centering \ding{51} & \centering \ding{55} & \multirow{2}{5.1em}{\centering \cite{r1-2504650}}\\
\cline{3-11}
 & & \centering 3.7 & \centering 40 & \centering \ding{55} & \centering \ding{55} & \centering \ding{51} & \centering \ding{55} & \centering \ding{55} & \centering \ding{51} & \centering \ding{55} &\\
\hline
\end{longtable}
\end{center}
\vspace{-4em}
Moreover, parameters for the absolute ToA were reported by ZTE and Sanechips based on RT simulations \cite{R1-2406128} and Ericsson \cite{r1-2409441}, Huawei, and HiSilicon \cite{r1-2409402} based on measurements. Additionally, Outdoor-to-Indoor (O2I) ASD and O2I and O2O cluster ASD values were reported by Ericsson \cite{r1-2406717, r1-2402613, r1-2404521, r1-2504650} based on measurements, whereas BUPT \cite{r1-2408095} provided values only for O2O cluster ASD based on measurements.

\subsection{SMa}
\begin{center}
\begin{longtable}{|p{5.5em}|p{2.2em}|p{2.7em}|p{3em}|p{3.5em}|p{1.2em}|p{1.8em}|p{1.8em}|p{1.8em}|p{1.8em}|p{4.5em}|p{5.1em}|}
\caption{Meas. and RT data at various Freq. and Band. by various sources for the outdoor to outdoor (O2O) SMa scenario. \ding{51} indicates that the parameter was reported whereas, \ding{55} indicates that the parameter was not reported. Some companies use continuous wave (CW) in measurement. `NA' indicates data was not provided.}
\label{tab:sma_measurements}
\\
\hline
\centering \textbf{Source} & 
\textbf{Type} & 
\textbf{Freq. (GHz)} & 
\textbf{Band. (MHz)} &
\textbf{PL/\- Excess PL/SF} & 
\textbf{DS} & 
\textbf{ASD} &
\textbf{ASA} & 
\textbf{ZSA} & 
\textbf{ZSD} & 
\centering \textbf{Number of Clusters} & 
\textbf{References} \\
\hline
\endfirsthead

\hline
\centering \textbf{Source} & 
\textbf{Type} & 
\textbf{Freq. (GHz)} & 
\textbf{Band. (MHz)} &
\textbf{PL/\- Excess PL/SF} & 
\textbf{DS} & 
\textbf{ASD} &
\textbf{ASA} & 
\textbf{ZSA} & 
\textbf{ZSD} & 
\centering \textbf{Number of Clusters} & 
\textbf{References} \\
\hline
\endhead
\hline
\endfoot

\hline
\endfoot
%\multicolumn{12}{l}{\textsuperscript{b}.} \\
\endlastfoot

\centering Nokia & Meas. & \centering 28 & \centering 0.02 & \centering \ding{51} & \centering \ding{55} & \centering \ding{55} & \centering \ding{55} & \centering \ding{55} & \centering \ding{55} & \centering \ding{55} & \multicolumn{1}{c|}{\cite{r1-2406139}}\\
\hline
\multirow{3}{5.5em}{\centering AT\&T} & \multirow{3}{2em}{\centering Meas.} & \centering 7 & \centering 400 & \centering \ding{51} & \centering \ding{51} & \centering \ding{55} & \centering \ding{51} & \centering \ding{51} & \centering \ding{55} & \centering \ding{55} & \multirow{3}{5.1em}{\centering \cite{r1-2501371,r1-2410655}} \\
\cline{3-11}
 & & \centering 8 & \centering 400 & \centering \ding{51} & \centering \ding{51} & \centering \ding{55} & \centering \ding{51} & \centering \ding{51} & \centering \ding{55} & \centering \ding{55} &  \\
\cline{3-11}
\centering{AT\&T} & Meas. & \centering 15 & \centering 400 & \centering \ding{51} & \centering \ding{51} & \centering \ding{55} & \centering \ding{51} & \centering \ding{51} & \centering \ding{55} & \centering \ding{55} & \multicolumn{1}{c|}{\cite{r1-2501371,r1-2410655}}\\
\hline
\centering Apple & \centering RT & \centering 8 & \centering NA & \centering \ding{51} & \centering \ding{51} & \centering \ding{51} & \centering \ding{51} & \centering \ding{51} & \centering \ding{55} & \centering \ding{51} & \multicolumn{1}{c|}{\cite{r1-2408484}}\\
\hline
\centering ZTE & \centering RT & \centering 7 & \centering NA & \centering \ding{51} & \centering \ding{55} & \centering \ding{55} & \centering \ding{55} & \centering \ding{55} & \centering \ding{55} & \centering \ding{55} & \multicolumn{1}{c|}{\cite{r1-2404212}}\\
\hline
\multirow{2}{5.5em}{\centering BT, Ericsson} & \multirow{2}{2em}{\centering Meas.} & \centering 3.56 & \centering 40 & \centering \ding{55} & \centering \ding{55} & \centering \ding{51} & \centering \ding{55} & \centering \ding{55} & \centering \ding{51} & \centering \ding{55} & \multirow{2}{5em}{ \centering \cite{r1-2504650}}\\
\cline{3-11}
 & & \centering 3.7 & \centering 40 & \centering \ding{55} & \centering \ding{55} & \centering \ding{51} & \centering \ding{55} & \centering \ding{55} & \centering \ding{51} & \centering \ding{55} &\\
\hline
\centering Vodafone, Ericsson & \vspace{0.05em} \centering Meas. & \vspace{0.05em} \centering 3.4 & \vspace{0.05em} \centering 80 & \vspace{0.05em} \centering \ding{55} & \vspace{0.05em} \centering \ding{55} & \vspace{0.05em} \centering \ding{51} & \vspace{0.05em} \centering \ding{55} & \vspace{0.05em} \centering \ding{55} & \vspace{0.05em} \centering \ding{51} & \vspace{0.05em} \centering \ding{55} & \multirow{2}{5em}{\centering \cite{r1-2407106}} \\
\hline
\multirow{6}{5.5em}{\centering Ericsson} & \multirow{6}{2em}{\centering Meas.} & \centering 0.87 & \centering CW & \centering \ding{51} & \centering \ding{55} & \centering \ding{55} & \centering \ding{55} & \centering \ding{55} & \centering \ding{55} & \centering \ding{55} & \multirow{6}{5.1em}{ \centering \cite{r1-2402613,10999643} }\\
\cline{3-11}
 & & \centering 2.01 & \centering CW  & \centering \ding{51} & \centering \ding{55} & \centering \ding{55} & \centering \ding{55} & \centering \ding{55} & \centering \ding{55} & \centering \ding{55} & \\
\cline{3-11}
 & & \centering 5.02 & \centering CW & \centering \ding{51} & \centering \ding{55} & \centering \ding{55} & \centering \ding{55} & \centering \ding{55} & \centering \ding{55} & \centering \ding{55} &\\
\cline{3-11}
 & & \centering 10.29 & \centering CW & \centering \ding{51} & \centering \ding{55} & \centering \ding{55} & \centering \ding{55} & \centering \ding{55} & \centering \ding{55} & \centering \ding{55} &\\
\cline{3-11}
 & & \centering 22 & \centering CW & \centering \ding{51} & \centering \ding{55} & \centering \ding{55} & \centering \ding{55} & \centering \ding{55} & \centering \ding{55} & \centering \ding{55} &\\
\cline{3-11}
 & & \centering 37 & \centering CW & \centering \ding{51} & \centering \ding{55} & \centering \ding{55} & \centering \ding{55} & \centering \ding{55} & \centering \ding{55} & \centering \ding{55} &\\
        \hline
        \centering Sony & 
        \centering Meas. & 
        \centering 15 &
        \centering 100 &
        \centering \ding{51} &
        \centering \ding{51} &
        \centering \ding{51} &
        \centering \ding{51} &
        \centering \ding{51} &
        \centering \ding{51} &
        \centering \ding{55} &
        \multirow{1}{5.1em}{\centering \cite{8012501,r1-2406485}}\\
    \hline
\end{longtable}
\end{center}
\vspace{-2.5em}
Moreover, parameters for the absolute ToA and both O2I and O2O cluster ASD were provided by Ericsson based on measurements \cite{r1-2502878, r1-2404521}. Additionally, K-factor values were reported by NTT DOCOMO based on measurements \cite{r1-2502777} and by ZTE and Sanechips based on RT simulations \cite{r1-2503633}.

\subsection{RMa}
\begin{center}
\begin{longtable}{|p{5.5em}|p{2.2em}|p{2.7em}|p{3em}|p{3.5em}|p{1.2em}|p{1.8em}|p{1.8em}|p{1.8em}|p{1.8em}|p{4.5em}|p{5.1em}|}
\caption{Meas. and RT data at various Freq. and Band. by various sources for the outdoor to outdoor (O2O) RMa scenario. \ding{51} indicates that the parameter was reported whereas, \ding{55} indicates that the parameter was not reported. Some companies use continuous wave (CW) in measurement. `NA' indicates data was not provided.}
\label{tab:rma_measurements}
\\
\hline
\centering \textbf{Source} & 
\textbf{Type} & 
\textbf{Freq. (GHz)} & 
\textbf{Band. (MHz)} &
\textbf{PL/\- Excess PL/SF} & 
\textbf{DS} & 
\textbf{ASD} &
\textbf{ASA} & 
\textbf{ZSA} & 
\textbf{ZSD} & 
\centering \textbf{Number of Clusters} & 
\textbf{References} \\
\hline
\endfirsthead

\hline
\textbf{Source} & 
\textbf{Type} & 
\textbf{Freq. (GHz)} & 
\textbf{Band. (MHz)} &
\textbf{PL/\- Excess PL/SF} & 
\textbf{DS} & 
\textbf{ASD} &
\textbf{ASA} & 
\textbf{ZSA} & 
\textbf{ZSD} & 
\textbf{Number of Clusters} & 
\textbf{References} \\
\hline
\endhead
\hline
\endfoot

\multirow{2}{5.5em}{\centering Qual\-comm} & \multirow{2}{2em}{\centering Meas.} & \centering 3.4 & \centering CW & \centering \ding{51} & \centering \ding{55} & \centering \ding{55} & \centering \ding{55} & \centering \ding{55} & \centering \ding{55} & \centering \ding{55} &  \multirow{2}{5.1em}{\centering \cite{r1-2405169}} \\
\cline{3-11}
& & \centering 13 & \centering CW & \centering \ding{51} & \centering \ding{55} & \centering \ding{55} & \centering \ding{55} & \centering \ding{55} & \centering \ding{55} & \centering \ding{55} & \\
\hline
\multirow{2}{5.5em}{\centering BT, Ericsson} & \multirow{2}{2em}{\centering Meas.} & \centering 3.56 & \centering 40 & \centering \ding{55} & \centering \ding{55} & \centering \ding{51} & \centering \ding{55} & \centering \ding{55} & \centering \ding{51} & \centering \ding{55} & \multirow{2}{5.1em}{\centering \cite{r1-2504650}}\\
\cline{3-11}
 & & \centering 3.7 & \centering 40 & \centering \ding{55} & \centering \ding{55} & \centering \ding{51} & \centering \ding{55} & \centering \ding{55} & \centering \ding{51} & \centering \ding{55} & \\
\end{longtable}
\end{center}
\vspace{-2.5em}
Moreover, parameters for the absolute ToA were reported by ZTE and Sanechips based on RT simulations \cite{r1-2501897}.

\subsection{InH}
\begin{center}
\begin{longtable}{|p{5.5em}|p{2.2em}|p{2.7em}|p{3em}|p{3.5em}|p{1.2em}|p{1.8em}|p{1.8em}|p{1.8em}|p{1.8em}|p{4.5em}|p{5.1em}|}
\caption{Meas. and RT data at various Freq. and Band. by various sources for the InH scenario. \ding{51} indicates that the parameter was reported whereas, \ding{55} indicates that the parameter was not reported. Some companies use continuous wave (CW) in measurement. `NA' indicates data was not provided.}
\label{tab:inh_measurements}
\\
\hline
\centering \textbf{Source} & 
\textbf{Type} & 
\textbf{Freq. (GHz)} & 
\textbf{Band. (MHz)} &
\textbf{PL/\- Excess PL/SF} & 
\textbf{DS} & 
\textbf{ASD} &
\textbf{ASA} & 
\textbf{ZSA} & 
\textbf{ZSD} & 
\centering \textbf{Number of Clusters} & 
\textbf{References} \\
\hline
\endfirsthead

\hline
\centering \textbf{Source} & 
\textbf{Type} & 
\textbf{Freq. (GHz)} & 
\textbf{Band. (MHz)} &
\textbf{PL/\- Excess PL/SF} & 
\textbf{DS} & 
\textbf{ASD} &
\textbf{ASA} & 
\textbf{ZSA} & 
\textbf{ZSD} & 
\centering \textbf{Number of Clusters} & 
\textbf{References} \\
\hline
\endhead
\hline
\endfoot
\multirow{2}{5.5em}{\centering Sharp, NYU, Nokia} & 
        \multirow{2}{3.6em}{Meas.} & 
        \centering 6.75 &
        \centering 1000 &
        \centering \ding{51} &
        \centering \ding{51} &
        \centering \ding{51} &
        \centering \ding{51} &
        \centering \ding{51} &
        \centering \ding{51} &
        \centering \ding{55} &
        \multirow{2}{5.1em}{\centering \cite{r1-2410319, shakya2024propagation, shakya2024comprehensive, poddar2025validation, poddar2025single}}\\
        \cline{3-11}
        & 
        & 
        \centering 16.95 &
        \centering 1000 &
        \centering \ding{51} &
        \centering \ding{51} &
        \centering \ding{51} &
        \centering \ding{51} &
        \centering \ding{51} &
        \centering \ding{51} &
        \centering \ding{55} &
        \\
        \hline
        \centering Rohde \& Schwarz & \vspace{0.05em} Meas. & \vspace{0.05em} \centering 14 & \vspace{0.05em} \centering 2000 & \vspace{0.05em} \centering \ding{51} &  \vspace{0.05em} \centering \ding{51} &  \vspace{0.05em} \centering \ding{51} &  \vspace{0.05em} \centering \ding{51} &  \vspace{0.05em} \centering \ding{51} &  \vspace{0.05em} \centering \ding{51} &  \vspace{0.05em} \centering \ding{51} & \multirow{2}{5.1em}{\centering \cite{r1-2410254,11190468}}  \\
        \hline
        \multirow{2}{5.5em}{\centering Sony} & 
        Meas. & 
        \centering 15 &
        \centering 100 &
        \centering \ding{51} &
        \centering \ding{51} &
        \centering \ding{55} &
        \centering \ding{55} &
        \centering \ding{55} &
        \centering \ding{55} &
        \centering \ding{55} &
        \multirow{2}{5.1em}{\centering \cite{r1-2404514,7915373}}
        \\
        \cline{2-11}
        & 
        \centering RT & 
        \centering 15 &
        \centering 100 &
        \centering \ding{51} &
        \centering \ding{51} &
        \centering \ding{55} &
        \centering \ding{55} &
        \centering \ding{55} &
        \centering \ding{55} &
        \centering \ding{55} &
        \\
        \hline
        \multirow{4}{5.5em}{\centering AT\&T} & \multirow{4}{2em}{\centering Meas.} & \centering 7 & \centering 400 & \centering \ding{51} & \centering \ding{51} & \centering \ding{55} & \centering \ding{51} & \centering \ding{51} & \centering \ding{55} & \centering \ding{55} & \multirow{4}{5.1em}{\centering \cite{r1-2410655,r1-2406869}}\\
        \cline{3-11}
 & & \centering 8 & \centering 400 & \centering \ding{51} & \centering \ding{51} & \centering \ding{55} & \centering \ding{51} & \centering \ding{51} & \centering \ding{55} & \centering \ding{55} &  \\
\cline{3-11}
 & & \centering 11 & \centering 400 & \centering \ding{51} & \centering \ding{51} & \centering \ding{55} & \centering \ding{51} & \centering \ding{51} & \centering \ding{55} & \centering \ding{55} & \\
\cline{3-11}
 & & \centering 15 & \centering 400 & \centering \ding{51} & \centering \ding{51} & \centering \ding{55} & \centering \ding{51} & \centering \ding{51} & \centering \ding{55} & \centering \ding{55} & \\
\hline
        \centering Apple & Meas. & \centering 13 & \centering 550  & \centering \ding{51} & \centering \ding{51} & \centering \ding{55} & \centering \ding{55} & \centering \ding{55} & \centering \ding{55} & \centering \ding{55} & \multicolumn{1}{c|}{ \cite{r1-2500798}} \\
        \hline
        \centering ZTE & Meas. & \centering 6-10 & \centering NA & \centering \ding{55} & \centering \ding{51} & \centering \ding{55} & \centering \ding{55} & \centering \ding{55} & \centering \ding{55} & \centering \ding{55} & \multicolumn{1}{c|}{ \cite{r1-2404212}} \\
        \hline
        \centering BUPT, Spark & 
        \centering Meas. & 
        \centering 13 &
        \centering 400 &
        \centering \ding{55} &
        \centering \ding{51}&
        \centering \ding{51}&
        \centering \ding{51}&
        \centering \ding{51}&
        \centering \ding{51}&
        \centering \ding{51}&
        \multicolumn{1}{c|}{ \cite{r1-2408095}}
        \\
        \hline
        \centering Keysight & 
        \centering Meas. & 
        \centering 10.1 &
        \centering 500 &
        \centering \ding{51}&
        \centering \ding{51}&
        \centering \ding{51}&
        \centering \ding{51}&
        \centering \ding{51}&
        \centering \ding{51}&
        \centering \ding{55}&
        \multicolumn{1}{c|}{ \cite{r1-2406393,7497530}}
        \\
        \hline
        \centering Huawei & 
        \centering Meas. & 
        \centering 10 &
        \centering 500 &
        \centering \ding{55}&
        \centering \ding{51}&
        \centering \ding{51}&
        \centering \ding{51}&
        \centering \ding{55}&
        \centering \ding{51}&
        \centering \ding{51}&
        \multicolumn{1}{c|}{ \cite{r1-2402009,r1-2403925}}
        \\
        \hline
\end{longtable}
\end{center}
\vspace{-2.5em}
Moreover, parameters for the absolute ToA based on RT simulations were reported by Ericsson \cite{r1-2500506}, ZTE, and Sanechips \cite{r1-2409724} and measurements from Huawei and HiSilicon \cite{r1-2409402}. Additionally, Keysight \cite{r1-2406393} and Rohde \& Schwarz \cite{r1-2410254} provide values for the K-factor based on measurement. 

\subsection{InF}
\begin{center}
\begin{longtable}{|p{5.5em}|p{2.2em}|p{2.7em}|p{3em}|p{3.5em}|p{1.2em}|p{1.8em}|p{1.8em}|p{1.8em}|p{1.8em}|p{4.5em}|p{5.1em}|}
\caption{Meas. and RT data at various Freq. and Band. by various sources for the InF scenario. \ding{51} indicates that the parameter was reported whereas, \ding{55} indicates that the parameter was not reported. Some companies use continuous wave (CW) in measurement. `NA' indicates data was not provided.}
\label{tab:inf_measurements}
\\
\hline
\centering \textbf{Source} & 
\textbf{Type} & 
\textbf{Freq. (GHz)} & 
\textbf{Band. (MHz)} &
\textbf{PL/\- Excess PL/SF} & 
\textbf{DS} & 
\textbf{ASD} &
\textbf{ASA} & 
\textbf{ZSA} & 
\textbf{ZSD} & 
\centering \textbf{Number of Clusters} & 
\textbf{References} \\
\hline
\endfirsthead

\hline
\textbf{Source} & 
\textbf{Type} & 
\textbf{Freq. (GHz)} & 
\textbf{Band. (MHz)} &
\textbf{PL/\- Excess PL/SF} & 
\textbf{DS} & 
\textbf{ASD} &
\textbf{ASA} & 
\textbf{ZSA} & 
\textbf{ZSD} & 
\textbf{Number of Clusters} & 
\textbf{References} \\
\hline
\endhead
\hline
\endfoot
\multirow{2}{5.5em}{\centering Sharp, NYU, Nokia} & 
        \multirow{2}{3.6em}{Meas.} & 
        \centering 6.75 &
        \centering 1000 &
        \centering \ding{51} &
        \centering \ding{51} &
        \centering \ding{51} &
        \centering \ding{51} &
        \centering \ding{51} &
        \centering \ding{51} &
        \centering \ding{55} &
        \multirow{2}{5.1em}{\centering \cite{r1-2410319, ying2024upper}}\\
        \cline{3-10}
        & 
        & 
        \centering 16.95 &
        \centering 1000 &
        \centering \ding{51} &
        \centering \ding{51} &
        \centering \ding{51} &
        \centering \ding{51} &
        \centering \ding{51} &
        \centering \ding{51} &
        \centering \ding{55} &
        \\
    \hline
        \multirow{3}{5.5em}{\centering Nokia, Anritsu} & 
        \multirow{3}{3.6em}{Meas.} & 
        \centering 3.5 &
        \centering 2000 &
        \centering \ding{51} &
        \centering \ding{51} &
        \centering \ding{55} &
        \centering \ding{55} &
        \centering \ding{55} &
        \centering \ding{55} &
        \centering \ding{55} &
        \multicolumn{1}{c|}{\multirow{3}{2em}{ \cite{r1-2403996}}}\\
        \cline{3-10}
        & 
        & 
        \centering 11 &
        \centering 2000 &
        \centering \ding{51} &
        \centering \ding{51} &
        \centering \ding{55} &
        \centering \ding{55} &
        \centering \ding{55} &
        \centering \ding{55} &
        \centering \ding{55} &
        \\
        \cline{3-10}
        & 
        & 
        \centering 29 &
        \centering 2000 &
        \centering \ding{51} &
        \centering \ding{51} &
        \centering \ding{55} &
        \centering \ding{55} &
        \centering \ding{55} &
        \centering \ding{55} &
        \centering \ding{55} &
        \\
        \hline
        \centering Apple & Meas. & \centering 13 & \centering 100 & \centering \ding{51} & \centering \ding{55} & \centering \ding{55} & \centering \ding{55} & \centering \ding{55} & \centering \ding{55}  & \centering \ding{55} & \multicolumn{1}{c|}{\cite{r1-2406858}} \\
\end{longtable}
\end{center}

\subsection{Penetration Loss Measurements}
\begin{longtable}{|p{5.5em}|p{2em}|p{3.5em}|p{2.5em}|p{5em}|p{2em}|p{4em}|p{3em}|p{3em}|p{5em}|}
\caption{Meas. and RT data at various Freq. and Band. by various sources for the penetration loss of various commonly found materials. \ding{51} indicates that the parameter was reported whereas, \ding{55} indicates that the parameter was not reported. Some companies use continuous wave (CW) in measurement. `NA' indicates data was not provided.}
\label{tab:inf_measurements}
\\
\hline
\centering \textbf{Source} & 
\textbf{Type} & 
\textbf{Freq. (GHz)} & 
\textbf{Band. (MHz)} &
\centering \textbf{Standard multi-pane glass} & 
\textbf{IRR glass} & 
\textbf{Concrete} &
\textbf{Wood}&
\textbf{Ply \-wood}&
\textbf{References} \\
\hline
\endfirsthead

\hline
\centering \textbf{Source} & 
\textbf{Type} & 
\textbf{Freq. (GHz)} & 
\textbf{Band. (MHz)} &
\centering \textbf{Standard multi-pane glass} & 
\textbf{IRR glass} & 
\textbf{Concrete} &
\textbf{Wood}& 
\textbf{Ply\-wood}&
\textbf{References} \\
\hline
\endhead
\hline
\endfoot

\multirow{2}{5.5em}{\centering Sharp, NYU} & 
        \multirow{2}{3.6em}{Meas.} & 
        \centering 6.75 &
        \centering 1000 &
        \centering \ding{51} &
        \centering \ding{51} &
        \centering \ding{51} &
        \centering \ding{51} &
        \centering \ding{51} &
        \multirow{2}{5.1em}{\centering \cite{r1-2405339,10901400}}\\
        \cline{3-9}
        & 
        & 
        \centering 16.95 &
        \centering 1000 &
        \centering \ding{51} &
        \centering \ding{51} &
        \centering \ding{51} &
        \centering \ding{51} &
        \centering \ding{51} &
        \\
    \hline

\multirow{2}{5.5em}{\centering Qualcomm} & 
        \multirow{2}{3.6em}{Meas.} & 
        \centering 3.4 &
        \centering CW &
        \centering \ding{51} &
        \centering \ding{51} &
        \centering \ding{55} &
        \centering \ding{55} &
        \centering \ding{55} &
        \multirow{2}{5.1em}{\centering \cite{r1-2405170}}\\
        \cline{3-9}
        & 
        & 
        \centering 13 &
        \centering CW &
        \centering \ding{51} &
        \centering \ding{51} &
        \centering \ding{55} &
        \centering \ding{55} &
        \centering \ding{55} &
        \\
    \hline

     \centering Nokia &  Meas. & \centering 28 & \centering 0.02 & \centering \ding{51} & \centering \ding{55} & \centering \ding{51} & \centering \ding{51} &
     \centering \ding{55} &
     \multirow{1}{5.1em}{\centering \cite{r1-2406139}} \\ 
     \hline  
    
    \vspace{1em} \centering Apple & 
    \vspace{1em}\centering Meas. & 
    \centering 7.5-30 (Step size: 0.5)&
    \vspace{1em}\centering NA &
    \vspace{1em}\centering \ding{51} &
    \vspace{1em}\centering \ding{51} &
    \vspace{1em}\centering \ding{51} &
    \vspace{1em}\centering \ding{51} &
    \vspace{1em}\centering \ding{51} &
     \multirow{4}{5.1em}{\centering \cite{r1-2409819}}
    \\
    \hline
    
    \vspace{0.05em} \centering ZTE & 
    \vspace{0.05em}  \centering RT & 
    \centering 6-24 (Step size: 1)&
    \vspace{0.05em} \centering NA &
    \vspace{0.05em} \centering \ding{51} &
    \vspace{0.05em} \centering \ding{51} &
    \vspace{0.05em} \centering \ding{51} &
    \vspace{0.05em} \centering \ding{51} &
    \vspace{0.05em} \centering \ding{55} &
     \multirow{3}{5.1em}{\centering \cite{r1-2404212}}
    \\
\end{longtable}
\label{tab:mat_pen_loss_meas}

\clearpage

%% file: meas_dataset.tex
The official measurement dataset used by 3GPP for the Rel-19 study can be accessed at:
\url{https://www.3gpp.org/ftp/tsg_ran/wg1_rl1/TSGR1_121/Inbox/R1-2504960.zip}
 The downloaded R1-2504960.zip archive contains the following two files:
\begin{itemize}
\item 2504960 7-24GHz data source.docx
\item CM data source\_v041.xlsx
\end{itemize}

\par The Word document titled “2504960 7-24GHz data source.docx” serves as a cover document for the Excel spreadsheet “CM data source\_v041.xlsx” and provides a high-level summary of the objective of the study focussed on validate using measurements the channel model of 3GPP TR 38.901~\cite{tr38901v18} at least for 7-24 GHz and presents which channel parameters were validated, updated, or determined to remain unchanged.

\par
\begin{figure}[p]
\centering
\includegraphics[width=\textwidth]{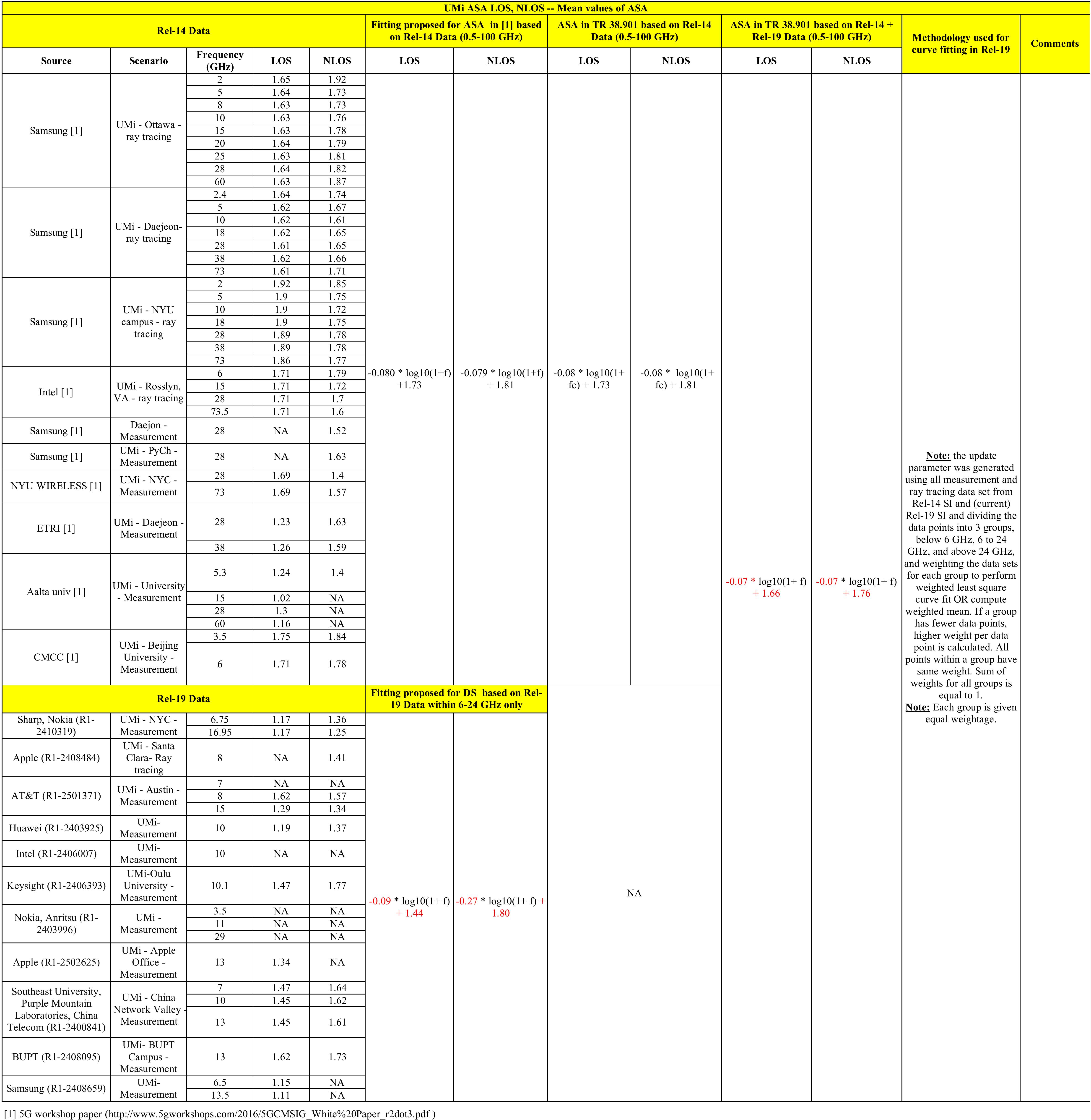}
\caption{Example of an Excel sheet corresponding to the data values for UMi Mean DS. Each source independently performed its own data processing and reported only the final statistics (e.g., mean delay spread) for the UMi scenario.}
\label{fig:sample_umi_meas_ds}
\end{figure}

 The Excel document “CM data source\_v041.xlsx” contains references to the measurement/simulation provided for the Rel-19 SI comprising two objectives as stated in Section \ref{sec:intro} and the datasets for only the channel parameters that resulted in updates to 3GPP TR 38.901~\cite{tr38901v18}. Although additional datasets were submitted for validation of other channel parameters, they did not lead to updates and are therefore not included in this Excel document. This Excel document contains 29 distinct sheets, and follow the naming format \textit{Scenario\_Statistic\_Parameter}, where \textit{Scenario} denotes the propagation scenario (e.g., UMi, UMa, SMa), \textit{Statistic} denotes the statistical measure such as the Mean or standard deviation (Std), and \textit{Parameter} denotes the channel parameter (e.g., DS, ASA, ASD, ZSA) or only the parameter name such as number of cluster (\#cluster) or plywood penetration loss. The datasets in each of the Excel sheets aggregate the reported statistics from multiple contributing sources, the final curve-fitting results over these statistics (with visual representations wherever applicable shown in Section~\ref{sec:curve_fit_results}), the methodology used for curve fitting, and any additional comments, all presented in a self-explanatory tabular format (an example is shown in Fig.~\ref{fig:sample_umi_meas_ds}). The sheets in the Excel document are listed in the following order.
\begin{enumerate}
    \item List of References: Contains detailed measurement and simulation descriptions by various sources for the overall study on channel modeling enhancements for the 7–24 GHz frequency range for NR~\cite{rp234018}. This study comprises of two objectives as described in Section \ref{sec:intro}. As this document only focuses on the objective related to validate using measurements the channel model of 3GPP TR 38.901~\cite{tr38901v18} at least for 7-24 GHz, the relevant information associated with this particular objective was extracted from this Excel sheet and summarized in Section \ref{sec:meas_camp}.
    \item UMi\_Mean\_DS
    \item UMi\_Std\_DS
    \item UMi\_Mean\_ASA
    \item UMi\_Std\_ASA
    \item UMa\_Mean\_DS
    \item UMa\_Std\_DS
    \item UMa\_Mean\_ASD
    \item UMa\_Std\_ASD
    \item UMa\_Mean\_ASA
    \item UMa\_Std\_ASA
    \item UMa\_Mean\_ZSA
    \item UMa\_Std\_ZSA
    \item UMi\_Mean\_ASD
    \item UMi\_Std\_ASD
    \item UMi\_Mean\_ZSA
    \item UMi\_Std\_ZSA
    \item \#cluster
    \item Plywood\_Pen\_Loss
    \item SMa\_Mean\_DS
    \item SMa\_Std\_DS
    \item SMa\_Mean\_ASA
    \item SMa\_Std\_ASA
    \item SMa\_Mean\_ZSA
    \item SMa\_Std\_ZSA
    \item SMa\_Mean\_ASD
    \item SMa\_Std\_ASD
    \item SMa\_Mean\_Cluster\_ASD
    \item UMa\_Mean\_Cluster\_ASD
\end{enumerate}

\clearpage

%% file: curve_fit_results.tex
% \clearpage

\subsection{UMi-Street Canyon DS}
\begin{figure*}[h]
    \centering
    \begin{subfigure}[b]{0.48\textwidth}
        \centering
        \includegraphics[width=\linewidth]{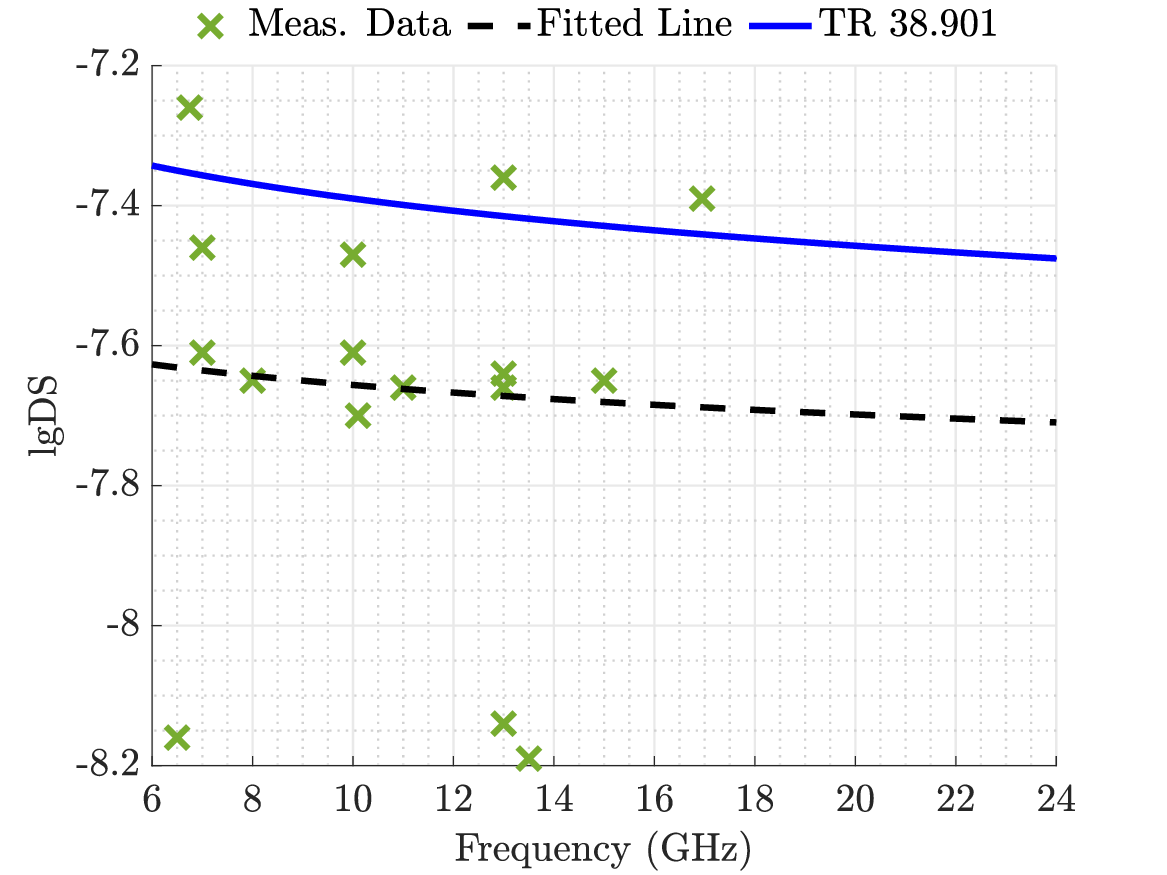}
        \caption{Meas. data for mean of lgDS in the UMi LOS scenario over 6–24 GHz from 3GPP Rel-19, with an OLS fitted line (-0.15 log$_{10}$(1+f) - 7.50~\cite{r1-2502415}) and the 3GPP TR 38.901 model (-0.24 log$_{10}$(1+f) - 7.14~\cite{tr38901v18}).}
        \label{fig:umi_los_mean_ds_rel19}
    \end{subfigure}
    \hfill
    \begin{subfigure}[b]{0.48\textwidth}
        \centering
        \includegraphics[width=\linewidth]{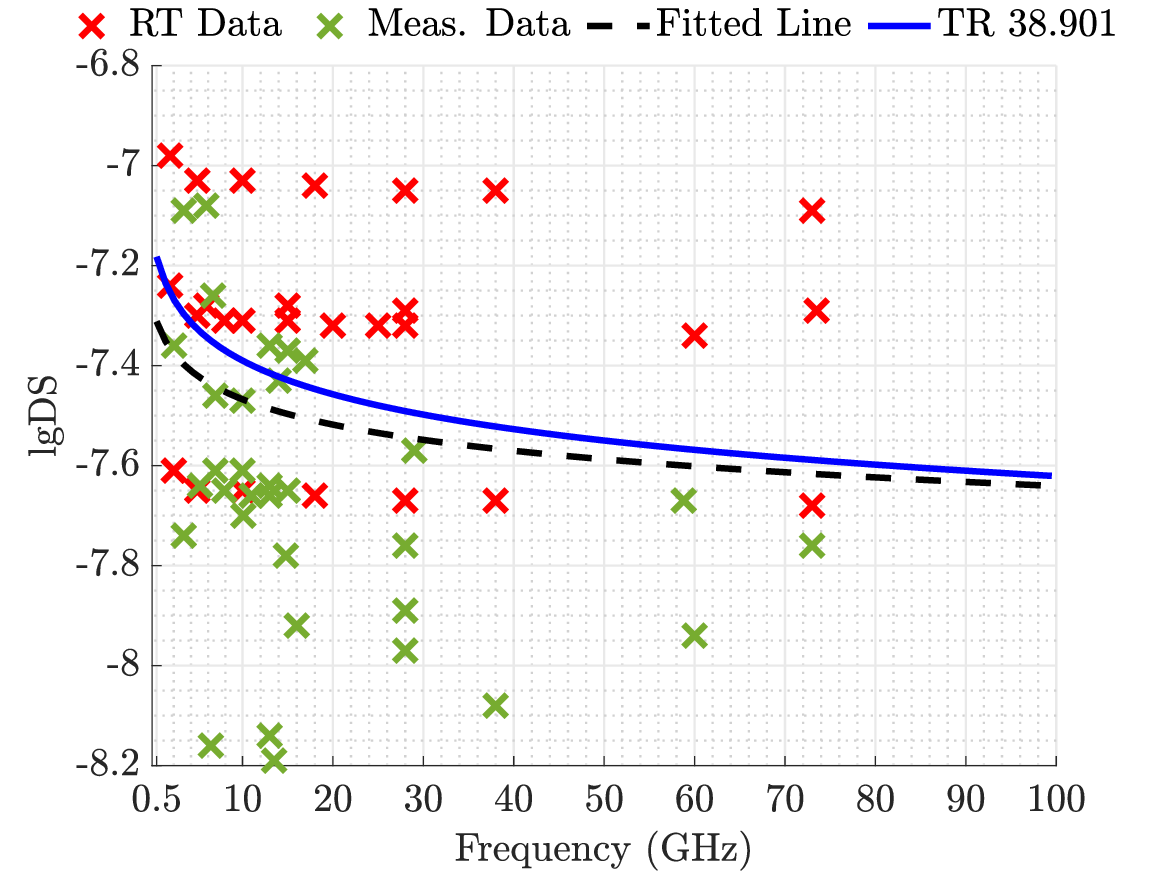}
        \caption{RT and Meas. data for mean of lgDS in the UMi LOS scenario over 0.5–100 GHz from 3GPP Rel-14 and Rel-19, with a WLS fitted line (-0.18 log$_{10}$(1+f) - 7.28~\cite{r1-2502415}) and the 3GPP TR 38.901 model (-0.24 log$_{10}$(1+f) - 7.14\cite{tr38901v18}).}
        \label{fig:umi_los_mean_ds_wls}
    \end{subfigure}
    
    \vspace{1em}

    \begin{subfigure}[b]{0.48\textwidth}
        \centering
        \includegraphics[width=\linewidth]{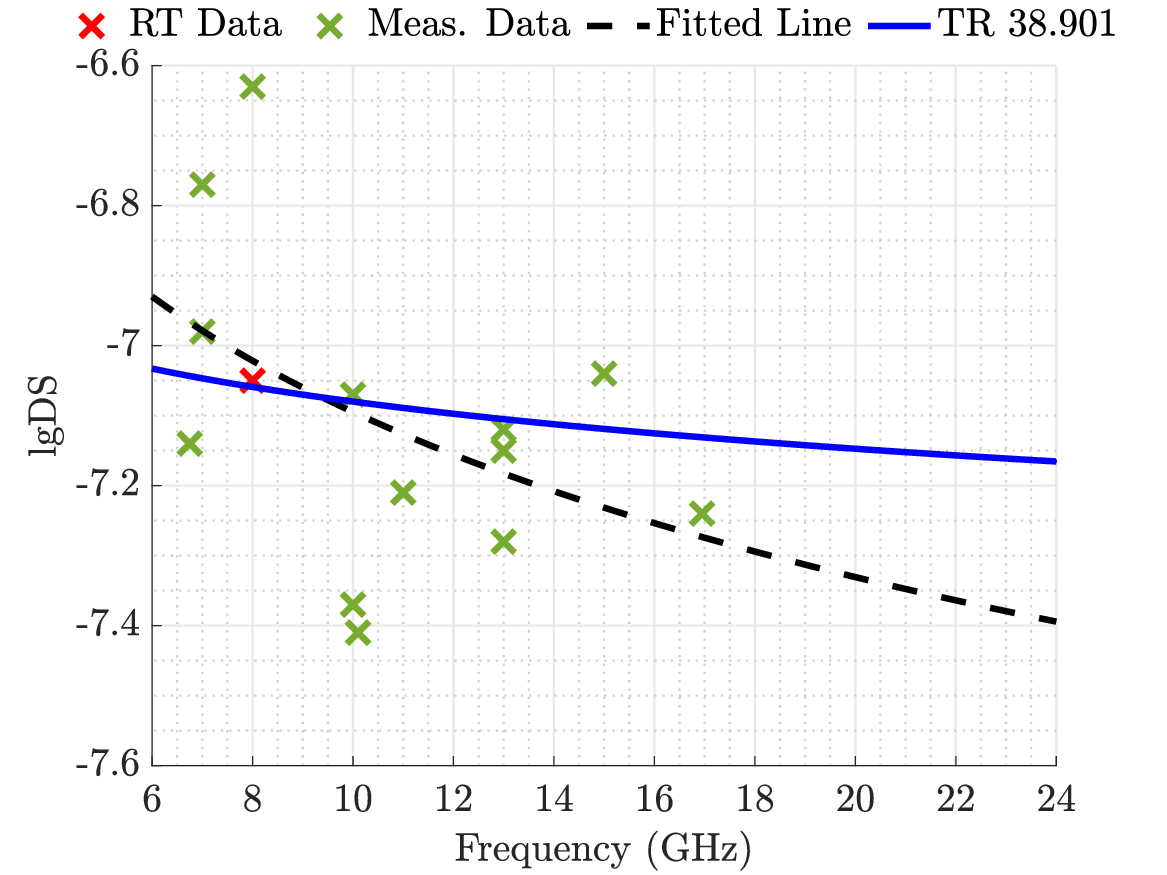}
        \caption{RT and Meas. data for mean of lgDS in the UMi NLOS scenario over 6–24 GHz from 3GPP Rel-19, with an OLS fitted line (-0.84 log$_{10}$(1+f) - 6.22~\cite{r1-2502415}) and the 3GPP TR 38.901 model (-0.24 log$_{10}$(1+f) - 6.83\cite{tr38901v18}).}
        \label{fig:umi_nlos_mean_ds_rel19}
    \end{subfigure}
    \hfill
    \begin{subfigure}[b]{0.48\textwidth}
        \centering
        \includegraphics[width=\linewidth]{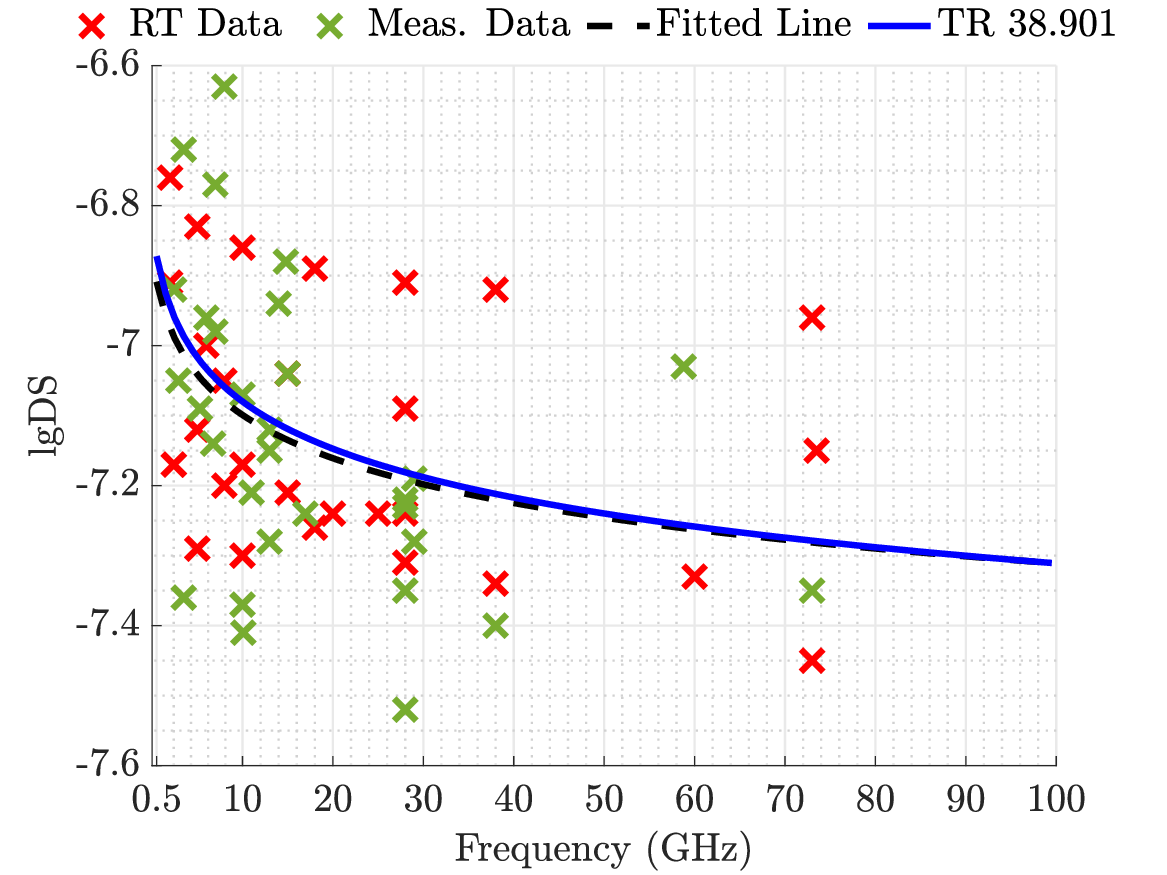}
        \caption{RT and Meas. data for mean of lgDS in the UMi NLOS scenario over 0.5–100 GHz from 3GPP Rel-14 and Rel-19, with a WLS fitted line (-0.22 log$_{10}$(1+f) - 6.87~\cite{r1-2502415}) and the 3GPP TR 38.901 model (-0.24 log$_{10}$(1+f) - 6.83\cite{tr38901v18}).}
        \label{fig:umi_nlos_mean_ds_wls}
    \end{subfigure}

    \caption{Curve fitting of RT and Meas. data for the mean of lgDS in UMi LOS and NLOS channel conditions OLS and WLS methods. (a) and (c) use data from 3GPP Rel-19 only (6–24 GHz), while (b) and (d) use combined data from Rel-14 and Rel-19 (0.5–100 GHz). All subfigures include a comparison with the existing 3GPP TR 38.901 model.}
    \label{fig:umi_mean_ds}
\end{figure*}

%%%%%%%%%%%% UMi DS STD %%%%%%%%%%%%%%
\begin{figure*}[h]
    \centering
    \begin{subfigure}[b]{0.48\textwidth}
        \centering
        \includegraphics[width=\linewidth]{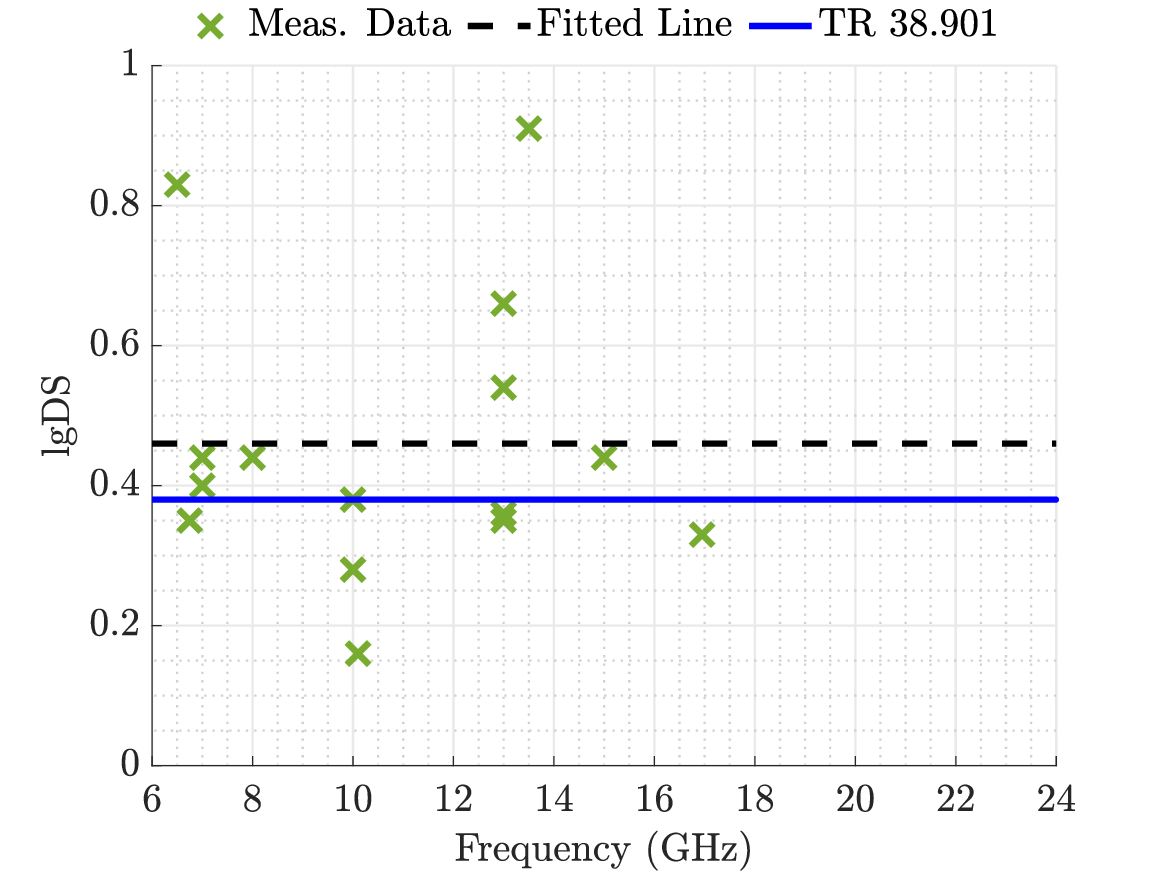}
        \caption{Meas. data for standard deviation of lgDS in the UMi LOS scenario over 6–24 GHz from 3GPP Rel-19, with an AM fitted line (0.46~\cite{r1-2502415}) and the 3GPP TR 38.901 model (0.38 \cite{tr38901v18}).}
        \label{fig:umi_los_std_ds_rel19}
    \end{subfigure}
    \hfill
    \begin{subfigure}[b]{0.48\textwidth}
        \centering
        \includegraphics[width=\linewidth]{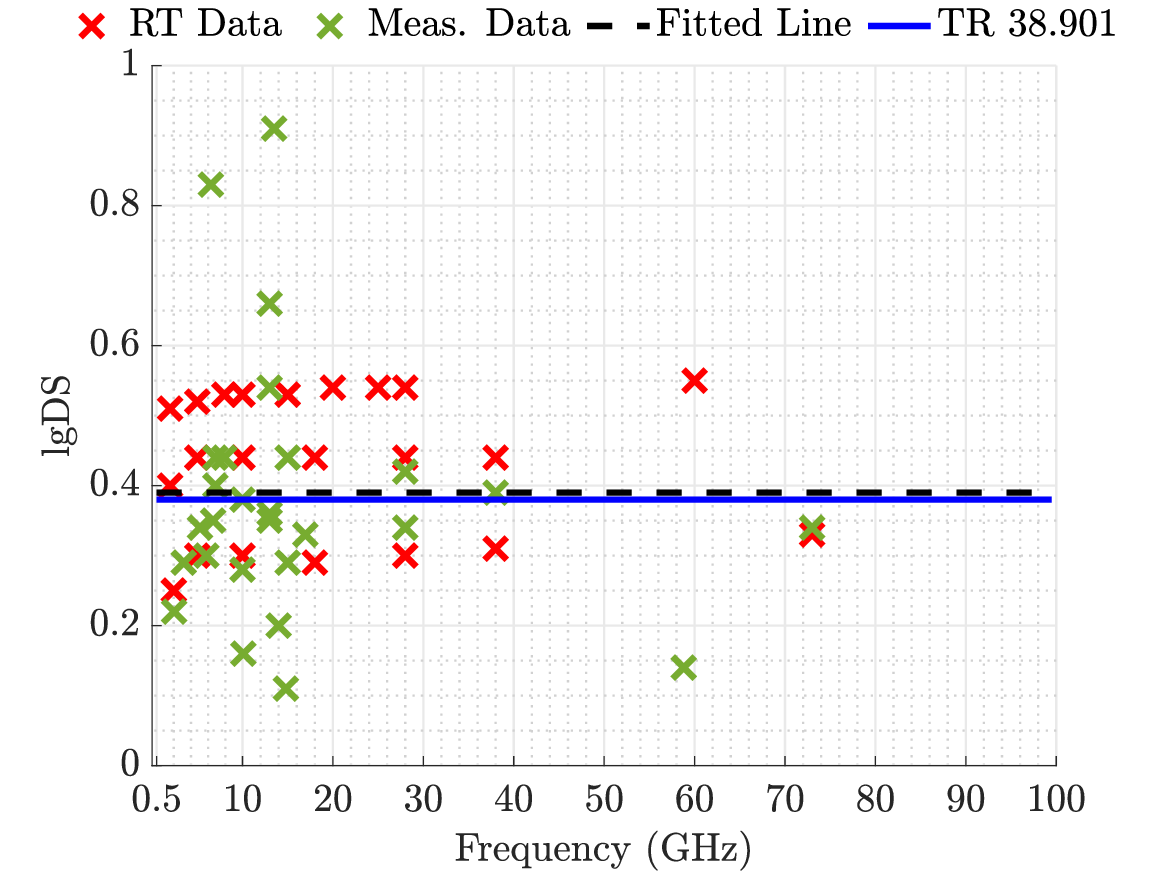}
        \caption{RT and Meas. data for standard deviation of lgDS in the UMi LOS scenario over 0.5-100 GHz from 3GPP Rel-14 and Rel-19, with a WM fitted line (0.39 \cite{r1-2502415}) and the 3GPP TR 38.901 model (0.38 \cite{tr38901v18}).}
        \label{fig:umi_los_std_ds_wls}
    \end{subfigure}
    
    \vspace{1em}

    \begin{subfigure}[b]{0.48\textwidth}
        \centering
        \includegraphics[width=\linewidth]{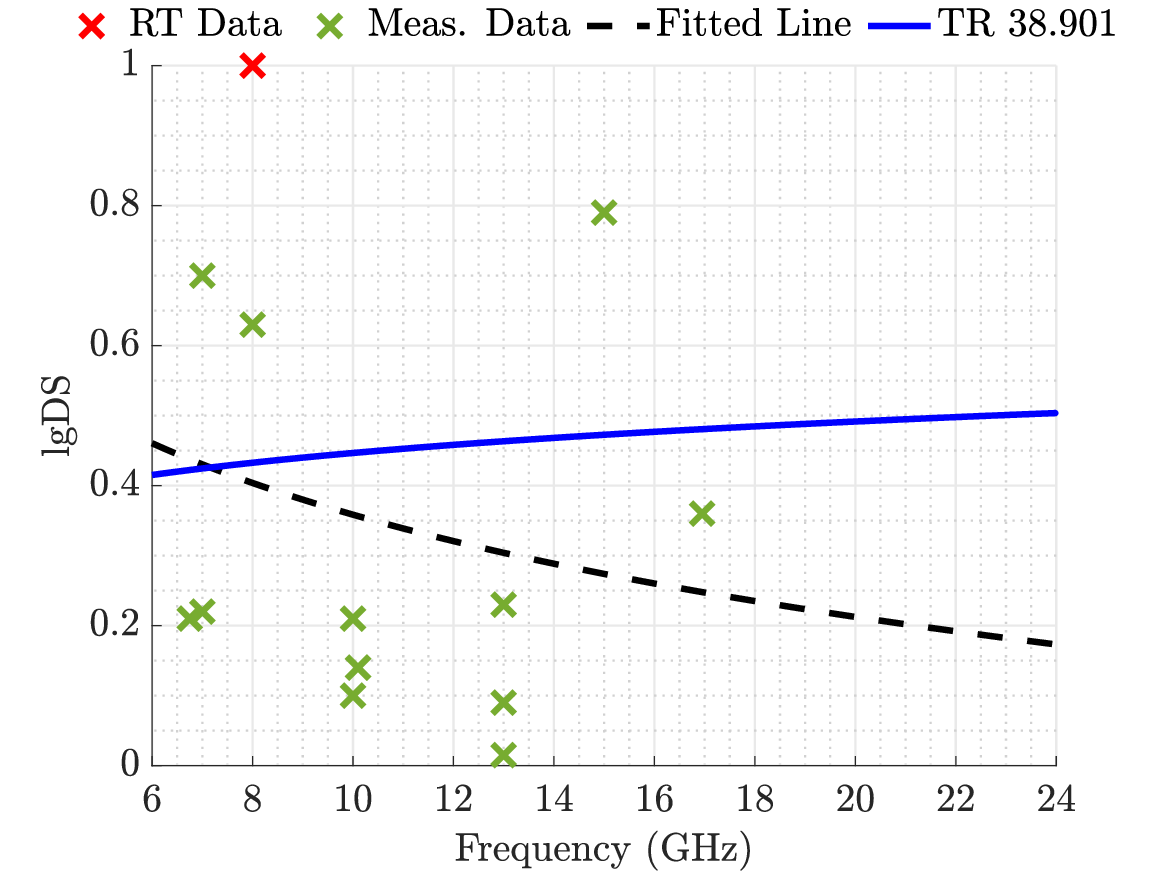}
        \caption{RT and Meas. data for standard deviation of lgDS in the UMi NLOS scenario over 6–24 GHz from 3GPP Rel-19, with an OLS fitted line (-0.52 log$_{10}$(1+f) + 0.90~\cite{r1-2502415}) and the 3GPP TR 38.901 model (0.16 log$_{10}$(1+f) + 0.28 \cite{tr38901v18}).}
        \label{fig:umi_nlos_std_ds_rel19}
    \end{subfigure}
    \hfill
    \begin{subfigure}[b]{0.48\textwidth}
        \centering
        \includegraphics[width=\linewidth]{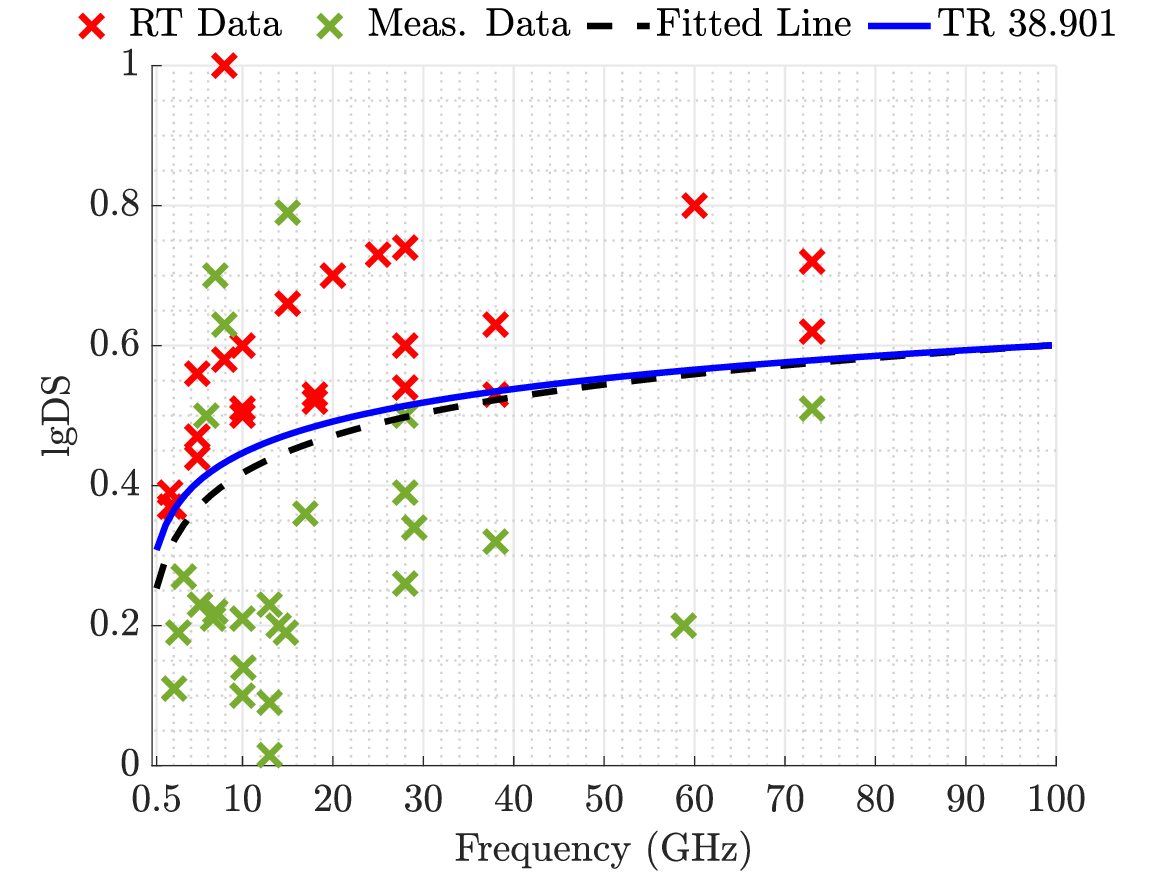}
        \caption{RT and Meas. data for standard deviation of lgDS in the UMi NLOS scenario over 0.5–100 GHz from 3GPP Rel-14 and Rel-19, with a WLS fitted line (0.19 log$_{10}$(1+f) + 0.22 \cite{r1-2502415}) and the 3GPP TR 38.901 model (0.16 log$_{10}$(1+f) + 0.28  \cite{tr38901v18}).}
        \label{fig:umi_nlos_std_ds_wls}
    \end{subfigure}

    \caption{Curve fitting of RT and Meas. data for the standard deviation of lgDS in UMi LOS and NLOS channel conditions using arithmetic mean (AM), weighted mean (WM), ordinary least squares (OLS) and weighted least squares (WLS) methods. (a) and (c) use data from 3GPP Rel-19 only (6–24 GHz), while (b) and (d) use combined data from Rel-14 and Rel-19 (0.5–100 GHz). All subfigures include a comparison with the existing 3GPP TR 38.901 model.}
    \label{fig:umi_std_ds}
\end{figure*}

\clearpage
\subsection{UMi-Street Canyon ASD}
\begin{figure*}[h]
    \centering
    \begin{subfigure}[b]{0.48\textwidth}
        \centering
        \includegraphics[width=\linewidth]{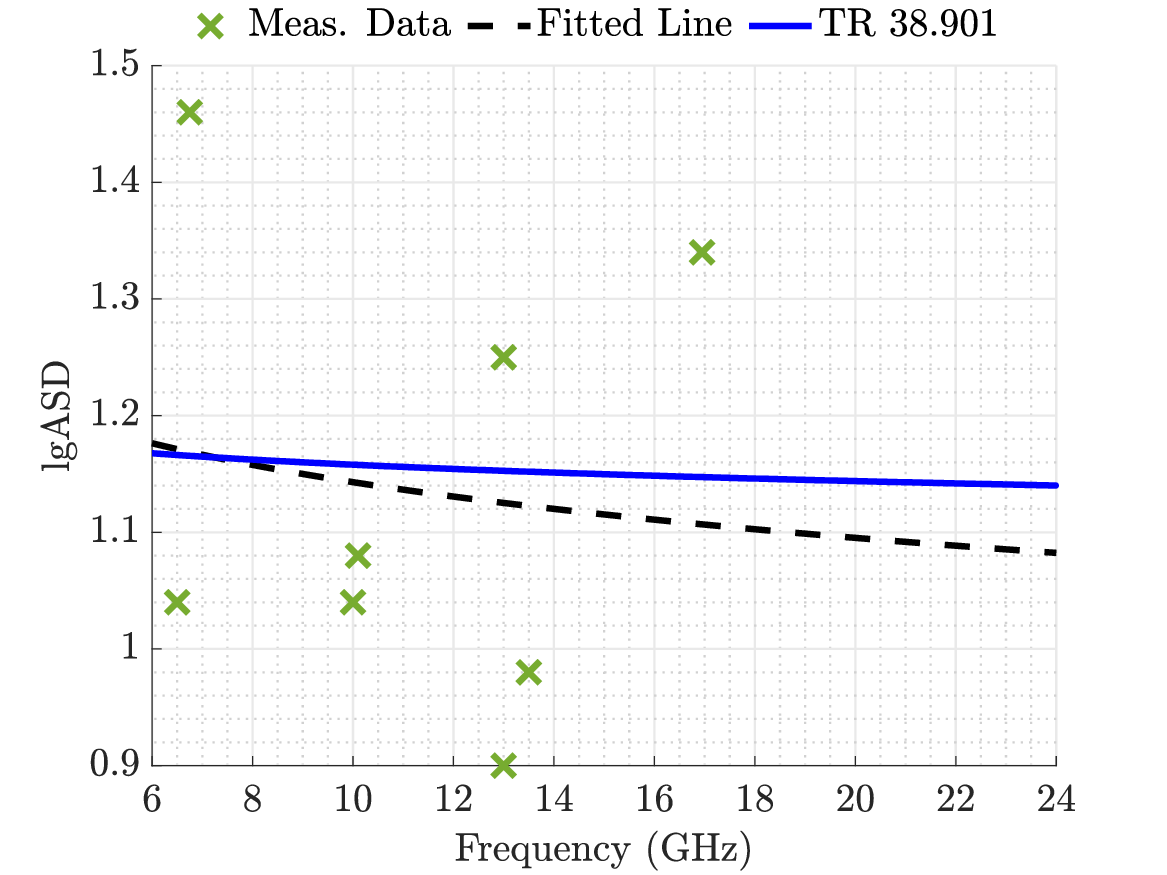}
        \caption{Meas. data for mean of lgASD in the UMi LOS scenario over 6–24 GHz from 3GPP Rel-19, with an OLS fitted line (-0.17 log$_{10}$(1+ f) + 1.32~\cite{r1-2502415}) and the 3GPP TR 38.901 model (-0.05 log$_{10}$(1+f) + 1.21~\cite{tr38901v18}).}
        \label{fig:umi_los_mean_asd_rel19}
    \end{subfigure}
    \hfill
    \begin{subfigure}[b]{0.48\textwidth}
        \centering
        \includegraphics[width=\linewidth]{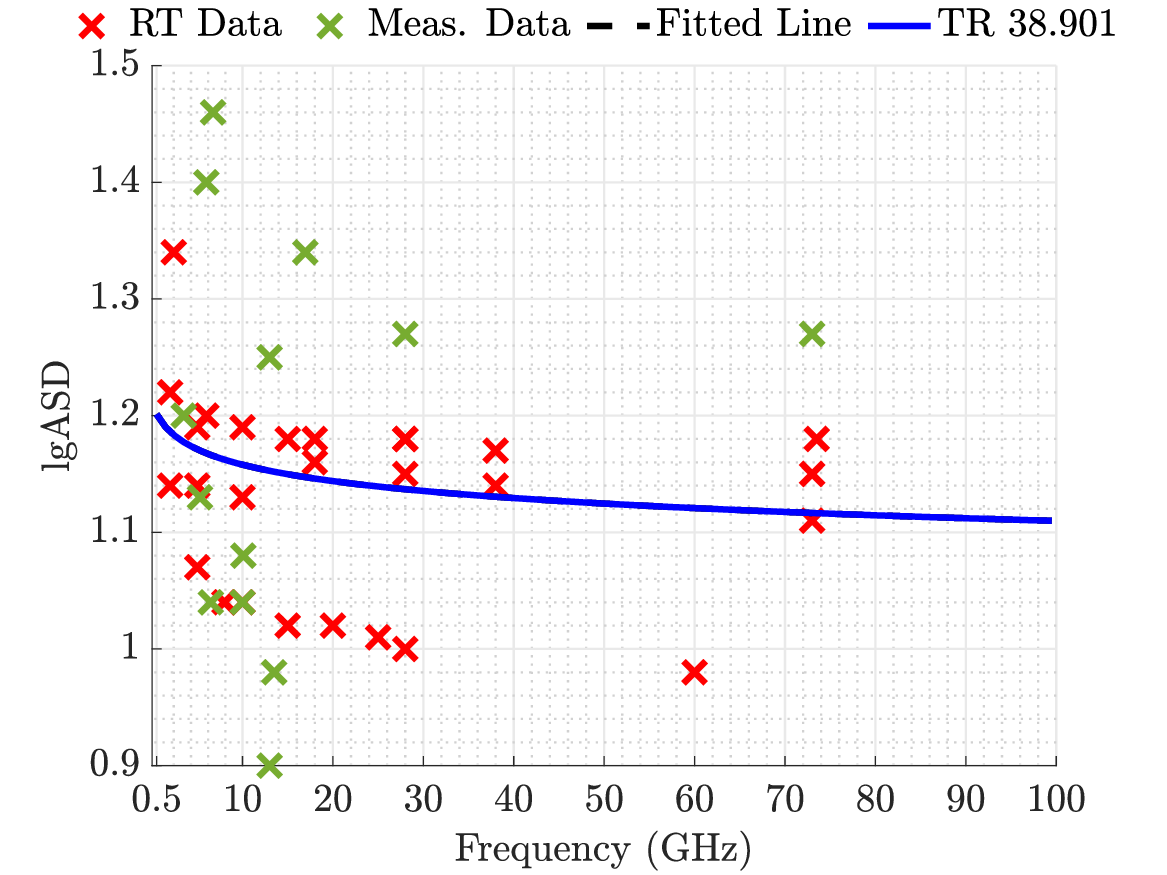}
        \caption{RT and Meas. data for mean of lgASD in the UMi LOS scenario over 0.5-100 GHz from 3GPP Rel-14 and Rel-19, with a WLS fitted line (-0.05 log$_{10}$(1+ f) + 1.21~\cite{r1-2502415}) and the 3GPP TR 38.901 model (-0.05 log$_{10}$(1+f) + 1.21~\cite{tr38901v18}).}
        \label{fig:umi_los_mean_asd_wls}
    \end{subfigure}
    
    \vspace{1em}

    \begin{subfigure}[b]{0.48\textwidth}
        \centering
        \includegraphics[width=\linewidth]{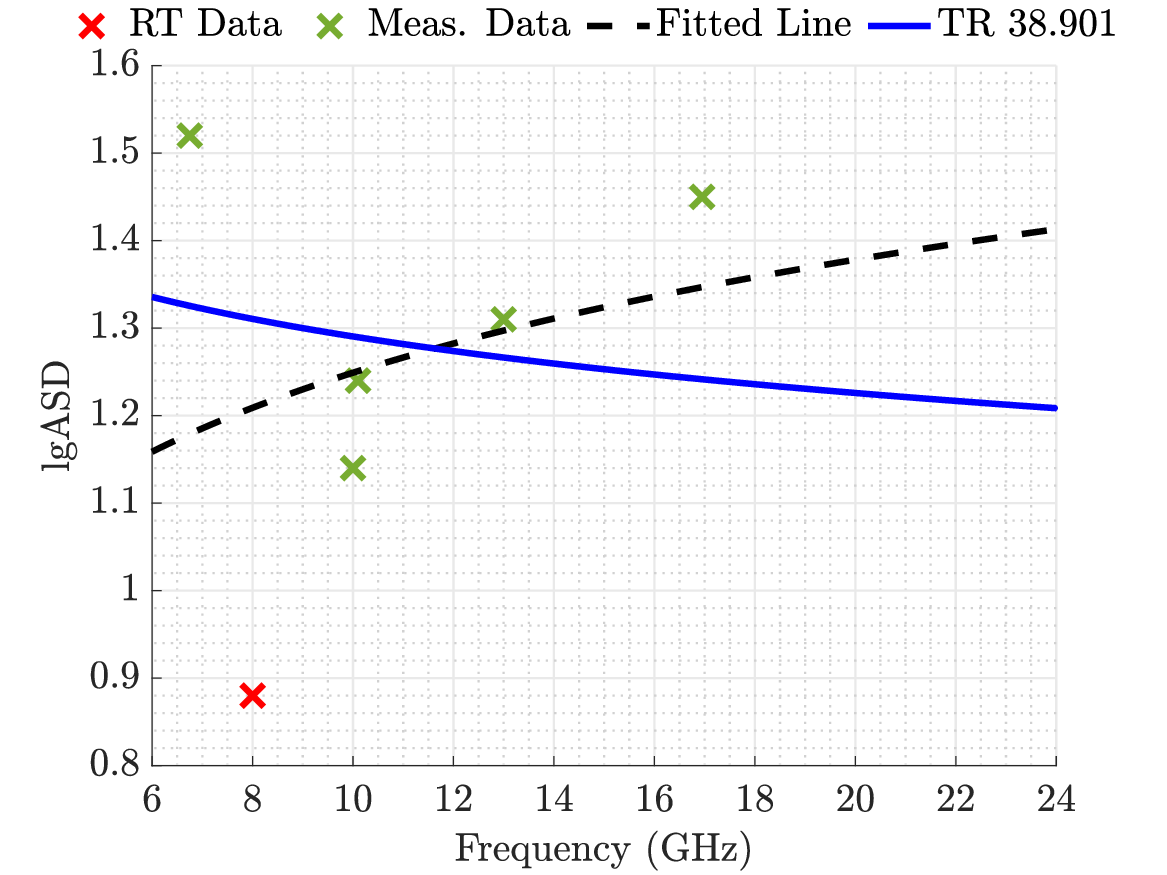}
        \caption{RT and Meas. data for mean of lgASD in the UMi NLOS scenario over 6–24 GHz from 3GPP Rel-19, with an OLS fitted line (0.46 log$_{10}$(1+ f) + 0.77~\cite{r1-2502415}) and the 3GPP TR 38.901 model (-0.23 log$_{10}$(1+f) + 1.53~\cite{tr38901v18}).}
        \label{fig:umi_nlos_mean_asd_rel19}
    \end{subfigure}
    \hfill
    \begin{subfigure}[b]{0.48\textwidth}
        \centering
        \includegraphics[width=\linewidth]{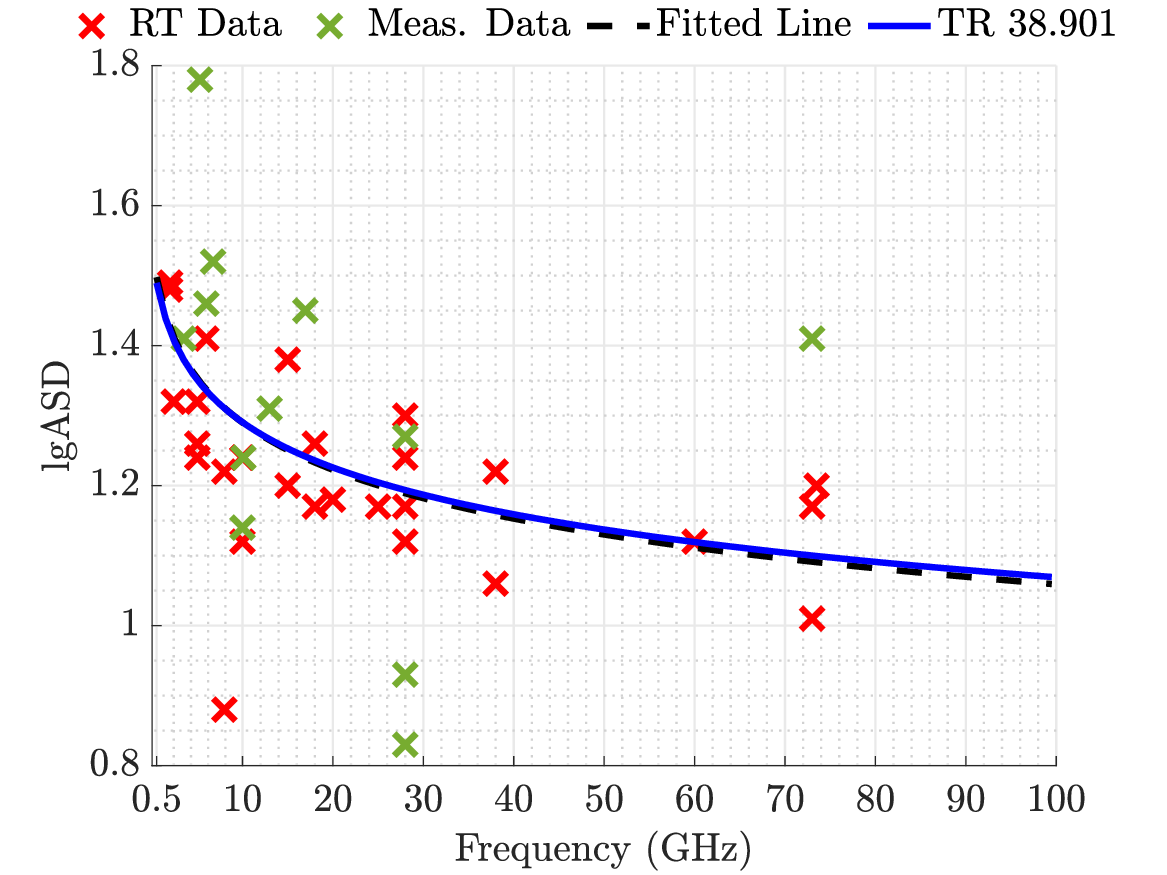}
        \caption{RT and Meas. data for mean of lgASD in the UMi NLOS scenario over 0.5-100 GHz from 3GPP Rel-14 and Rel-19, with a WLS fitted line (-0.24 log$_{10}$(1+f) + 1.54~\cite{r1-2502415}) and the 3GPP TR 38.901 model (-0.23 log$_{10}$(1+f) + 1.53~\cite{tr38901v18}).}
        \label{fig:umi_nlos_mean_asd_wls}
    \end{subfigure}

    \caption{Curve fitting of RT and Meas. data for the mean of lgASD in UMi LOS and NLOS channel conditions OLS and WLS methods. (a) and (c) use data from 3GPP Rel-19 only (6–24 GHz), while (b) and (d) use combined data from Rel-14 and Rel-19 (0.5–100 GHz). All subfigures include a comparison with the existing 3GPP TR 38.901 model.}
    \label{fig:umi_mean_asd}
\end{figure*}

%%%%%%%%%%%% UMi ASD STD %%%%%%%%%%%%
\begin{figure*}[h]
    \centering
    \begin{subfigure}[b]{0.48\textwidth}
        \centering
        \includegraphics[width=\linewidth]{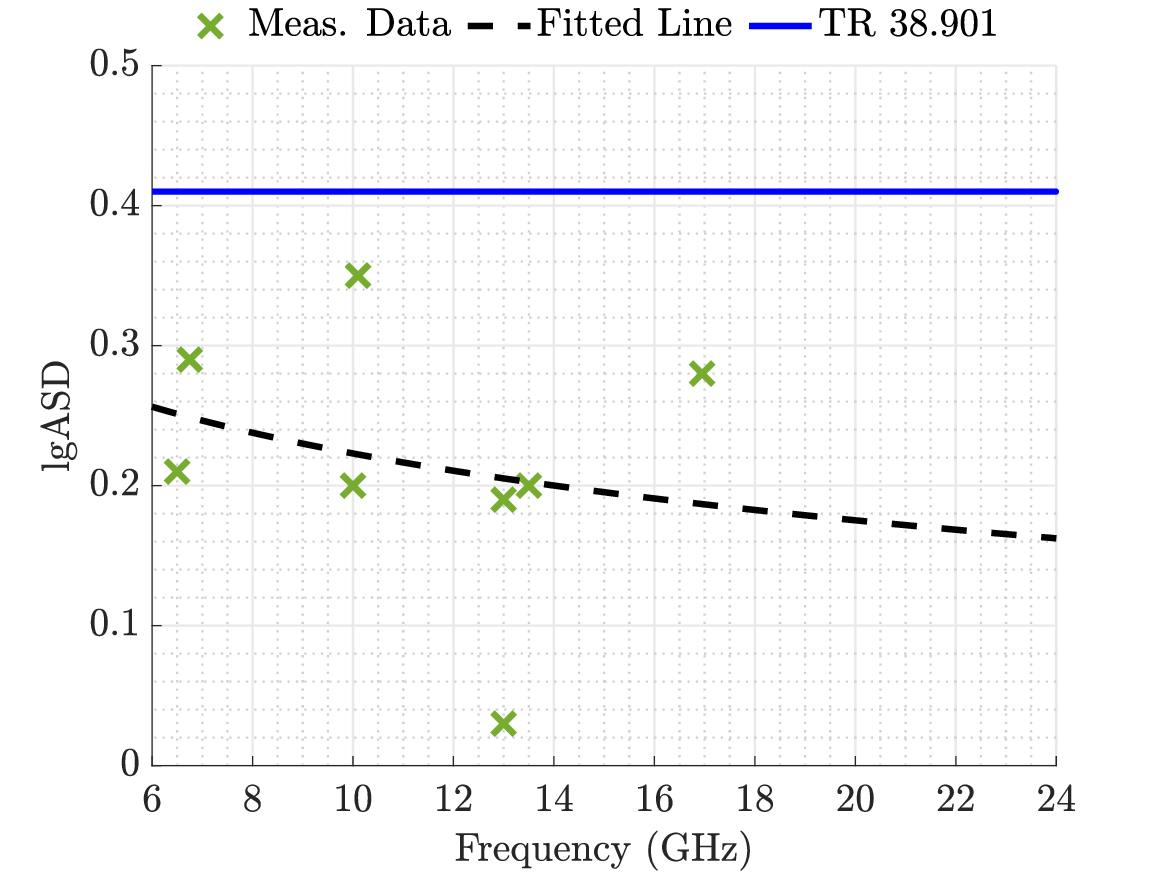}
        \caption{Meas. data for standard deviation of lgASD in the UMi LOS scenario over 6–24 GHz from 3GPP Rel-19, with an OLS fitted line (-0.17 log$_{10}$(1+f) + 0.40~\cite{r1-2502415}) and the 3GPP TR 38.901 model (0.41~\cite{tr38901v18}).}
        \label{fig:umi_los_std_asd_rel19}
    \end{subfigure}
    \hfill
    \begin{subfigure}[b]{0.48\textwidth}
        \centering
        \includegraphics[width=\linewidth]{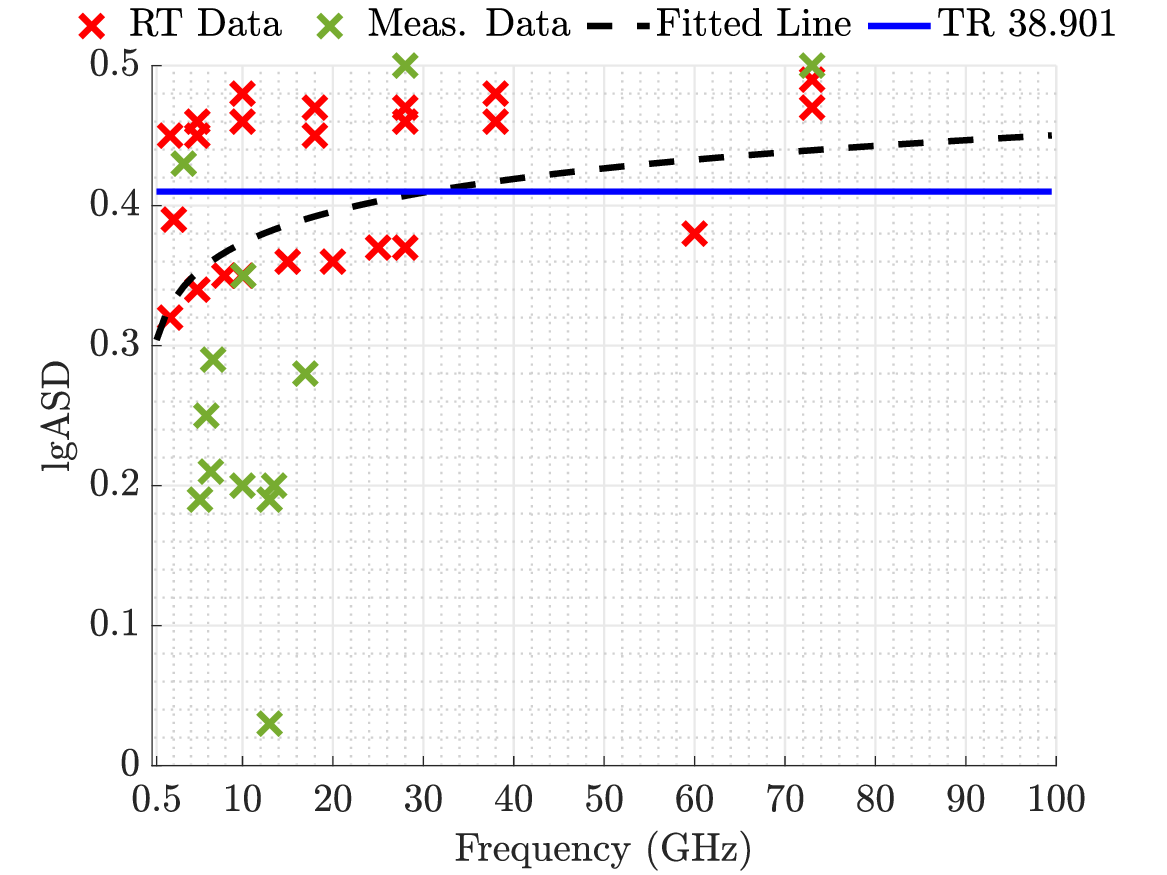}
        \caption{RT and Meas. data for standard deviation of lgASD in the UMi LOS scenario over 0.5-100 GHz from 3GPP Rel-14 and Rel-19, with a WLS fitted line (0.08 log$_{10}$(1+f) + 0.29~\cite{r1-2502415}) and the 3GPP TR 38.901 model (0.41~\cite{tr38901v18}).}
        \label{fig:umi_los_std_asd_wls}
    \end{subfigure}
    
    \vspace{1em}

    \begin{subfigure}[b]{0.48\textwidth}
        \centering
        \includegraphics[width=\linewidth]{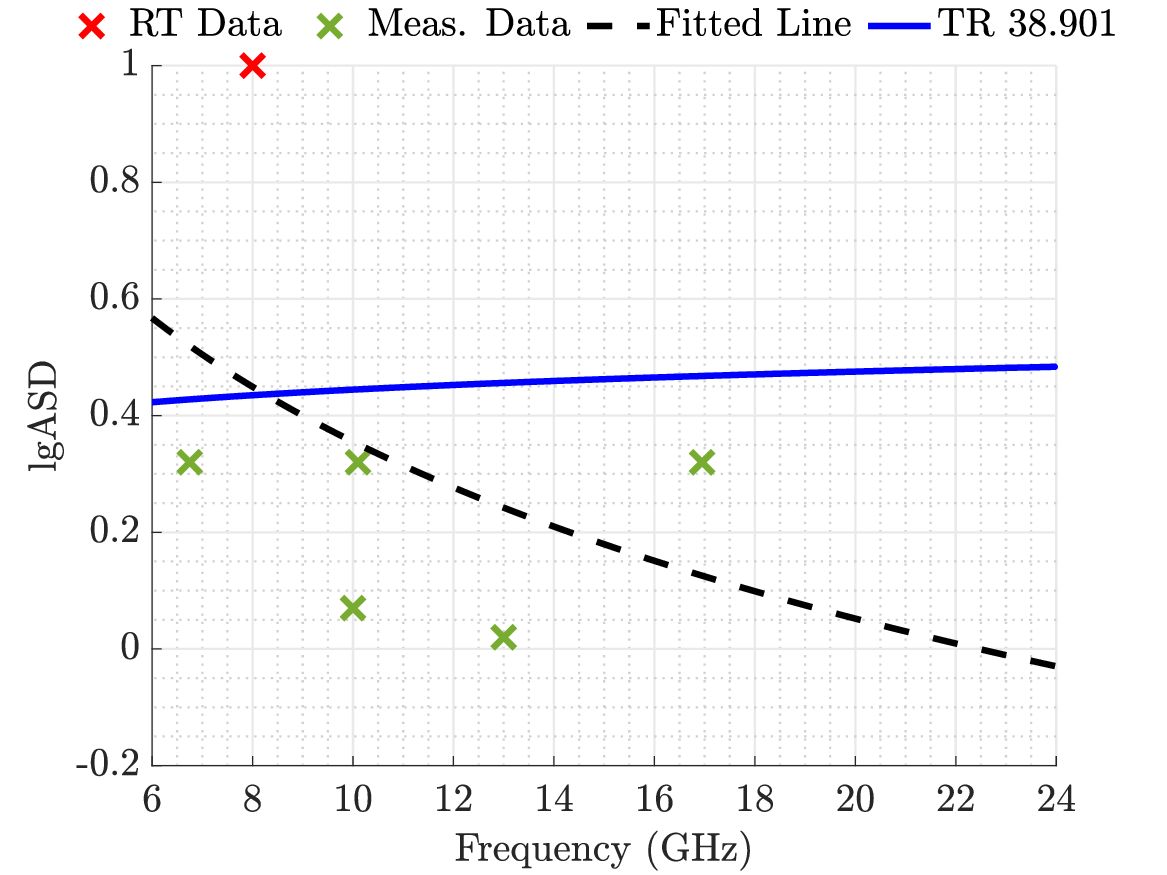}
        \caption{RT and Meas. data for standard deviation of lgASD in the UMi NLOS scenario over 6–24 GHz from 3GPP Rel-19, with an OLS fitted line (-1.08 log$_{10}$(1+f) + 1.48~\cite{r1-2502415}) and the 3GPP TR 38.901 model (0.11 log$_{10}$(1+f) + 0.33~\cite{tr38901v18}).}
        \label{fig:umi_nlos_std_asd_rel19}
    \end{subfigure}
    \hfill
    \begin{subfigure}[b]{0.48\textwidth}
        \centering
        \includegraphics[width=\linewidth]{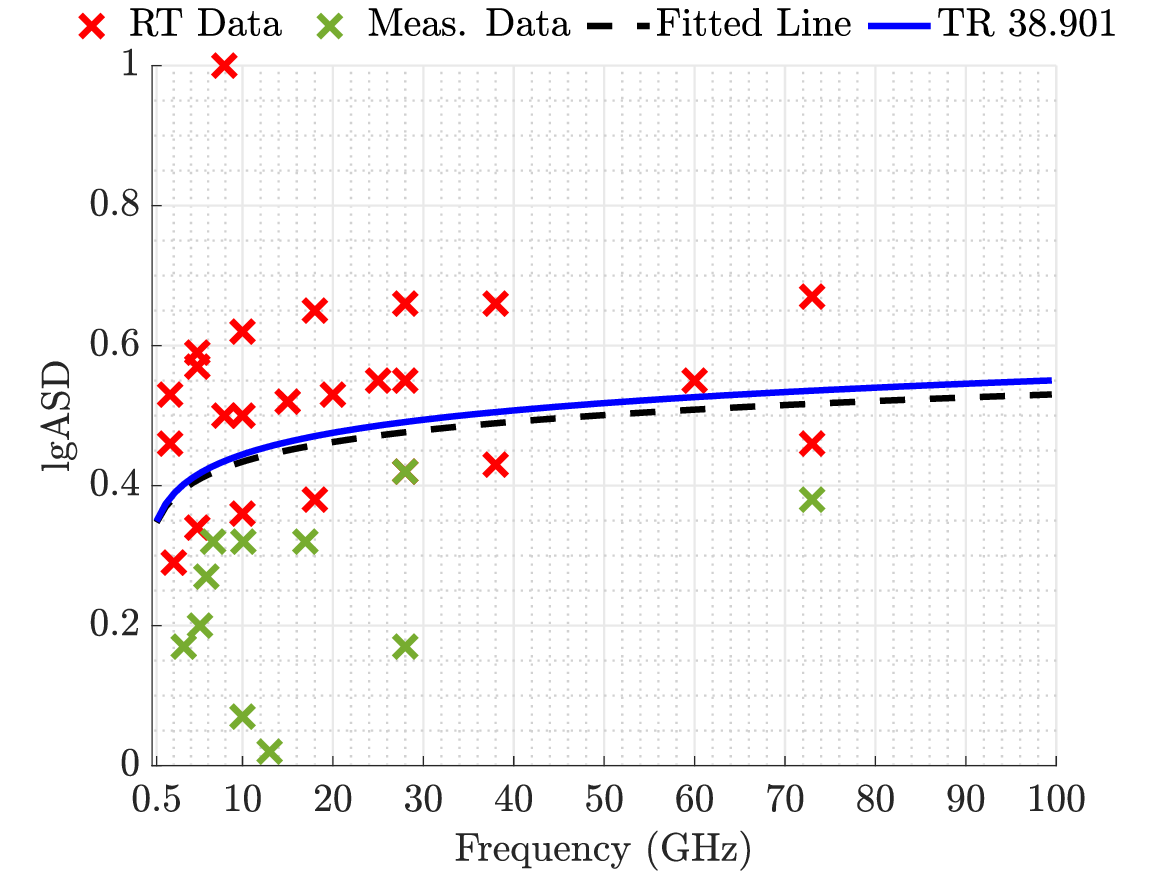}
        \caption{RT and Meas. data for standard deviation of lgASD in the UMi NLOS scenario over 0.5-100 GHz from 3GPP Rel-14 and Rel-19, with a WLS fitted line (0.10 log$_{10}$(1+f) + 0.33~\cite{r1-2502415}) and the 3GPP TR 38.901 model (0.11 log$_{10}$(1+f) + 0.33~\cite{tr38901v18}).}
        \label{fig:umi_nlos_std_asd_wls}
    \end{subfigure}

    \caption{Curve fitting of RT and Meas. data for the standard deviation of lgASD in UMi LOS and NLOS channel conditions OLS and WLS methods. (a) and (c) use data from 3GPP Rel-19 only (6–24 GHz), while (b) and (d) use combined data from Rel-14 and Rel-19 (0.5–100 GHz). All subfigures include a comparison with the existing 3GPP TR 38.901 model.}
    \label{fig:umi_std_asd}
\end{figure*}

\clearpage
\subsection{UMi-Street Canyon ASA}
\begin{figure*}[h]
    \centering
    \begin{subfigure}[b]{0.48\textwidth}
        \centering
        \includegraphics[width=\linewidth]{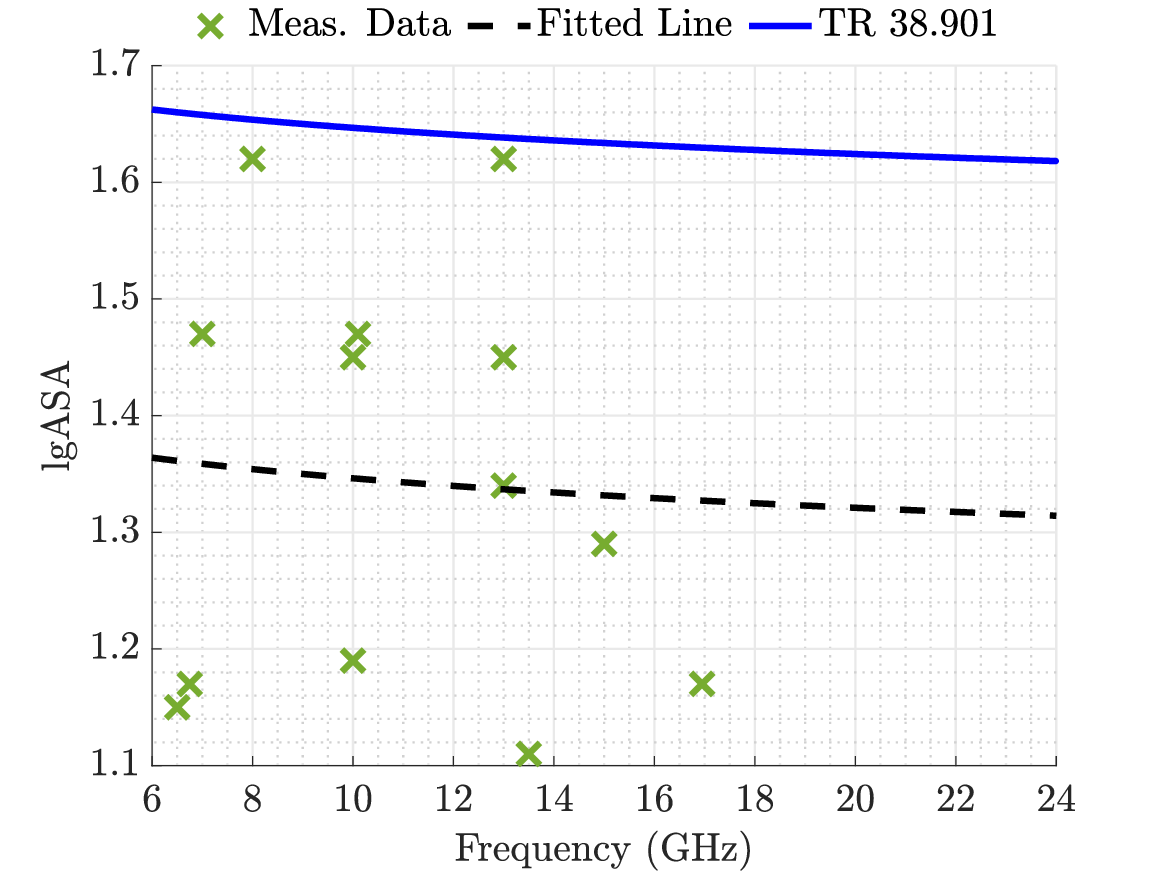}
        \caption{Meas. data for mean of lgASA in the UMi LOS scenario over 6–24 GHz from 3GPP Rel-19, with an OLS fitted line (-0.09 log$_{10}$(1+f) + 1.44~\cite{r1-2502415}) and the 3GPP TR 38.901 model (-0.08 log$_{10}$(1+f) + 1.73~\cite{tr38901v18}).}
        \label{fig:umi_los_mean_asa_rel19}
    \end{subfigure}
    \hfill
    \begin{subfigure}[b]{0.48\textwidth}
        \centering
        \includegraphics[width=\linewidth]{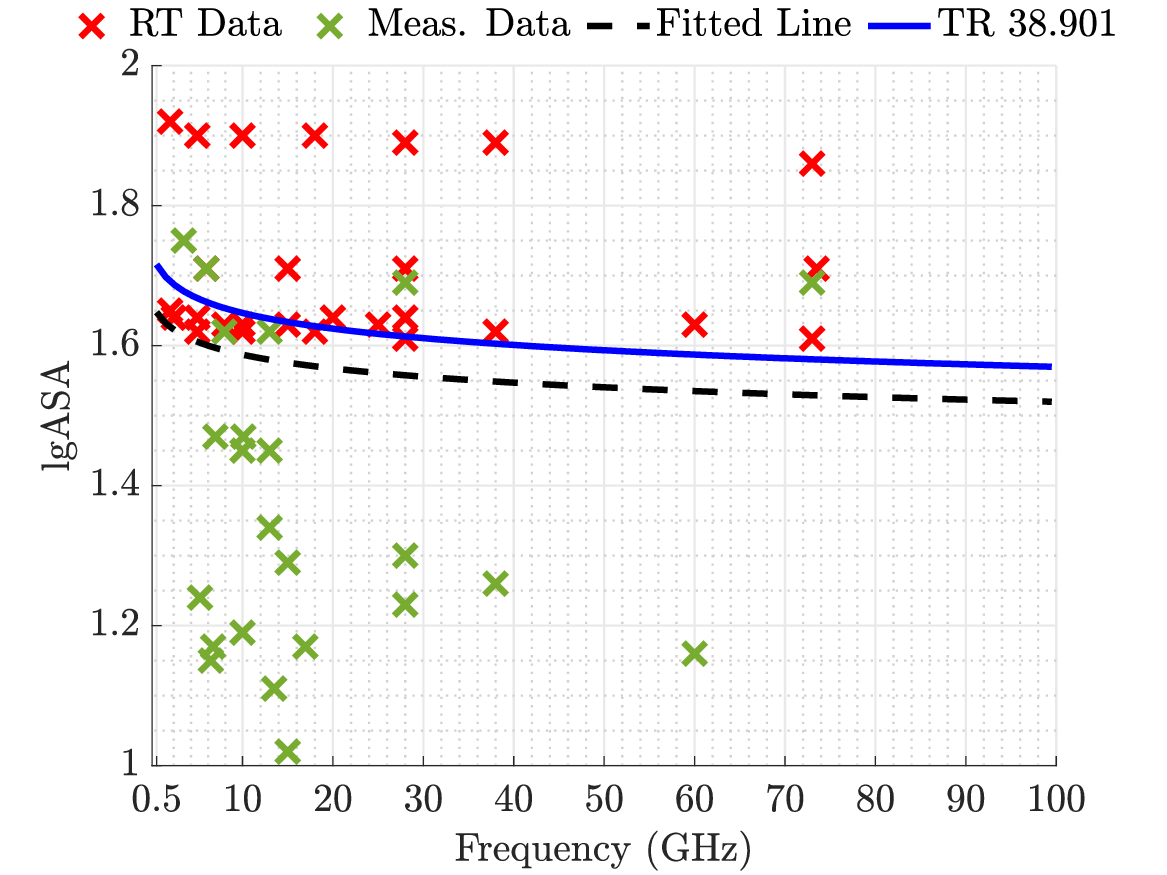}
        \caption{RT and Meas. data for mean of lgASA in the UMi LOS scenario over 0.5-100 GHz from 3GPP Rel-14 and Rel-19, with a WLS fitted line (-0.07 log$_{10}$(1+f) + 1.66~\cite{r1-2502415}) and the 3GPP TR 38.901 model (-0.08 log$_{10}$(1+f) + 1.73~\cite{tr38901v18}).}
        \label{fig:umi_los_mean_asa_wls}
    \end{subfigure}
    
    \vspace{1em}

    \begin{subfigure}[b]{0.48\textwidth}
        \centering
        \includegraphics[width=\linewidth]{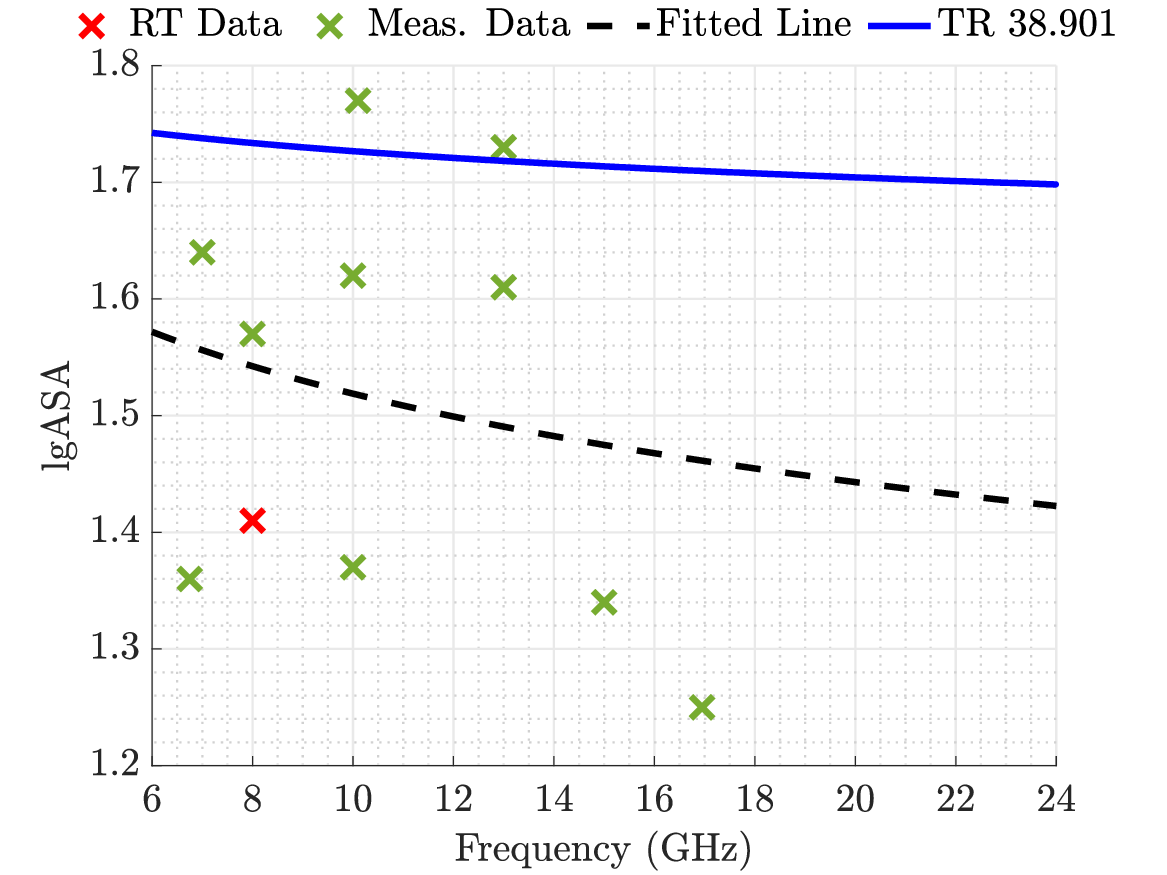}
        \caption{RT and Meas. data for mean of lgASA in the UMi NLOS scenario over 6–24 GHz from 3GPP Rel-19, with an   OLS fitted line (-0.27 log$_{10}$(1+f) + 1.80~\cite{r1-2502415}) and the 3GPP TR 38.901 model (-0.08 log$_{10}$(1+f) + 1.81~\cite{tr38901v18}).}
        \label{fig:umi_nlos_mean_asa_rel19}
    \end{subfigure}
    \hfill
    \begin{subfigure}[b]{0.48\textwidth}
        \centering
        \includegraphics[width=\linewidth]{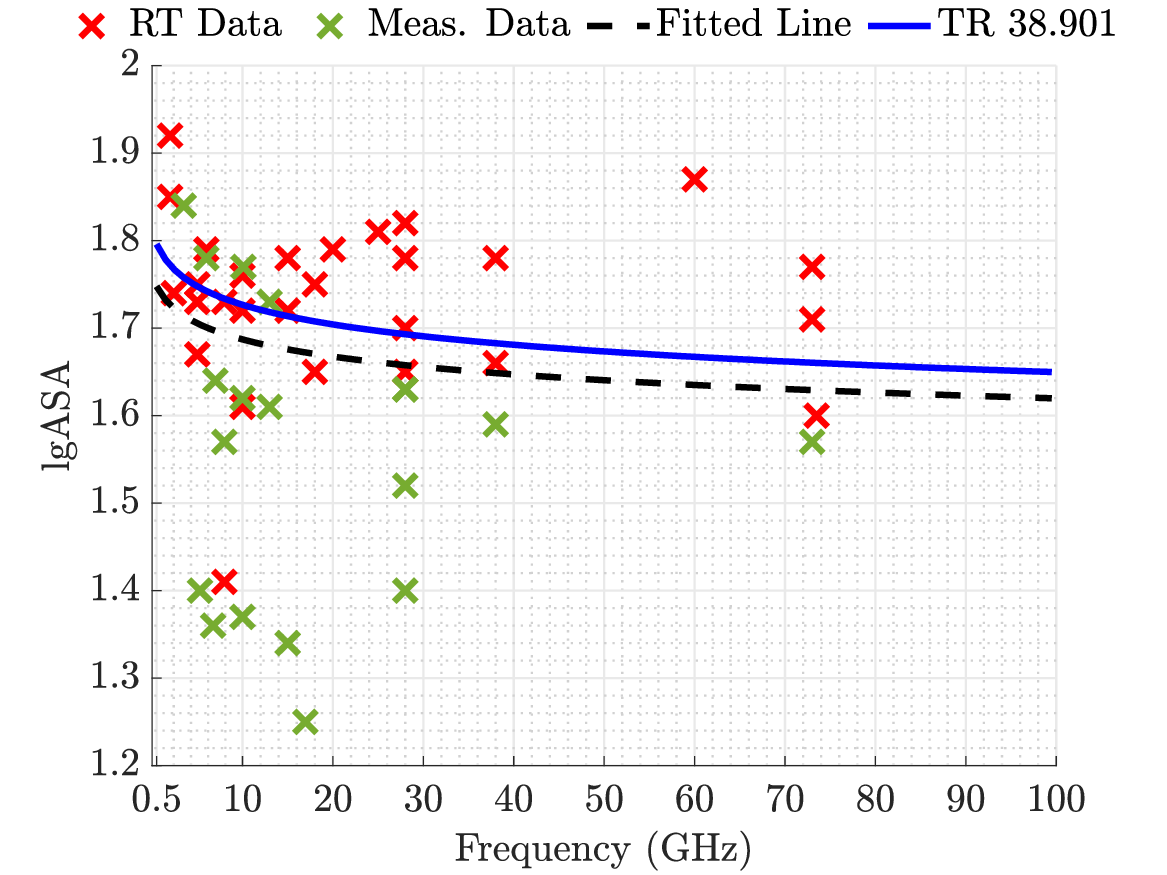}
        \caption{RT and Meas. data for mean of lgASA in the UMi NLOS scenario over 0.5-100 GHz from 3GPP Rel-14 and Rel-19, with a WLS fitted line (-0.07 log$_{10}$(1+f) + 1.76\cite{r1-2502415}) and the 3GPP TR 38.901 model (-0.08 log$_{10}$(1+f) + 1.81\cite{tr38901v18}).}
        \label{fig:umi_nlos_mean_asa_wls}
    \end{subfigure}

    \caption{Curve fitting of RT and Meas. data for the standard deviation of lgASA in UMi LOS and NLOS channel conditions OLS and WLS methods. (a) and (c) use data from 3GPP Rel-19 only (6–24 GHz), while (b) and (d) use combined data from Rel-14 and Rel-19 (0.5–100 GHz). All subfigures include a comparison with the existing 3GPP TR 38.901 model.}
    \label{fig:umi_mean_asa}
\end{figure*}

%%%%%%%%% UMi ASA STD %%%%%%%%
\begin{figure*}[h]
    \centering
    \begin{subfigure}[b]{0.48\textwidth}
        \centering
        \includegraphics[width=\linewidth]{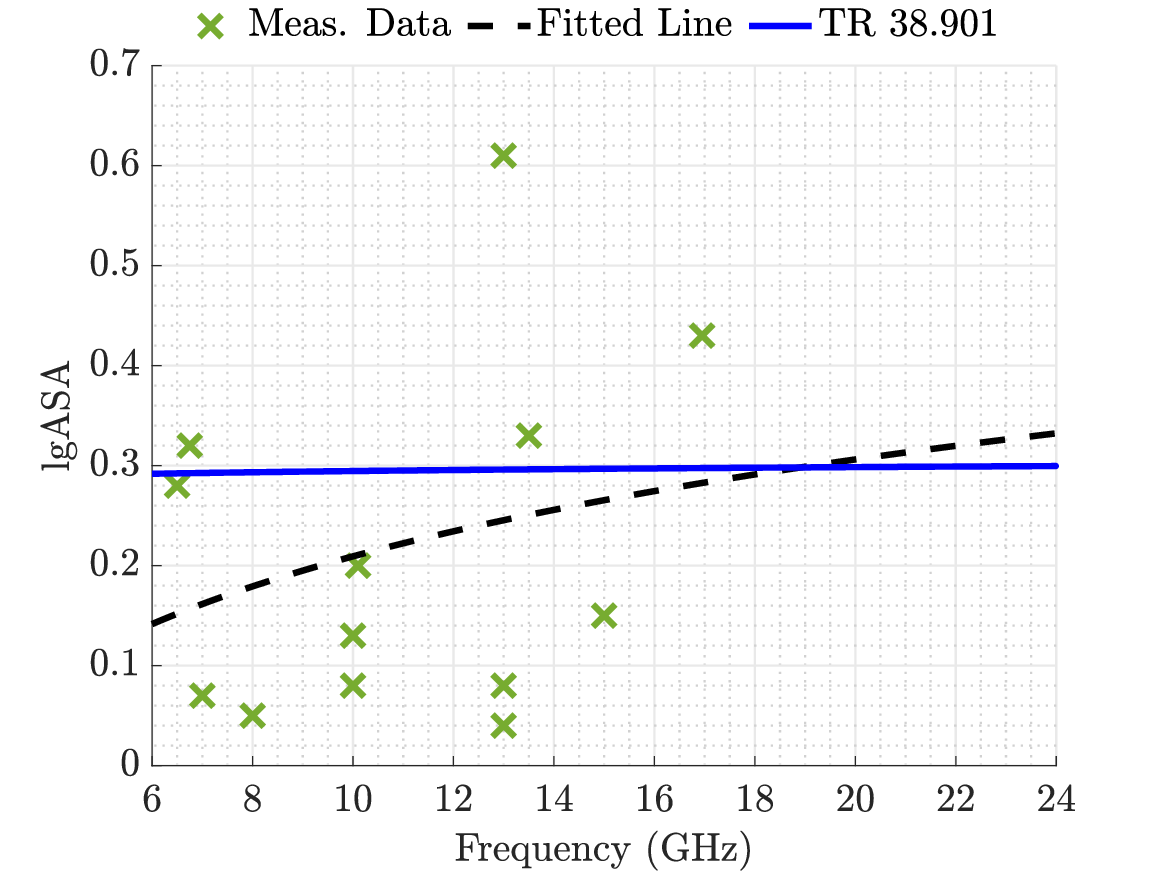}
        \caption{Meas. data for standard deviation of lgASA in the UMi LOS scenario over 6–24 GHz from 3GPP Rel-19, with an OLS fitted line (0.345 log$_{10}$(1+f) - 0.15~\cite{r1-2502415}) and the 3GPP TR 38.901 model (0.014 log$_{10}$(1+f) + 0.28~\cite{tr38901v18}).}
        \label{fig:umi_los_std_asa_rel19}
    \end{subfigure}
    \hfill
    \begin{subfigure}[b]{0.48\textwidth}
        \centering
        \includegraphics[width=\linewidth]{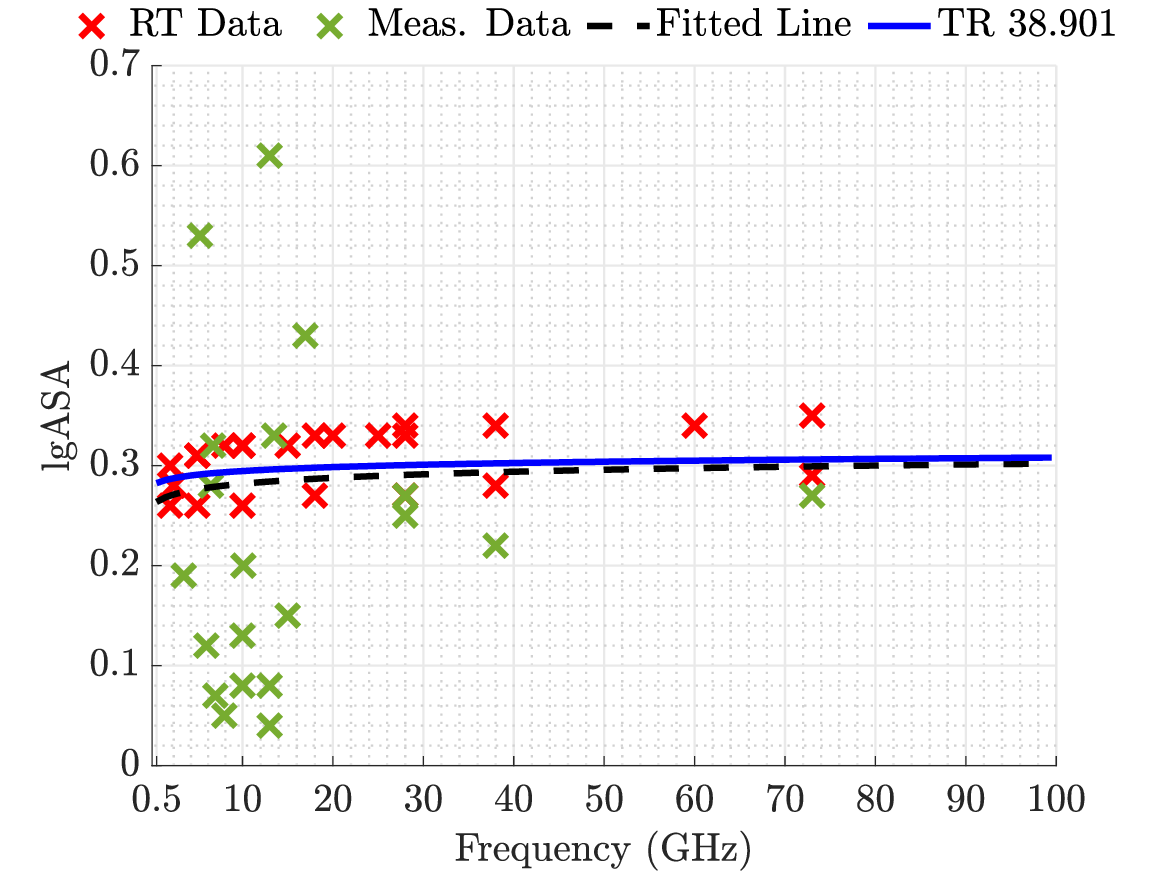}
        \caption{RT and Meas. data for standard deviation of lgASA in the UMi LOS scenario over 0.5-100 GHz from 3GPP Rel-14 and Rel-19, with a WLS fitted line (0.021 log$_{10}$(1+f) + 0.26~\cite{r1-2502415}) and the 3GPP TR 38.901 model (0.014 log$_{10}$(1+f) + 0.28 \cite{tr38901v18}).}
        \label{fig:umi_los_std_asa_wls}
    \end{subfigure}
    
    \vspace{1em}

    \begin{subfigure}[b]{0.48\textwidth}
        \centering
        \includegraphics[width=\linewidth]{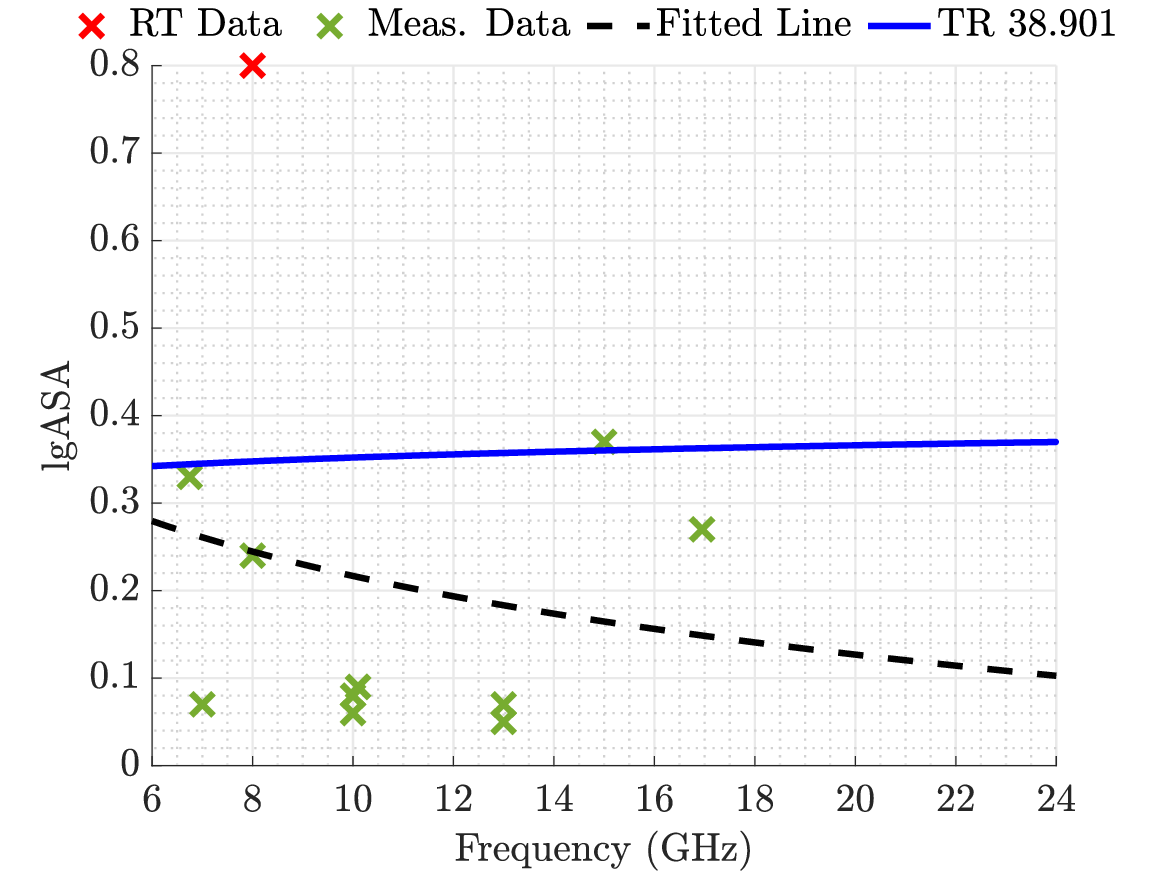}
        \caption{RT and Meas. data for standard deviation of lgASA in the UMi NLOS scenario over 6–24 GHz from 3GPP Rel-19, with an OLS fitted line (-0.32 log$_{10}$(1+f) + 0.55~\cite{r1-2502415}) and the 3GPP TR 38.901 model (0.05 log$_{10}$(1+f) + 0.3~\cite{tr38901v18}).}
    \label{fig:umi_nlos_std_asa_rel19}
    \end{subfigure}
    \hfill
    \begin{subfigure}[b]{0.48\textwidth}
        \centering
        \includegraphics[width=\linewidth]{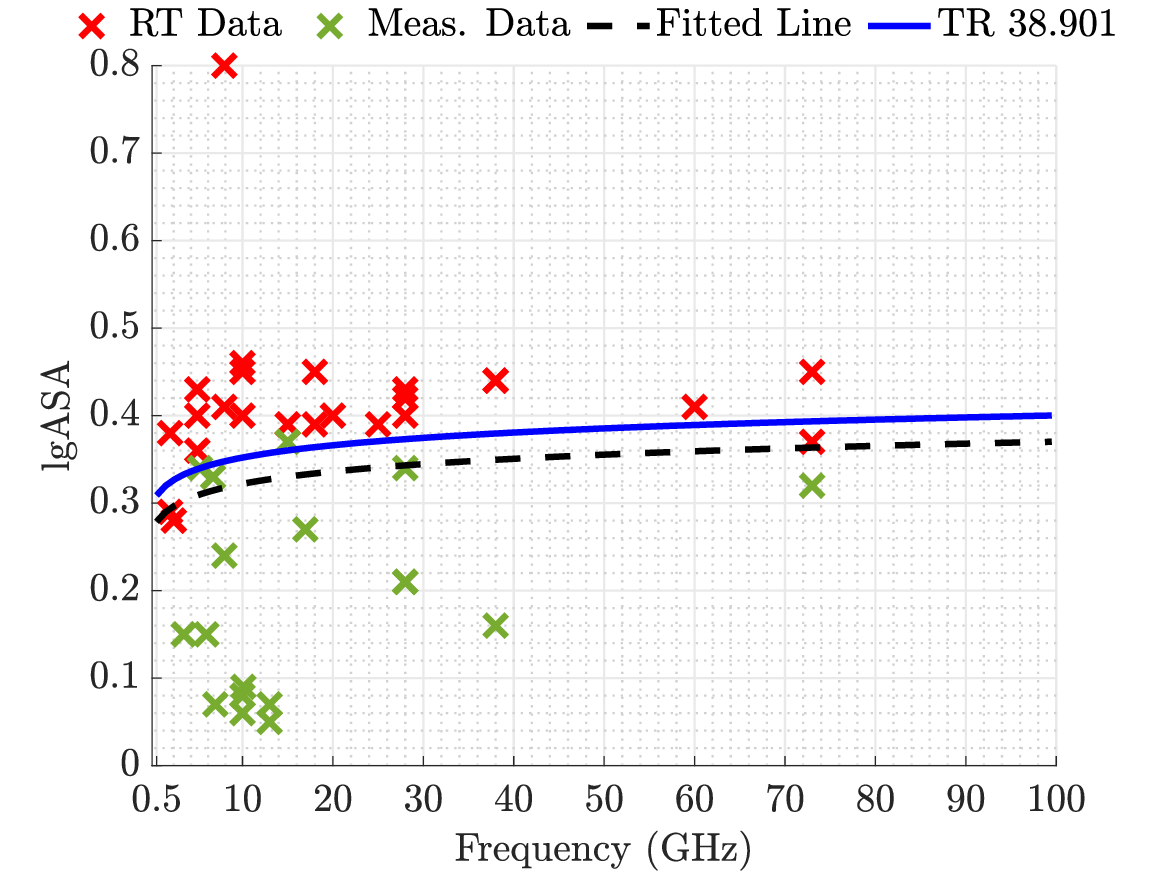}
        \caption{RT and Meas. data for standard deviation of lgASA in the UMi NLOS scenario over 0.5-100 GHz from 3GPP Rel-14 and Rel-19, with a WLS fitted line (0.05 log$_{10}$(1+f) + 0.27~\cite{r1-2502415}) and the 3GPP TR 38.901 model (0.05 log$_{10}$(1+f) + 0.3~\cite{tr38901v18}).}
        \label{fig:umi_nlos_std_asa_wls}
    \end{subfigure}

    \caption{Curve fitting of RT and Meas. data for the standard deviation of lgASA in UMi LOS and NLOS channel conditions OLS and WLS methods. (a) and (c) use data from 3GPP Rel-19 only (6–24 GHz), while (b) and (d) use combined data from Rel-14 and Rel-19 (0.5–100 GHz). All subfigures include a comparison with the existing 3GPP TR 38.901 model.}
    \label{fig:umi_std_asa}
\end{figure*}

\clearpage
\subsection{UMi-Street Canyon ZSA}
\begin{figure*}[h]
    \centering
    \begin{subfigure}[b]{0.48\textwidth}
        \centering
        \includegraphics[width=\linewidth]{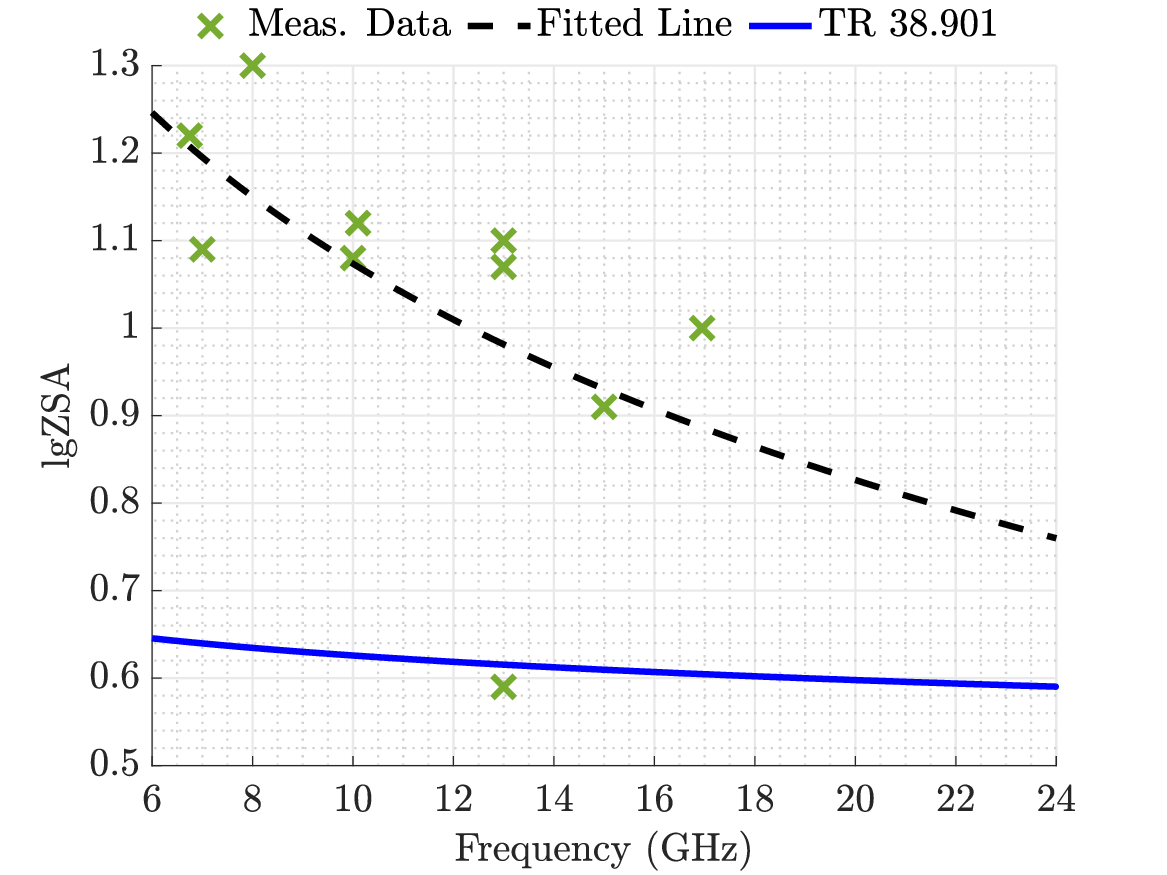}
        \caption{Meas. data for mean of lgZSA in the UMi LOS scenario over 6–24 GHz from 3GPP Rel-19, with an OLS fitted line (-0.88 log$_{10}$(1+f) + 1.99~\cite{r1-2502415}) and the 3GPP TR 38.901 model (-0.10 log$_{10}$(1+f) + 0.73~\cite{tr38901v18}).}
        \label{fig:umi_los_mean_zsa_rel19}
    \end{subfigure}
    \hfill
    \begin{subfigure}[b]{0.48\textwidth}
        \centering
        \includegraphics[width=\linewidth]{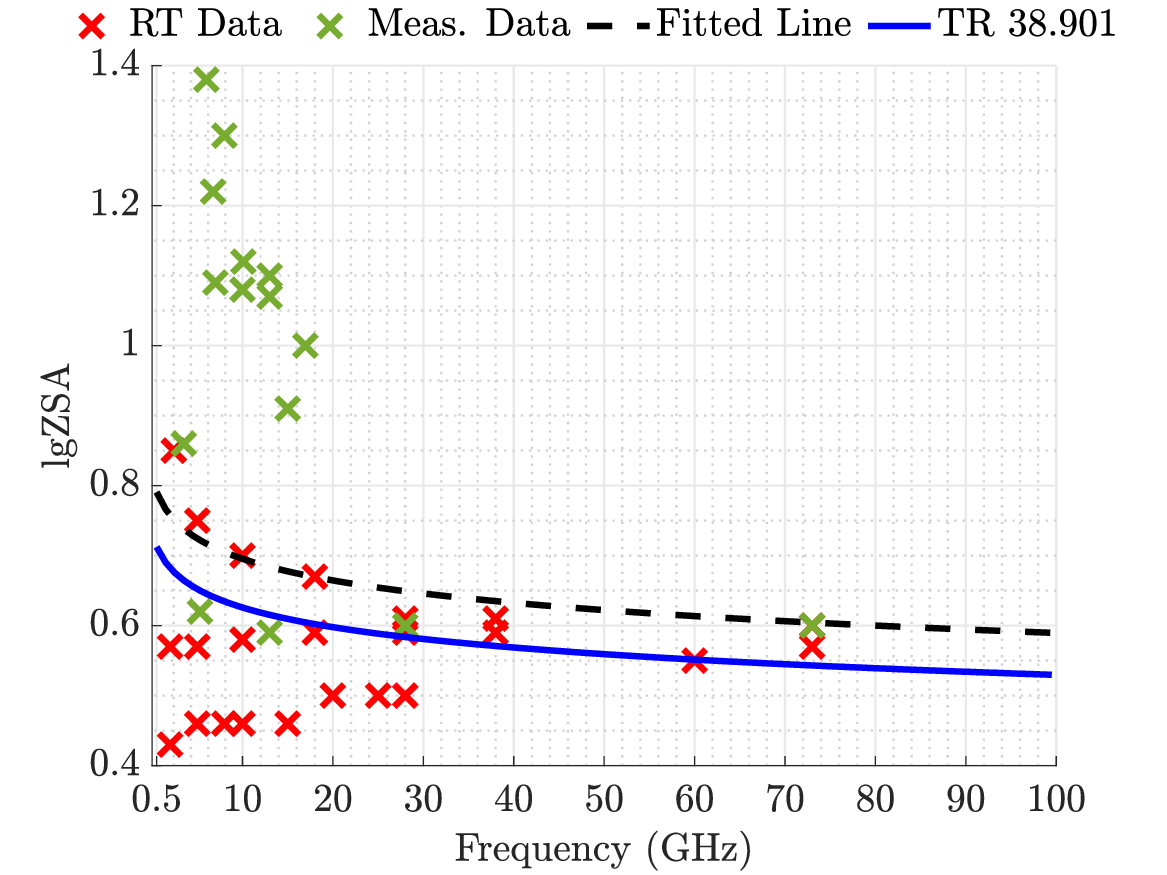}
        \caption{RT and Meas. data for mean of lgZSA in the UMi LOS scenario over 0.5-100 GHz from 3GPP Rel-14 and Rel-19, with a WLS fitted line (-0.11 log$_{10}$(1+f) + 0.81~\cite{r1-2502415}) and the 3GPP TR 38.901 model (-0.10 log$_{10}$(1+f) + 0.73~\cite{tr38901v18}).}
        \label{fig:umi_los_mean_zsa_wls}
    \end{subfigure}
    
    \vspace{1em}

    \begin{subfigure}[b]{0.48\textwidth}
        \centering
        \includegraphics[width=\linewidth]{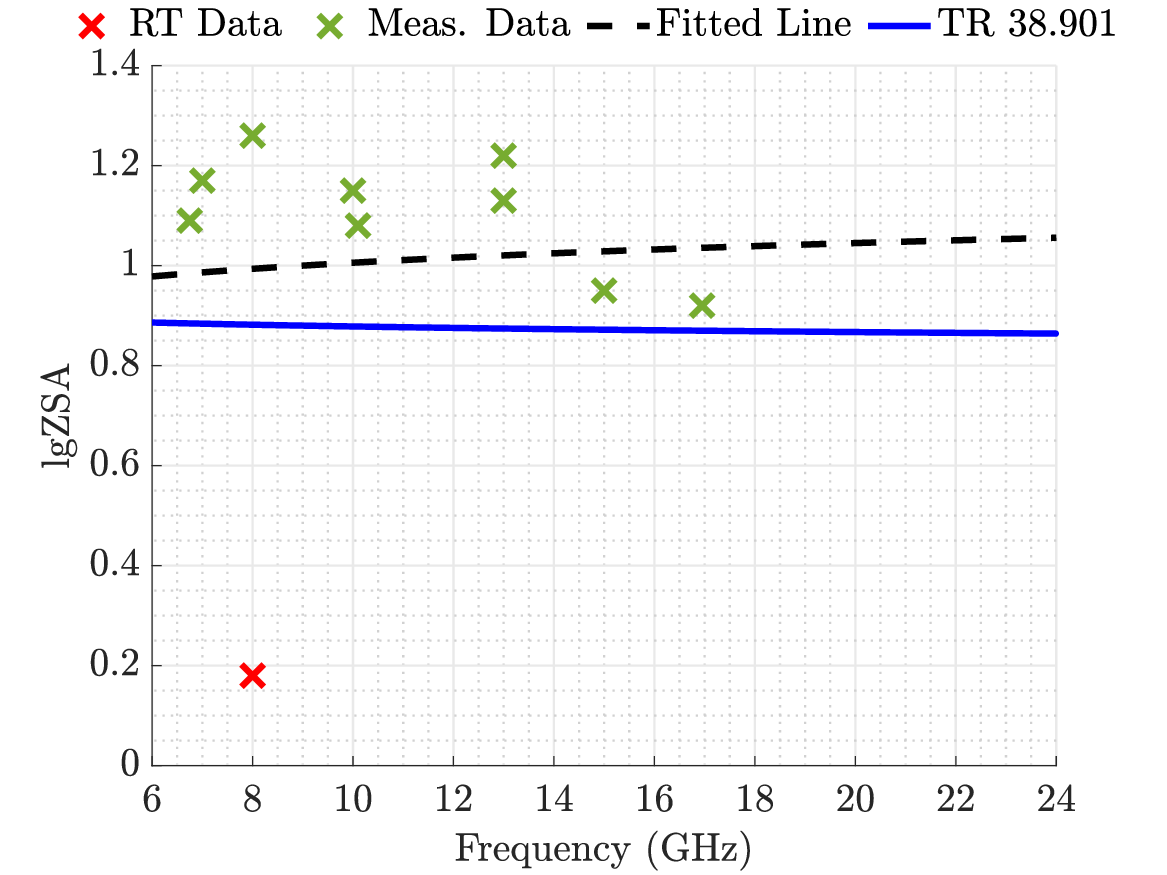}
        \caption{RT and Meas. data for mean of lgZSA in the UMi NLOS scenario over 6–24 GHz from 3GPP Rel-19, with an OLS fitted line (0.14 log$_{10}$(1+f) + 0.86~\cite{r1-2502415}) and the 3GPP TR 38.901 model (-0.04 log$_{10}$(1+f) + 0.92 \cite{tr38901v18}).}
        \label{fig:umi_nlos_mean_zsa_rel19}
    \end{subfigure}
    \hfill
    \begin{subfigure}[b]{0.48\textwidth}
        \centering
        \includegraphics[width=\linewidth]{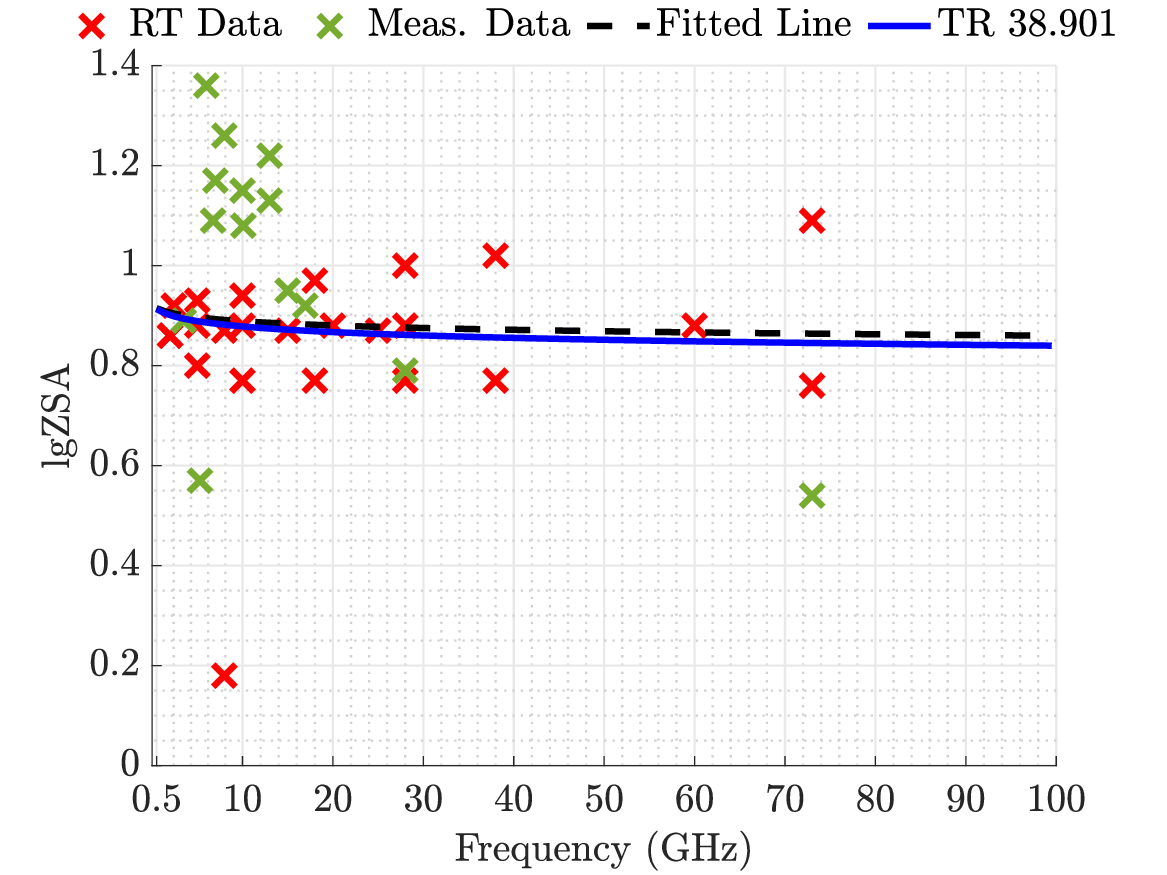}
        \caption{RT and Meas. data for mean of lgZSA in the UMi NLOS scenario over 0.5-100 GHz from 3GPP Rel-14 and Rel-19, with a WLS fitted line (-0.03 log$_{10}$(1+f) + 0.92~\cite{r1-2502415}) and the 3GPP TR 38.901 model (-0.04 log$_{10}$(1+f) + 0.92~\cite{tr38901v18}).}
        \label{fig:umi_nlos_mean_zsa_wls}
    \end{subfigure}

    \caption{Curve fitting of RT and Meas. data for the mean of lgZSA in UMi LOS and NLOS channel conditions OLS and WLS methods. (a) and (c) use data from 3GPP Rel-19 only (6–24 GHz), while (b) and (d) use combined data from Rel-14 and Rel-19 (0.5–100 GHz). All subfigures include a comparison with the existing 3GPP TR 38.901 model.}
    \label{fig:umi_mean_zsa}
\end{figure*}

%%%%%%%%% UMi ZSA STD %%%%%%%%
\begin{figure*}[h]
    \centering
    \begin{subfigure}[b]{0.48\textwidth}
        \centering
        \includegraphics[width=\linewidth]{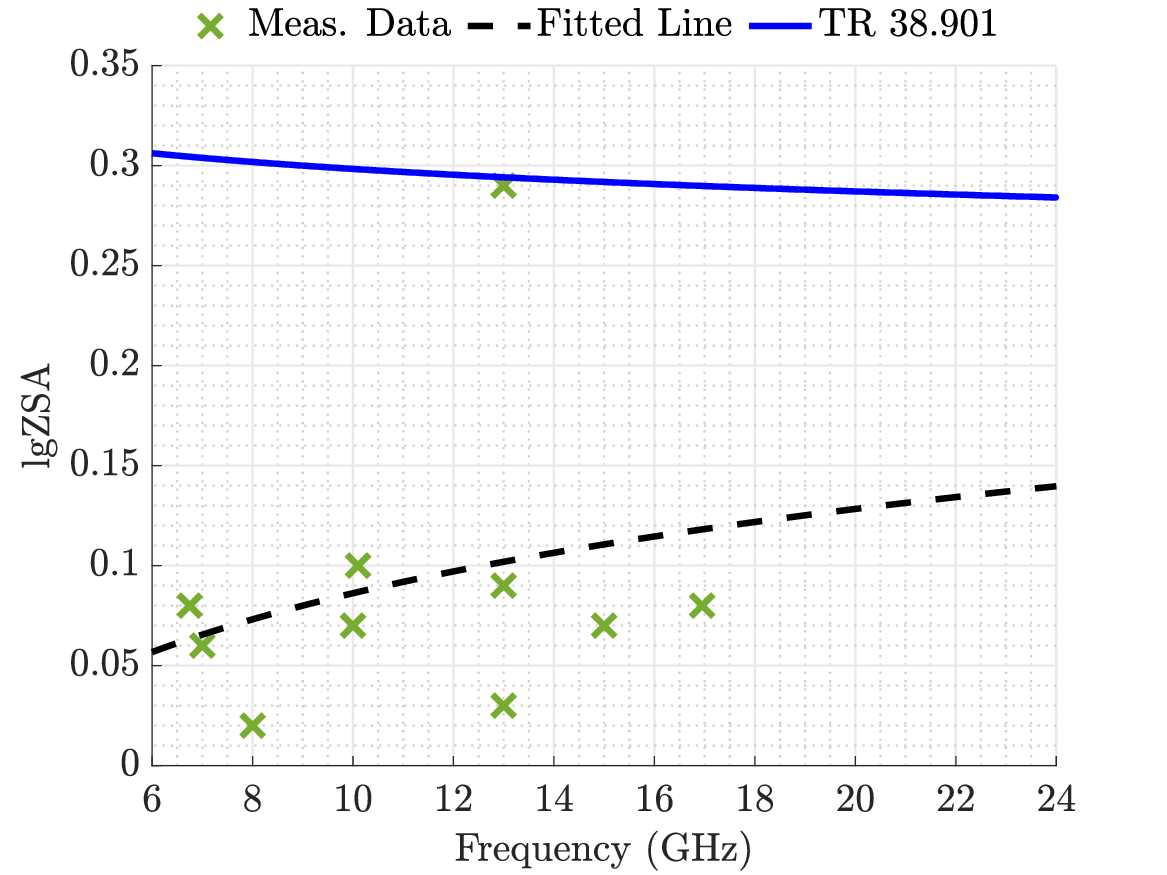}
        \caption{Meas. data for standard deviation of lgZSA in the UMi LOS scenario over 6–24 GHz from 3GPP Rel-19, with an OLS fitted line (0.15 log$_{10}$(1+f) - 0.07~\cite{r1-2502415}) and the 3GPP TR 38.901 model (-0.04 log$_{10}$(1+f) + 0.34~\cite{tr38901v18}).}
        \label{fig:umi_los_std_zsa_rel19}
    \end{subfigure}
    \hfill
    \begin{subfigure}[b]{0.48\textwidth}
        \centering
        \includegraphics[width=\linewidth]{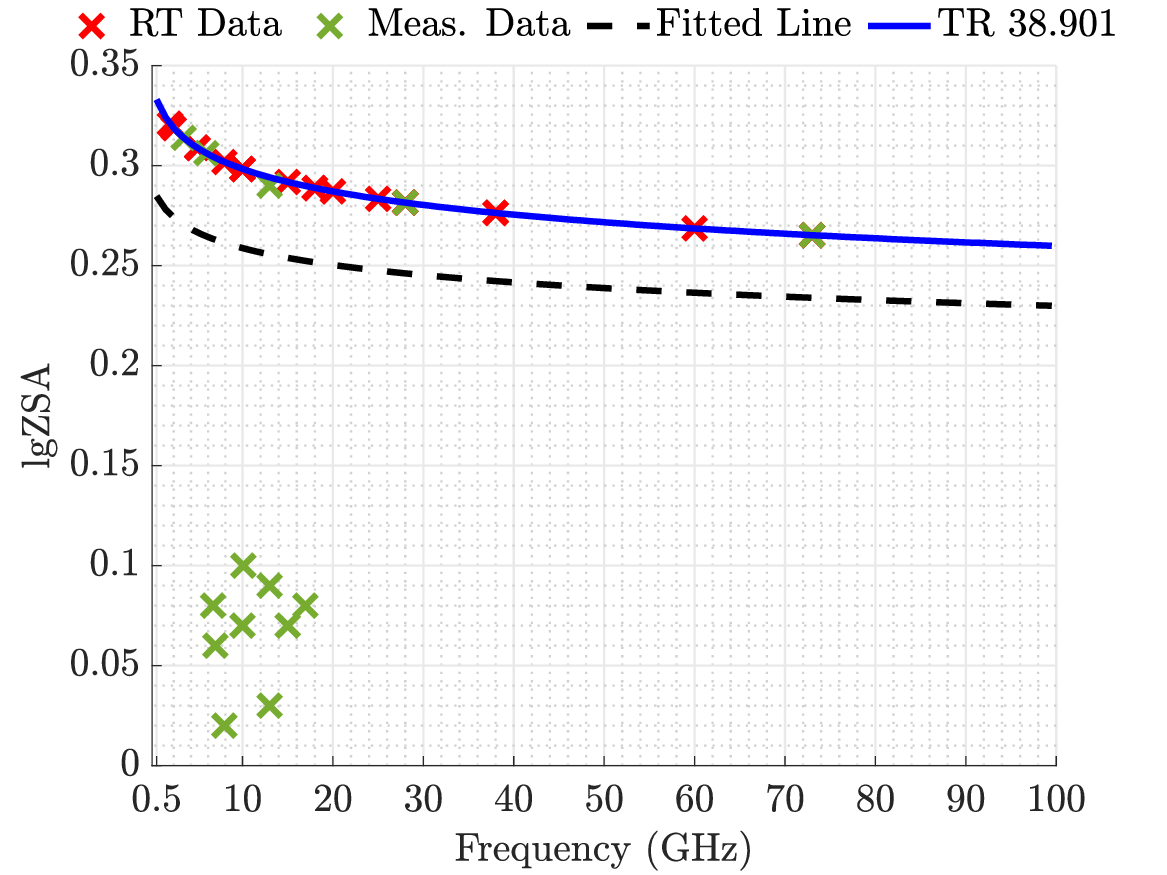}
        \caption{RT and Meas. data for standard deviation of lgZSA in the UMi LOS scenario over 0.5-100 GHz from 3GPP Rel-14 and Rel-19, with a WLS fitted line (-0.03 log$_{10}$(1+f) + 0.29~\cite{r1-2502415}) and the 3GPP TR 38.901 model (-0.04 log$_{10}$(1+f) + 0.34~\cite{tr38901v18}).}
        \label{fig:umi_los_std_zsa_wls}
    \end{subfigure}
    
    \vspace{1em}

    \begin{subfigure}[b]{0.48\textwidth}
        \centering
        \includegraphics[width=\linewidth]{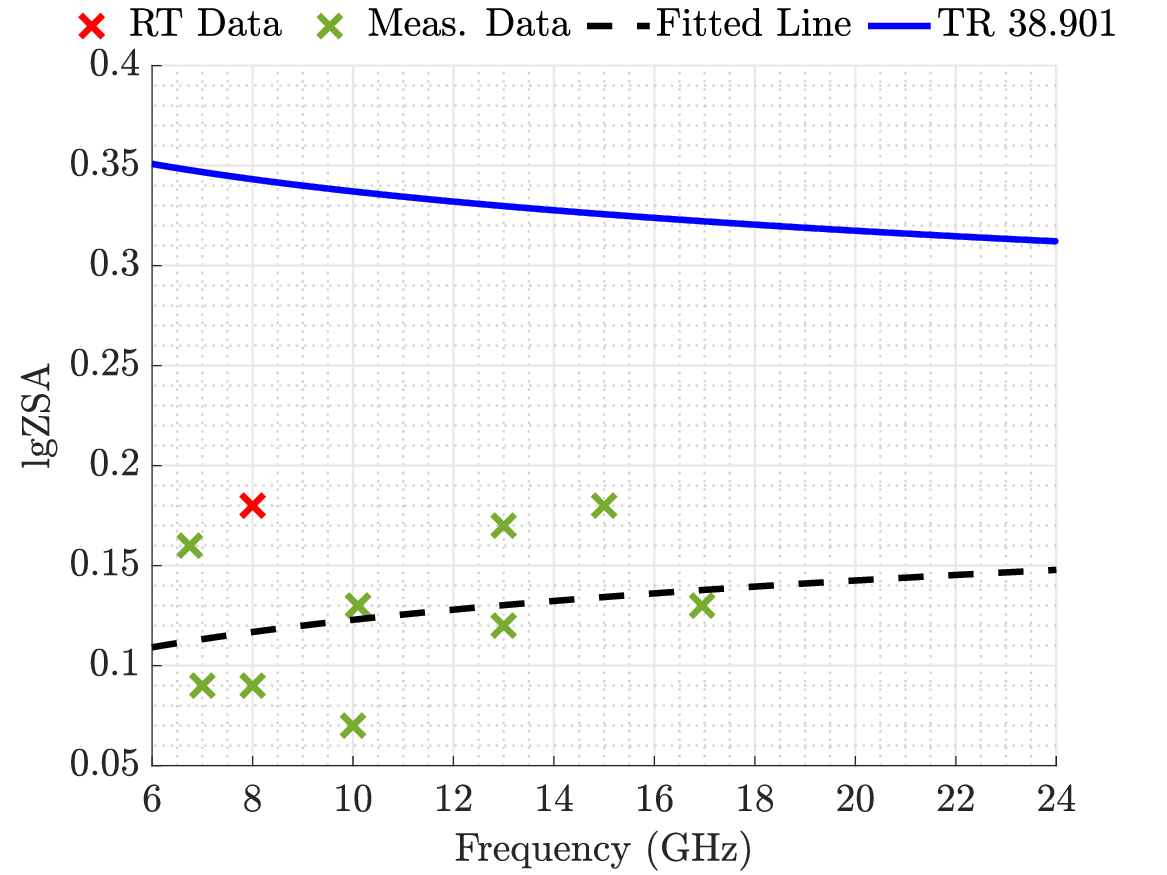}
        \caption{RT and Meas. data for standard deviation of lgZSA in the UMi NLOS scenario over 6–24 GHz from 3GPP Rel-19, with an OLS fitted line (0.07 log$_{10}$(1+f) + 0.05~\cite{r1-2502415}) and the 3GPP TR 38.901 model (-0.07 log$_{10}$(1+f) + 0.41~\cite{tr38901v18}).}
        \label{fig:umi_nlos_std_zsa_rel19}
    \end{subfigure}
    \hfill
    \begin{subfigure}[b]{0.48\textwidth}
        \centering
        \includegraphics[width=\linewidth]{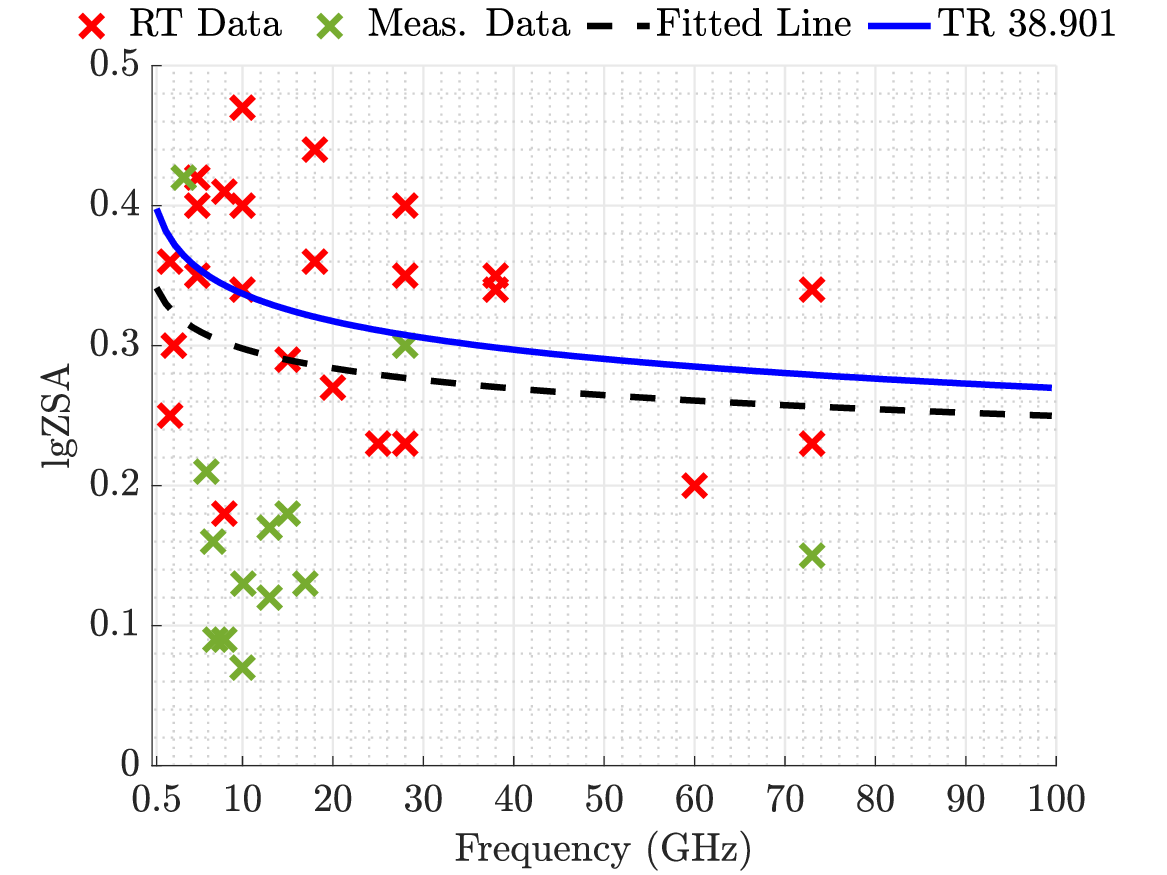}
        \caption{RT and Meas. data for standard deviation of lgZSA in the UMi NLOS scenario over 0.5-100 GHz from 3GPP Rel-14 and Rel-19, with a WLS fitted line (-0.05 log$_{10}$(1+f) + 0.35~\cite{r1-2502415}) and the 3GPP TR 38.901 model (-0.07 log$_{10}$(1+f) + 0.41~\cite{tr38901v18}).}
        \label{fig:umi_nlos_std_zsa_wls}
    \end{subfigure}

    \caption{Curve fitting of RT and Meas. data for the standard deviation of lgZSA in UMi LOS and NLOS channel conditions OLS and WLS methods. (a) and (c) use data from 3GPP Rel-19 only (6–24 GHz), while (b) and (d) use combined data from Rel-14 and Rel-19 (0.5–100 GHz). All subfigures include a comparison with the existing 3GPP TR 38.901 model.}
    \label{fig:umi_std_zsa}
\end{figure*}

\clearpage
\subsection{UMa DS}
\begin{figure*}[h]
    \centering
    \begin{subfigure}[b]{0.48\textwidth}
        \centering
        \includegraphics[width=\linewidth]{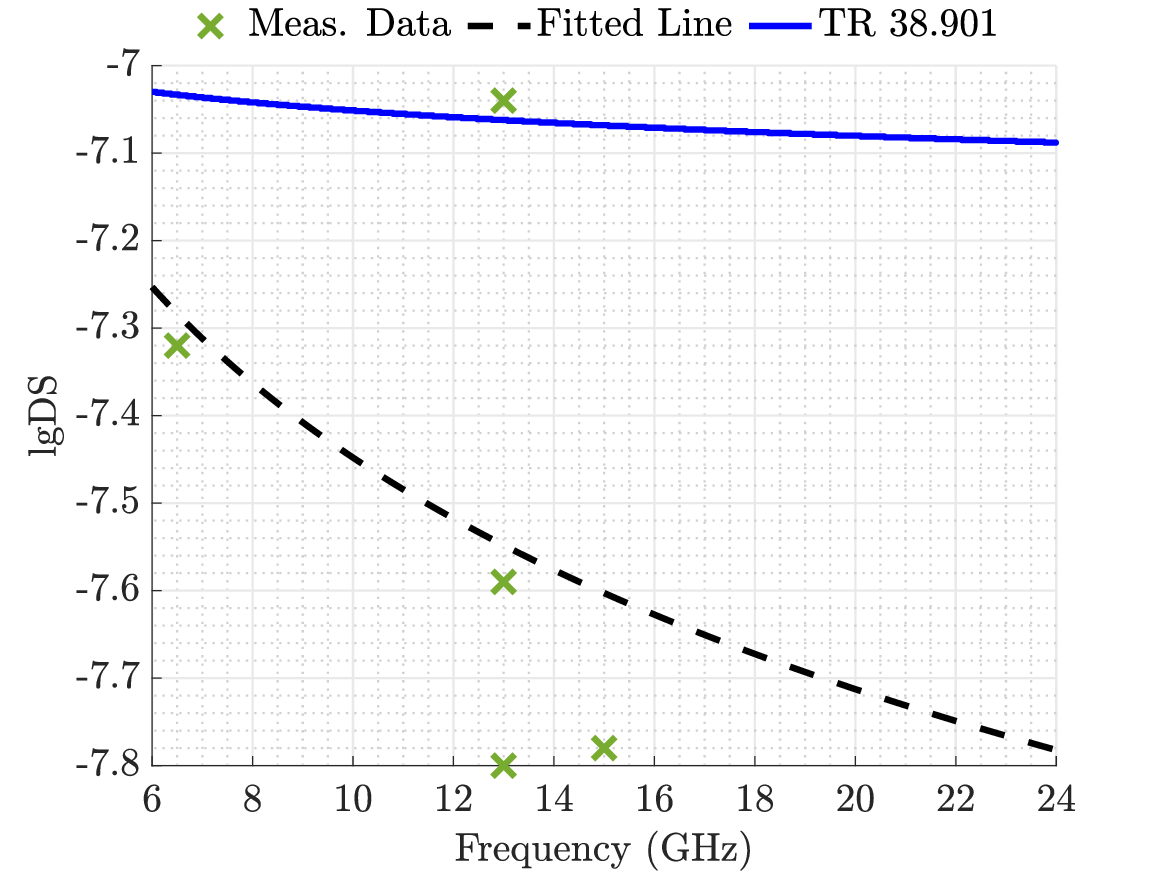}
        \caption{Meas. data for mean of lgDS in the UMa LOS scenario over 6–24 GHz from 3GPP Rel-19, with an OLS fitted line (-0.8790 log$_{10}$(f) - 6.569~\cite{r1-2502415}) and the 3GPP TR 38.901 model (-0.0963 log$_{10}$(f) - 6.955~\cite{tr38901v18}).}
        \label{fig:uma_los_mean_ds_rel19}
    \end{subfigure}
    \hfill
    \begin{subfigure}[b]{0.48\textwidth}
        \centering
        \includegraphics[width=\linewidth]{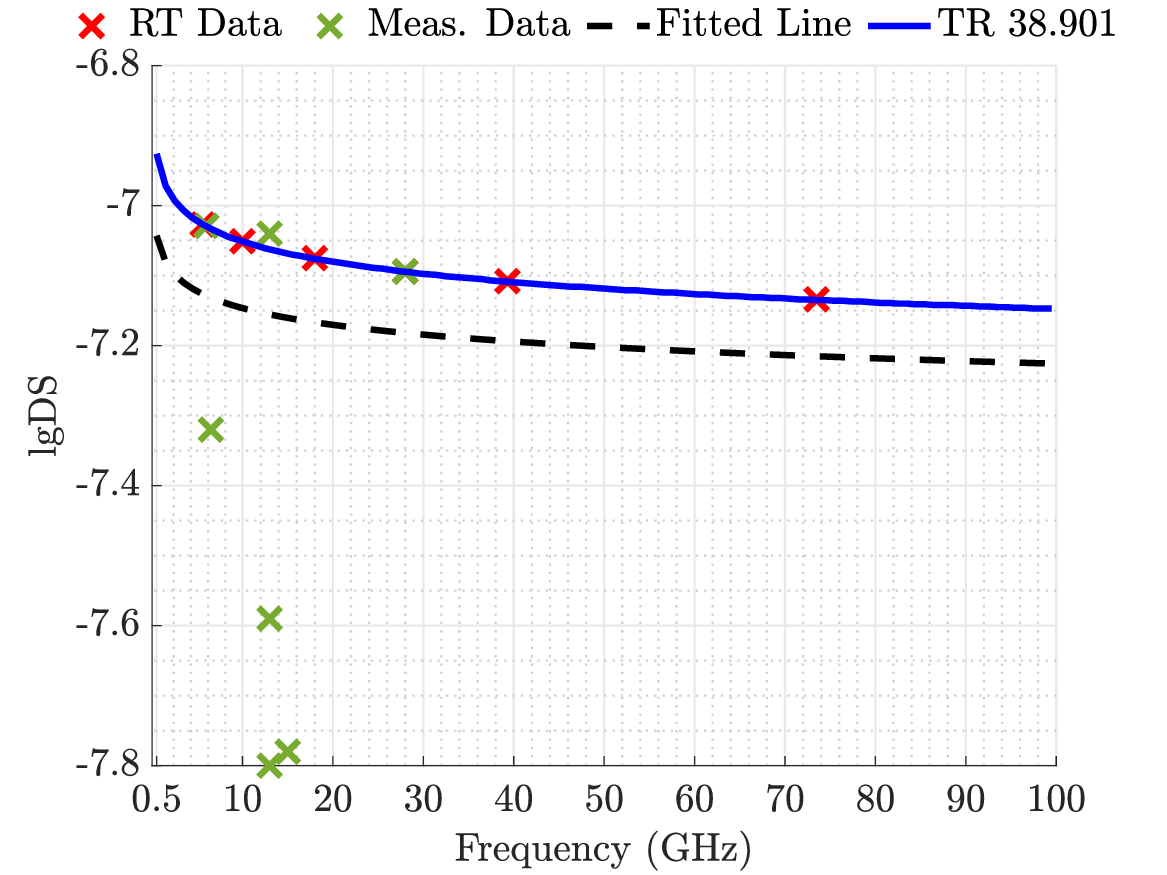}
        \caption{RT and Meas. data for mean of lgDS in the UMa LOS scenario over 0.5-100 GHz from 3GPP Rel-14 and Rel-19, with a WLS fitted line (-0.0794 log$_{10}$(f) - 7.067~\cite{r1-2502415}) and the 3GPP TR 38.901 model (-0.0963 log$_{10}$(f) - 6.955~\cite{tr38901v18}).}
        \label{fig:uma_los_mean_ds_wls}
    \end{subfigure}
    
    \vspace{1em}

    \begin{subfigure}[b]{0.48\textwidth}
        \centering
        \includegraphics[width=\linewidth]{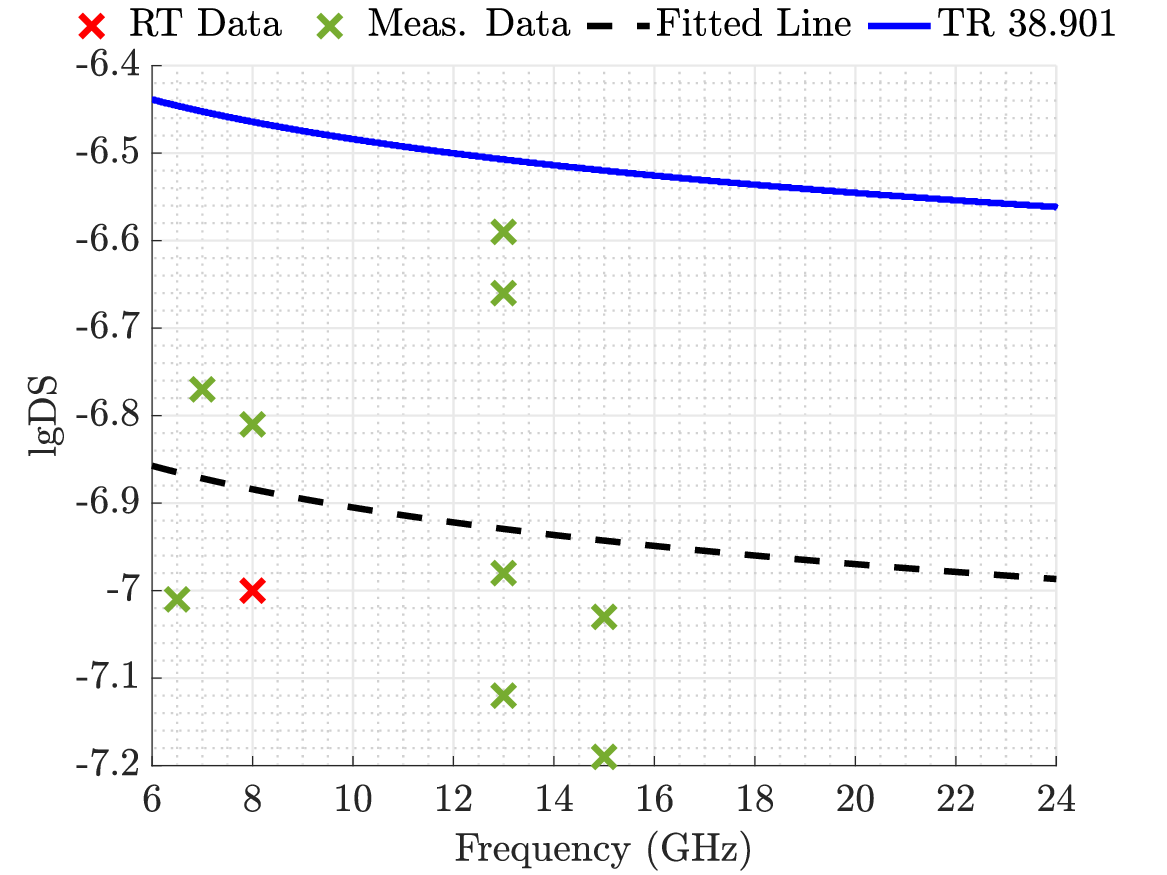}
        \caption{RT and Meas. data for mean of lgDS in the UMa NLOS scenario over 6–24 GHz from 3GPP Rel-19, with an OLS fitted line (-0.215 log$_{10}$(f) - 6.69~\cite{r1-2502415}) and the 3GPP TR 38.901 model (-0.204 log$_{10}$(f) - 6.28~\cite{tr38901v18}).}
        \label{fig:uma_nlos_mean_ds_rel19}
    \end{subfigure}
    \hfill
    \begin{subfigure}[b]{0.48\textwidth}
        \centering
        \includegraphics[width=\linewidth]{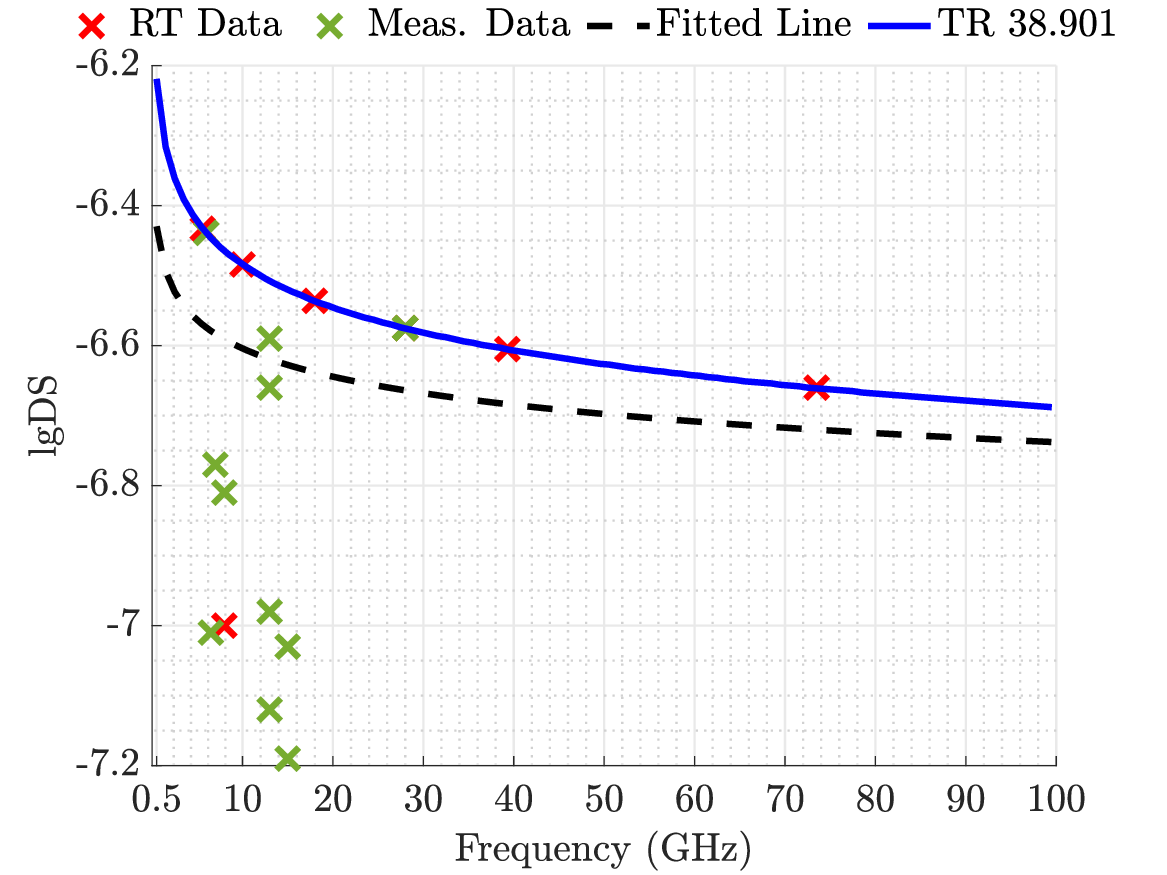}
        \caption{RT and Meas. data for mean of lgDS in the UMa NLOS scenario over 0.5-100 GHz from 3GPP Rel-14 and Rel-19, with a WLS fitted line (-0.134 log$_{10}$(f) - 6.47~\cite{r1-2502415}) and the 3GPP TR 38.901 model (-0.204 log$_{10}$(f) - 6.28~\cite{tr38901v18}).}
        \label{fig:uma_nlos_mean_ds_wls}
    \end{subfigure}

    \caption{Curve fitting of RT and Meas. data for the mean of lgDS in UMa LOS and NLOS channel conditions OLS and WLS methods. (a) and (c) use data from 3GPP Rel-19 only (6–24 GHz), while (b) and (d) use combined data from Rel-14 and Rel-19 (0.5–100 GHz). All subfigures include a comparison with the existing 3GPP TR 38.901 model.}
    \label{fig:uma_mean_ds}
\end{figure*}

%%%%%%%% UMa STD DS %%%%%%%%%%%%
\begin{figure*}[h]
    \centering
    \begin{subfigure}[b]{0.48\textwidth}
        \centering
        \includegraphics[width=\linewidth]{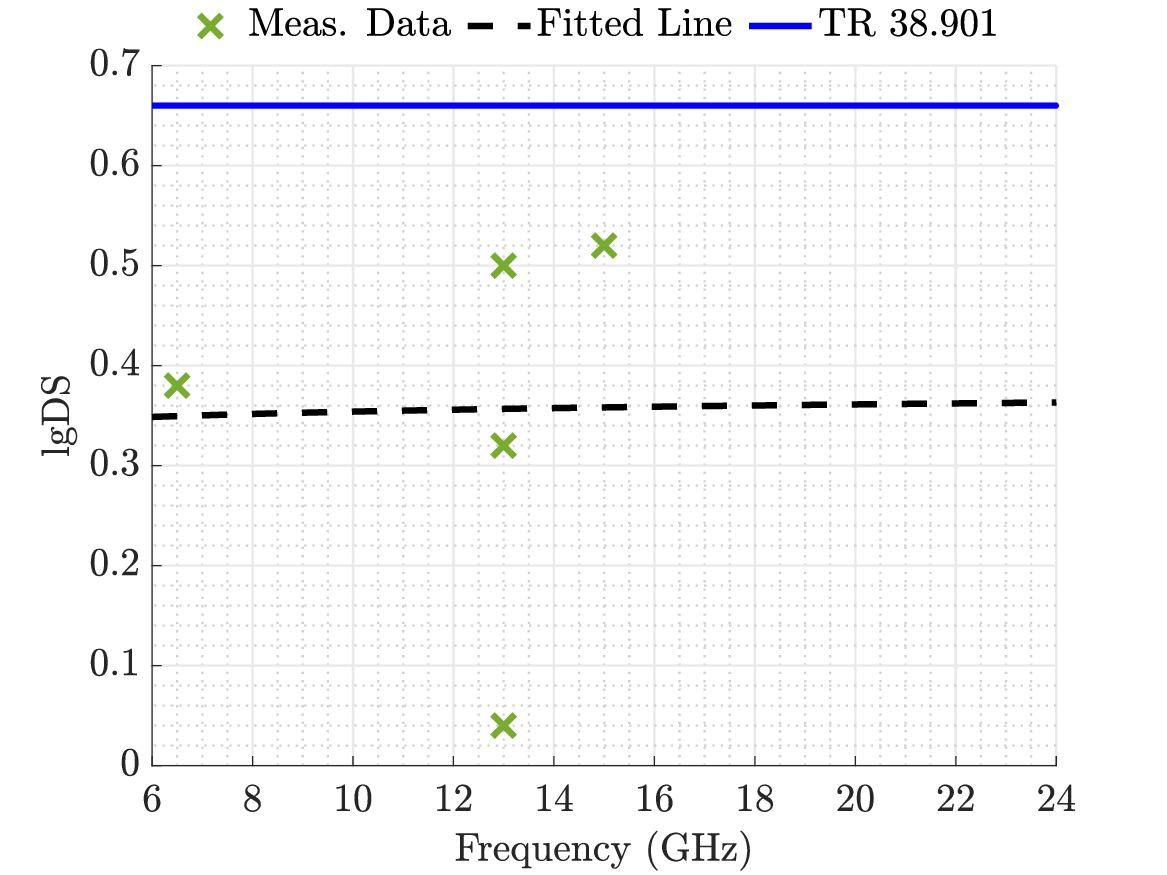}
        \caption{Meas. data for standard deviation of lgDS in the UMa LOS scenario over 6–24 GHz from 3GPP Rel-19, with an OLS fitted line (0.024 log$_{10}$(f) + 0.33~\cite{r1-2502415}) and the 3GPP TR 38.901 model (0.66~\cite{tr38901v18}).}
        \label{fig:uma_los_std_ds_rel19}
    \end{subfigure}
    \hfill
    \begin{subfigure}[b]{0.48\textwidth}
        \centering
        \includegraphics[width=\linewidth]{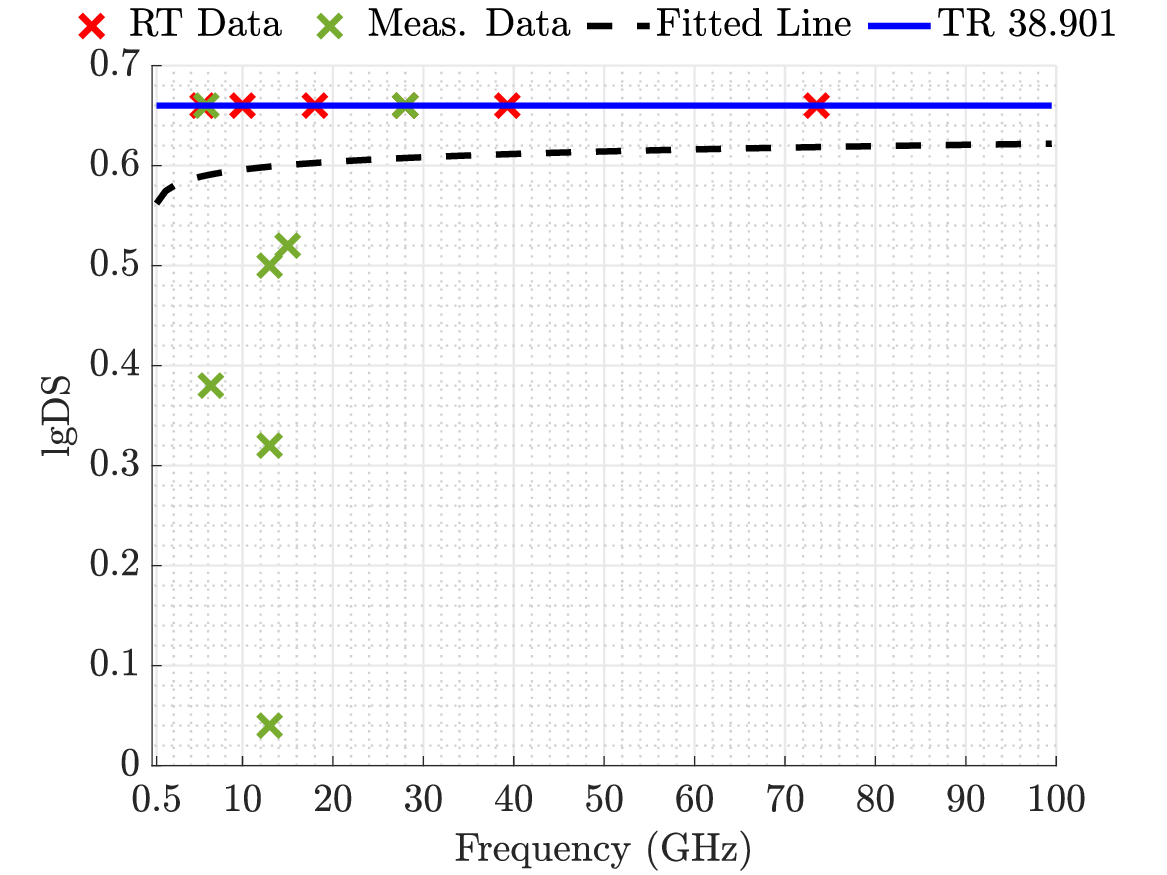}
        \caption{RT and Meas. data for standard deviation of lgDS in the UMa LOS scenario over 0.5-100 GHz from 3GPP Rel-14 and Rel-19, with a WLS fitted line (0.026 log$_{10}$(f) + 0.57~\cite{r1-2502415}) and the 3GPP TR 38.901 model (0.66~\cite{tr38901v18}).}
        \label{fig:uma_los_std_ds_wls}
    \end{subfigure}
    
    \vspace{1em}

    \begin{subfigure}[b]{0.48\textwidth}
        \centering
        \includegraphics[width=\linewidth]{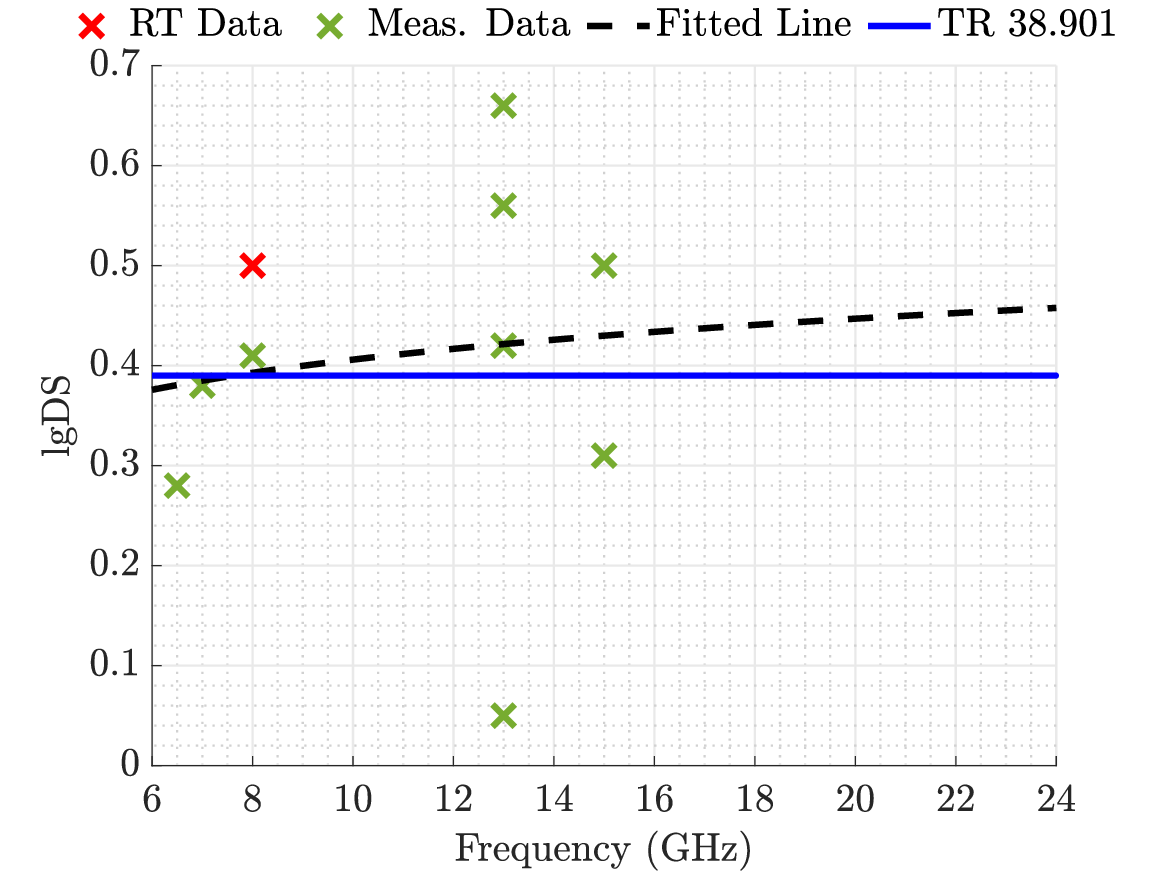}
        \caption{RT and Meas. data for standard deviation of lgDS in the UMa NLOS scenario over 6–24 GHz from 3GPP Rel-19, with an OLS fitted line (0.136 log$_{10}$(f) + 0.27~\cite{r1-2502415}) and the 3GPP TR 38.901 model (0.39~\cite{tr38901v18}).}
        \label{fig:uma_nlos_std_ds_rel19}
    \end{subfigure}
    \hfill
    \begin{subfigure}[b]{0.48\textwidth}
        \centering
        \includegraphics[width=\linewidth]{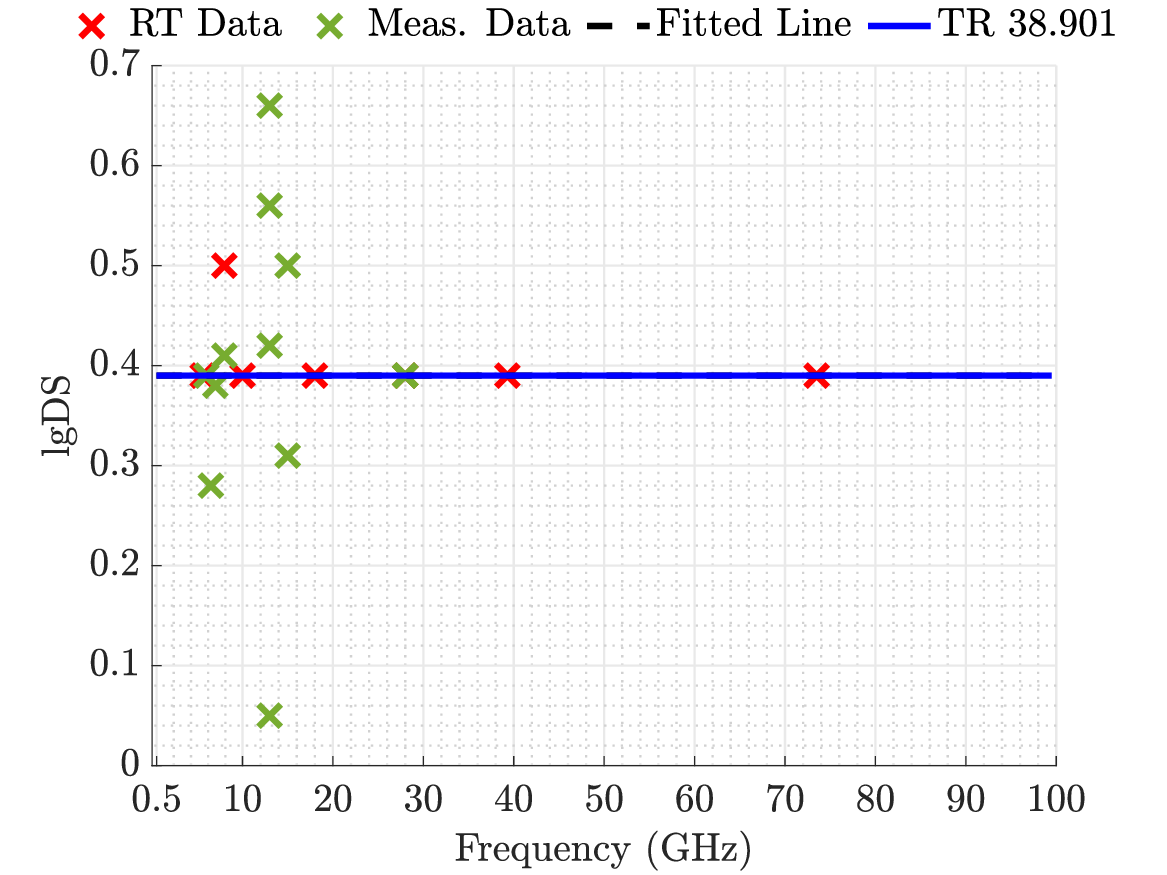}
        \caption{RT and Meas. data for standard deviation of lgDS in the UMa NLOS scenario over 0.5-100 GHz from 3GPP Rel-14 and Rel-19, with a WM fitted line (0.39~\cite{r1-2502415}) and the 3GPP TR 38.901 model (0.39~\cite{tr38901v18}).}
        \label{fig:uma_nlos_std_ds_wls}
    \end{subfigure}

    \caption{Curve fitting of RT and Meas. data for the standard deviation of lgDS in UMa LOS and NLOS channel conditions using ordinary least squares (OLS), weighted least squares (WLS) methods and  weighted mean (WM) methods. (a) and (c) use data from 3GPP Rel-19 only (6–24 GHz), while (b) and (d) use combined data from Rel-14 and Rel-19 (0.5–100 GHz). All subfigures include a comparison with the existing 3GPP TR 38.901 model.}
    \label{fig:uma_std_ds}
\end{figure*}

\clearpage
\subsection{UMa ASD}
\begin{figure*}[h]
    \centering
    \begin{subfigure}[b]{0.48\textwidth}
        \centering
        \includegraphics[width=\linewidth]{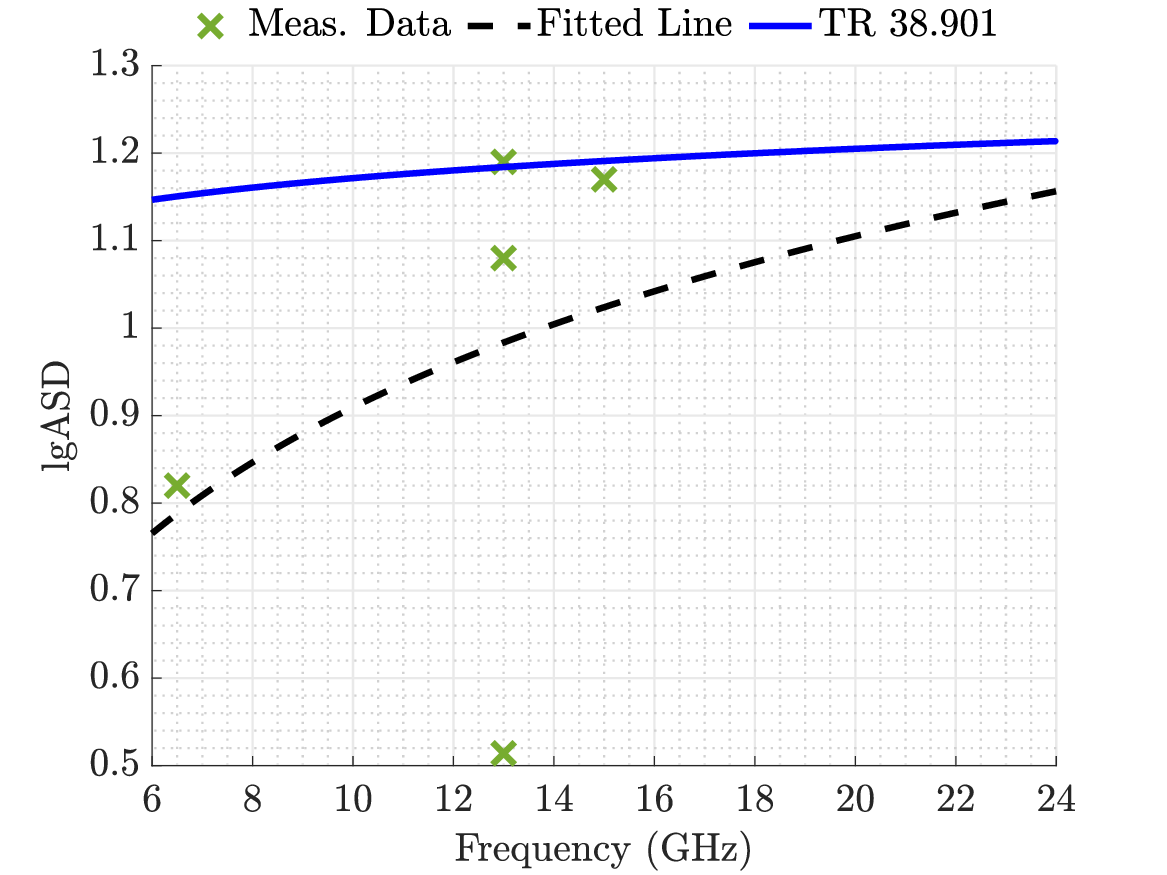}
        \caption{Meas. data for mean of lgASD in the UMa LOS scenario over 6–24 GHz from 3GPP Rel-19, with an OLS fitted line (0.6495 log$_{10}$(f) + 0.26~\cite{r1-2502415}) and the 3GPP TR 38.901 model (0.1114 log$_{10}$(f) + 1.06~\cite{tr38901v18}).}
        \label{fig:uma_los_mean_asd_rel19}
    \end{subfigure}
    \hfill
    \begin{subfigure}[b]{0.48\textwidth}
        \centering
        \includegraphics[width=\linewidth]{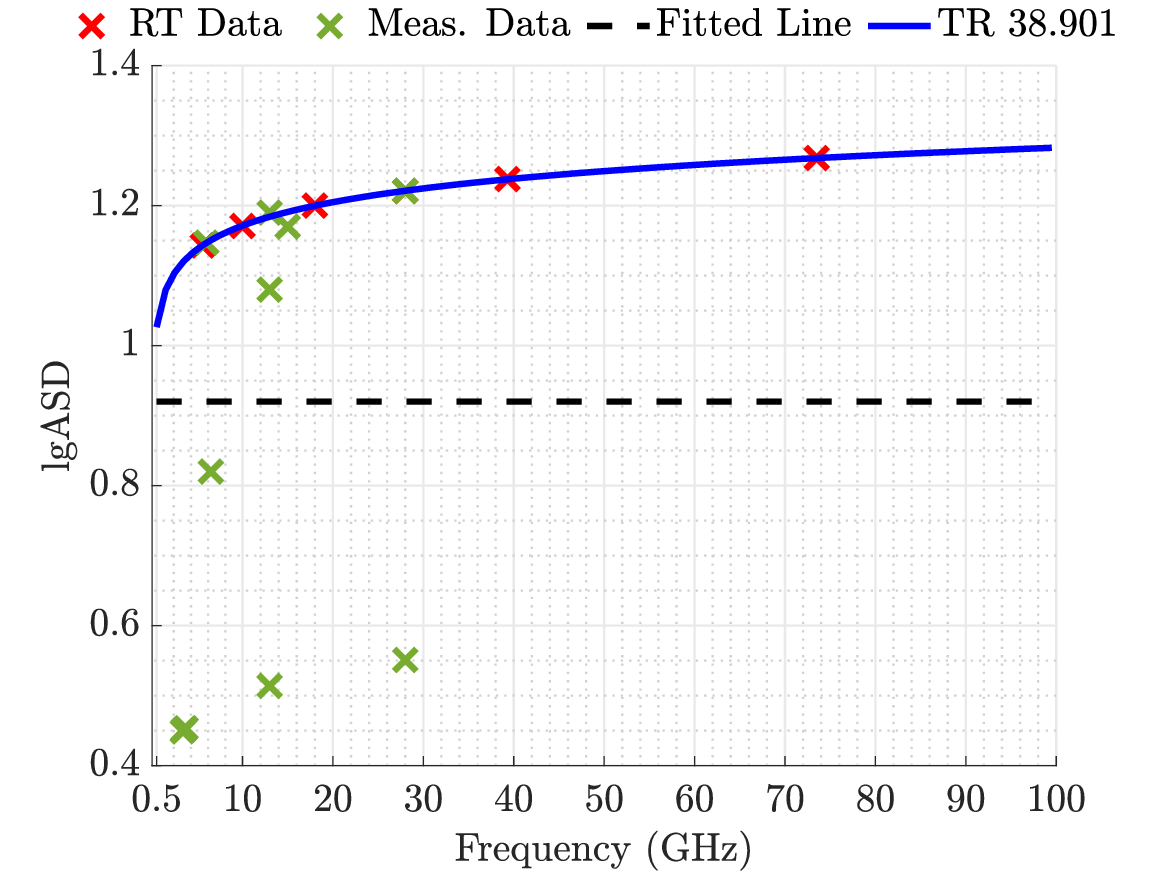}
        \caption{RT and Meas. data for mean of lgASD in the UMa LOS scenario over 0.5-100 GHz from 3GPP Rel-14 and Rel-19, with a WM fitted line (0.92~\cite{r1-2502415}) and the 3GPP TR 38.901 model (0.1114 log$_{10}$(f) + 1.06~\cite{tr38901v18}).}
        \label{fig:uma_los_mean_asd_wls}
    \end{subfigure}
    
    \vspace{1em}

    \begin{subfigure}[b]{0.48\textwidth}
        \centering
        \includegraphics[width=\linewidth]{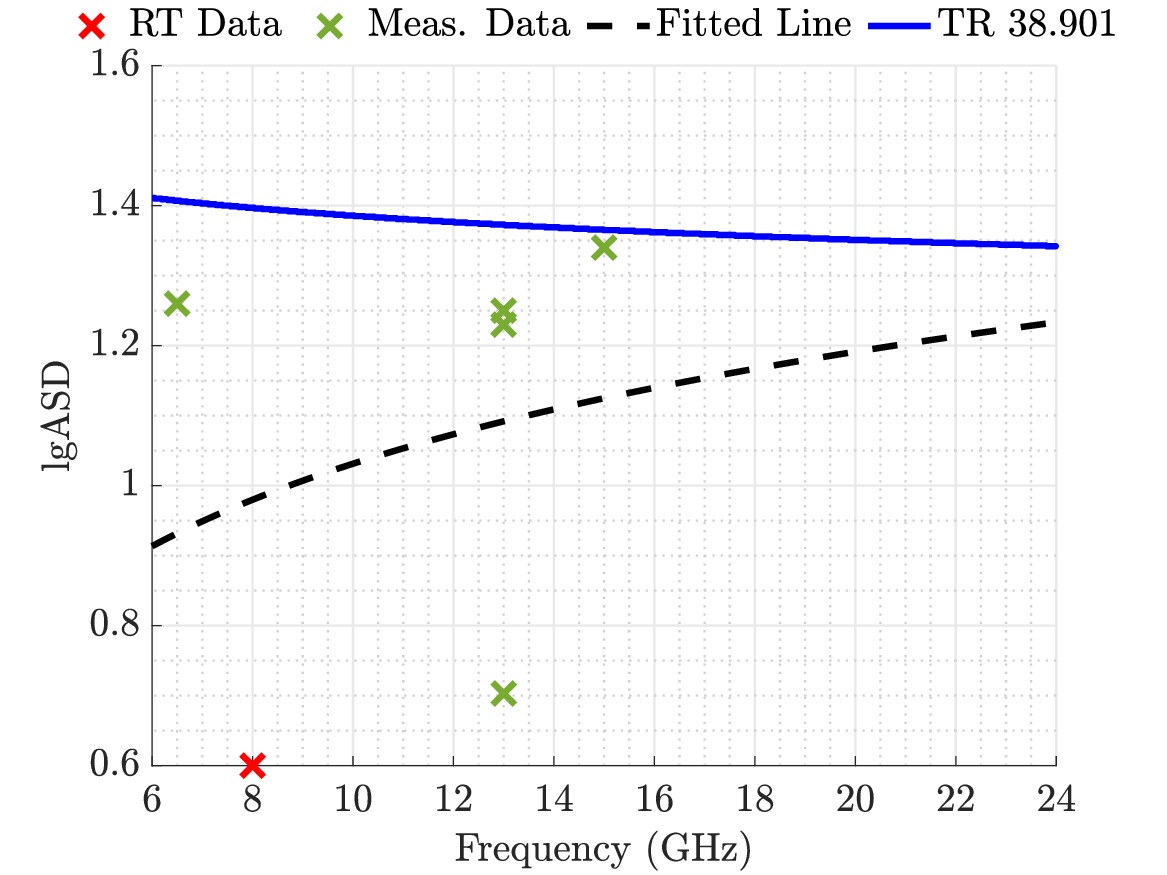}
        \caption{RT and Meas. data for mean of lgASD in the UMa NLOS scenario over 6–24 GHz from 3GPP Rel-19, with an OLS fitted line (0.5313 log$_{10}$(f) + 0.5~\cite{r1-2502415}) and the 3GPP TR 38.901 model (-0.1144 log$_{10}$(f) + 1.5~\cite{tr38901v18}).}
        \label{fig:uma_nlos_mean_asd_rel19}
    \end{subfigure}
    \hfill
    \begin{subfigure}[b]{0.48\textwidth}
        \centering
        \includegraphics[width=\linewidth]{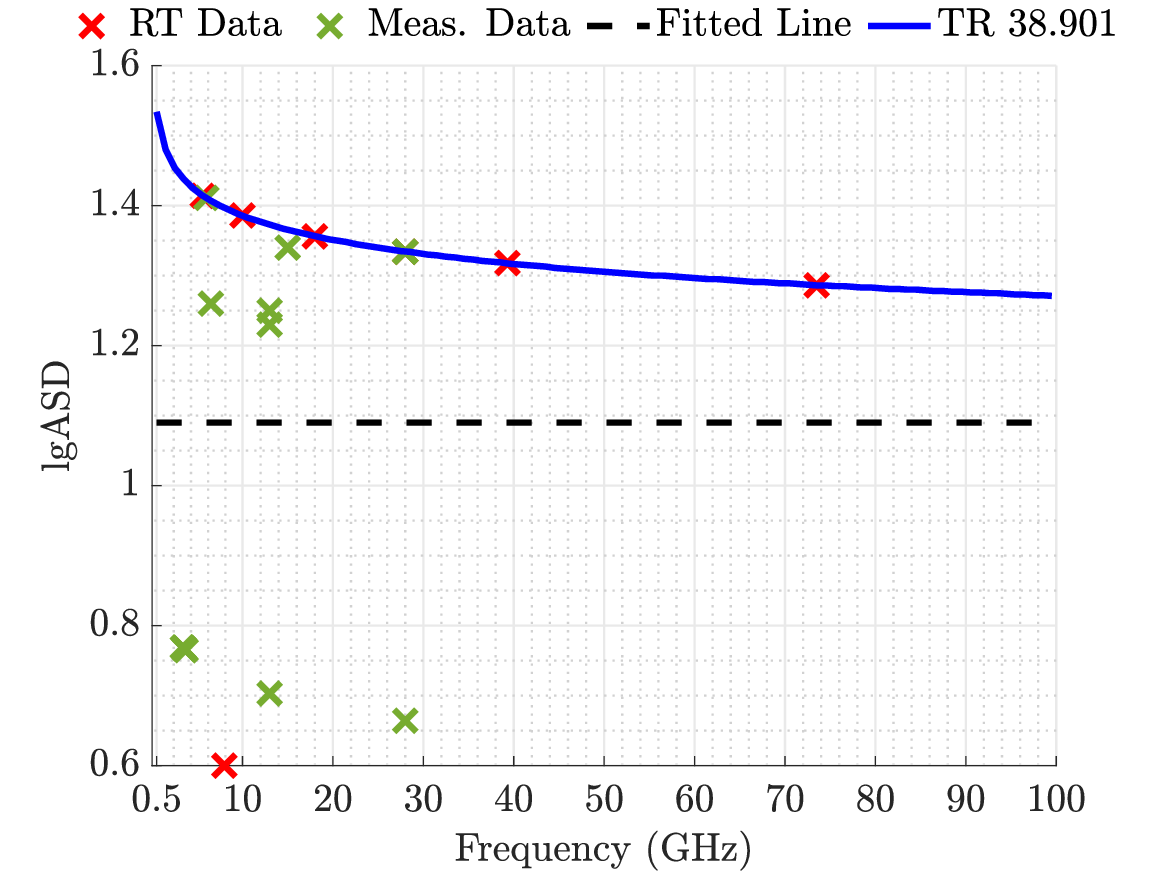}
        \caption{RT and Meas. data for mean of lgASD in the UMa NLOS scenario over 0.5-100 GHz from 3GPP Rel-14 and Rel-19, with a WM fitted line (1.09~\cite{r1-2502415}) and the 3GPP TR 38.901 model (-0.1144 log$_{10}$(f) + 1.5~\cite{tr38901v18}).}
        \label{fig:uma_nlos_mean_asd_wls}
    \end{subfigure}

    \caption{Curve fitting of RT and Meas. data for the mean of lgASD in UMa LOS and NLOS channel conditions using ordinary least squares (OLS) and weighted mean (WM) methods. (a) and (c) use data from 3GPP Rel-19 only (6–24 GHz), while (b) and (d) use combined data from Rel-14 and Rel-19 (0.5–100 GHz). All subfigures include a comparison with the existing 3GPP TR 38.901 model.}
    \label{fig:uma_mean_asd}
\end{figure*}

%%%%%%%%% UMa ASD STD %%%%%%%%%%%%%
\begin{figure*}[h]
    \centering
    \begin{subfigure}[b]{0.48\textwidth}
        \centering
        \includegraphics[width=\linewidth]{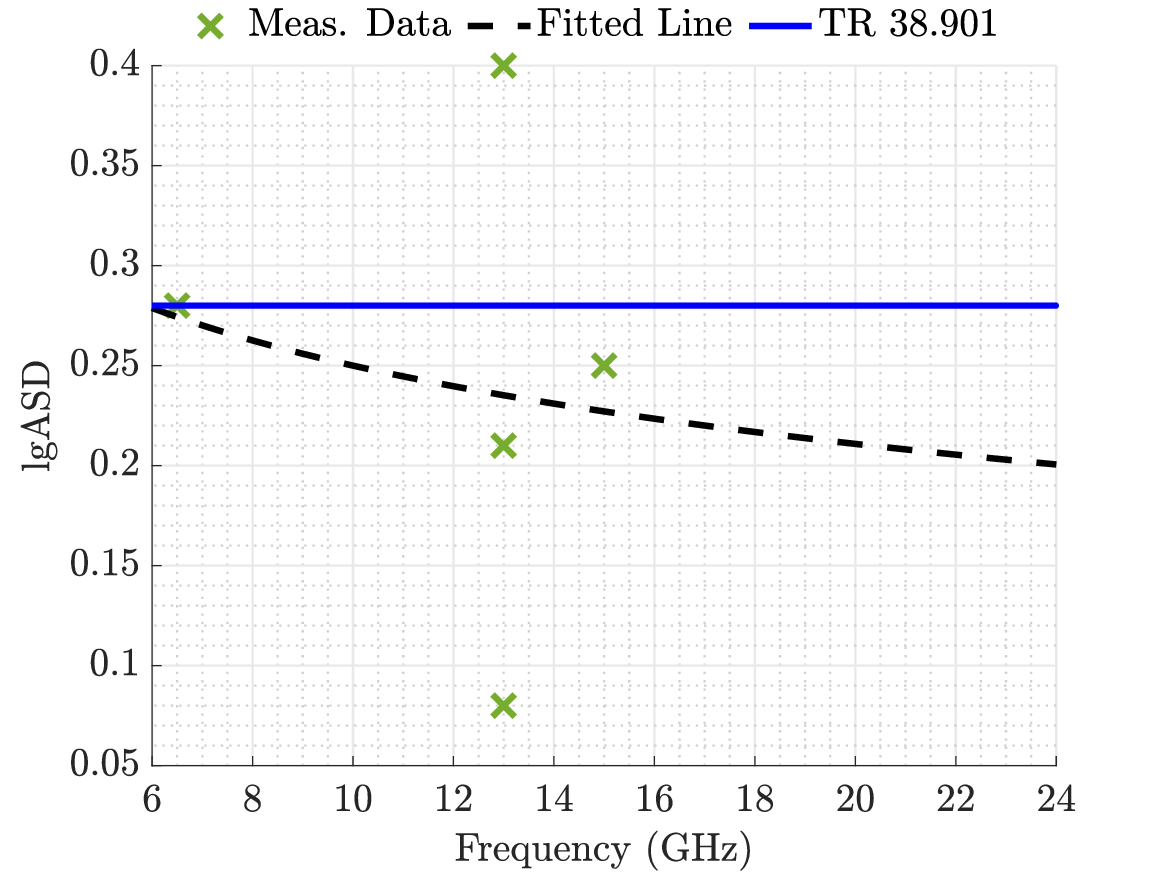}
        \caption{Meas. data for standard deviation of lgASD in the UMa LOS scenario over 6–24 GHz from 3GPP Rel-19, with an OLS fitted line (-0.13 log$_{10}$(f) + 0.38~\cite{r1-2502415}) and the 3GPP TR 38.901 model (0.28~\cite{tr38901v18}).}
        \label{fig:uma_los_std_asd_rel19}
    \end{subfigure}
    \hfill
    \begin{subfigure}[b]{0.48\textwidth}
        \centering
        \includegraphics[width=\linewidth]{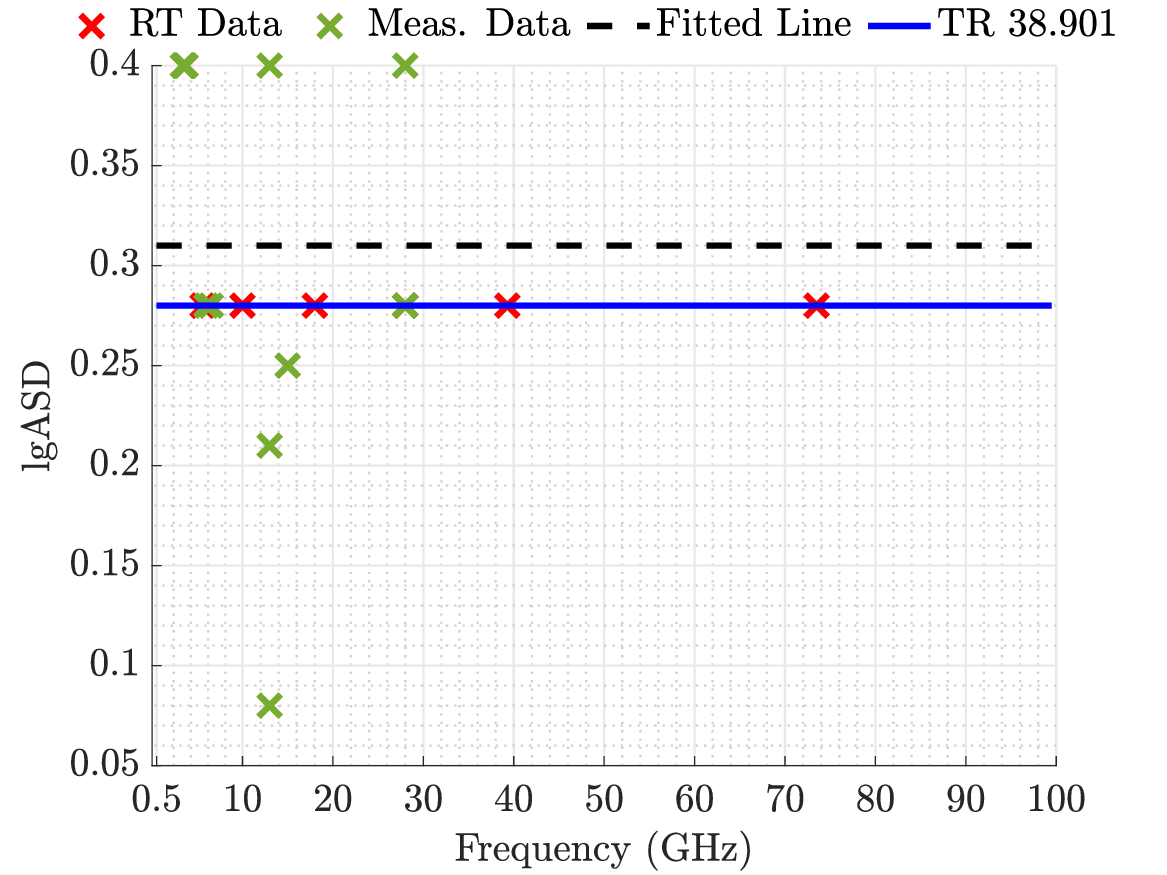}
        \caption{RT and Meas. data for standard deviation of lgASD in the UMa LOS scenario over 0.5-100 GHz from 3GPP Rel-14 and Rel-19, with a WM fitted line (0.31~\cite{r1-2502415}) and the 3GPP TR 38.901 model (0.28~\cite{tr38901v18}).}
        \label{fig:uma_los_std_asd_wls}
    \end{subfigure}
    
    \vspace{1em}

    \begin{subfigure}[b]{0.48\textwidth}
        \centering
        \includegraphics[width=\linewidth]{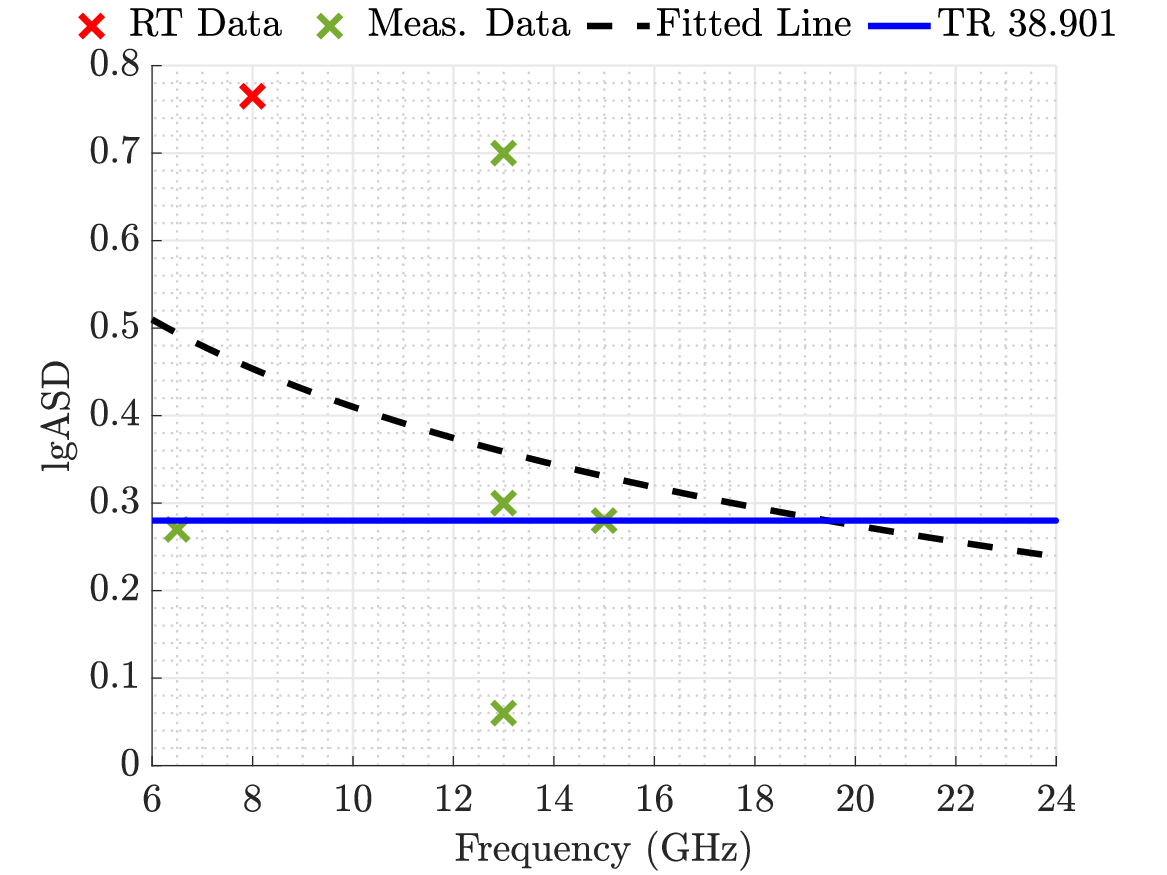}
        \caption{RT and Meas. data for mean of lgASD in the UMa NLOS scenario over 6–24 GHz from 3GPP Rel-19, with an OLS fitted line (-0.45 log$_{10}$(f) + 0.86~\cite{r1-2502415}) and the 3GPP TR 38.901 model (0.28~\cite{tr38901v18}).}
        \label{fig:uma_nlos_std_asd_rel19}
    \end{subfigure}
    \hfill
    \begin{subfigure}[b]{0.48\textwidth}
        \centering
        \includegraphics[width=\linewidth]{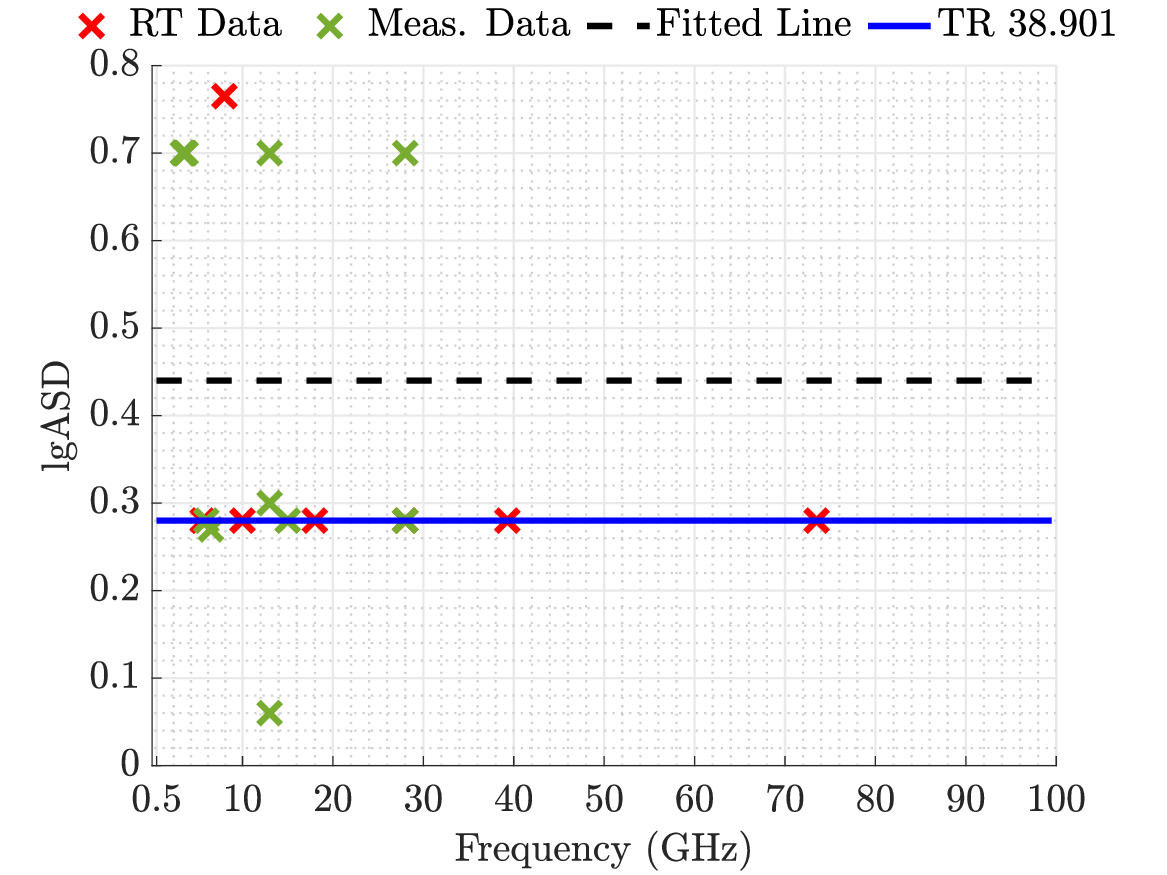}
        \caption{RT and Meas. data for mean of lgASD in the UMa NLOS scenario over 0.5-100 GHz from 3GPP Rel-14 and Rel-19, with a WM fitted line (0.44~\cite{r1-2502415}) and the 3GPP TR 38.901 model (0.28~\cite{tr38901v18}).}
        \label{fig:uma_nlos_std_asd_wls}
    \end{subfigure}

    \caption{Curve fitting of RT and Meas. data for the standard deviation of lgASD in UMa LOS and NLOS channel conditions using ordinary least squares (OLS) and weighted mean (WM) methods. (a) and (c) use data from 3GPP Rel-19 only (6–24 GHz), while (b) and (d) use combined data from Rel-14 and Rel-19 (0.5–100 GHz). All subfigures include a comparison with the existing 3GPP TR 38.901 model.}
    \label{fig:uma_std_asd}
\end{figure*}

\clearpage
\subsection{UMa ASA}
\begin{figure*}[h]
    \centering
    \begin{subfigure}[b]{0.48\textwidth}
        \centering
        \includegraphics[width=\linewidth]{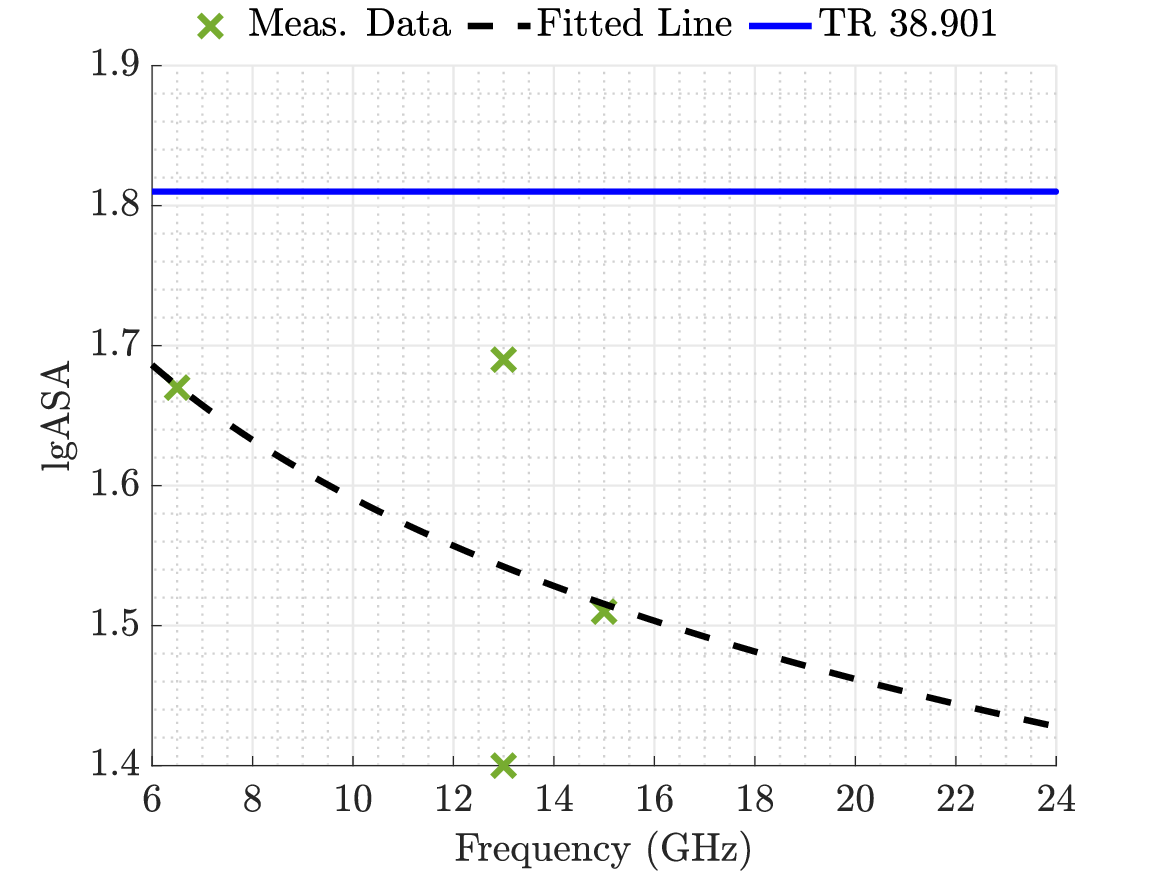}
        \caption{Meas. data for mean of lgASA in the UMa LOS scenario over 6–24 GHz from 3GPP Rel-19, with an OLS fitted line (-0.429 log$_{10}$(f) + 2.02~\cite{r1-2502415}) and the 3GPP TR 38.901 model (1.81 \cite{tr38901v18}).}
        \label{fig:uma_los_mean_asa_rel19}
    \end{subfigure}
    \hfill
    \begin{subfigure}[b]{0.48\textwidth}
        \centering
        \includegraphics[width=\linewidth]{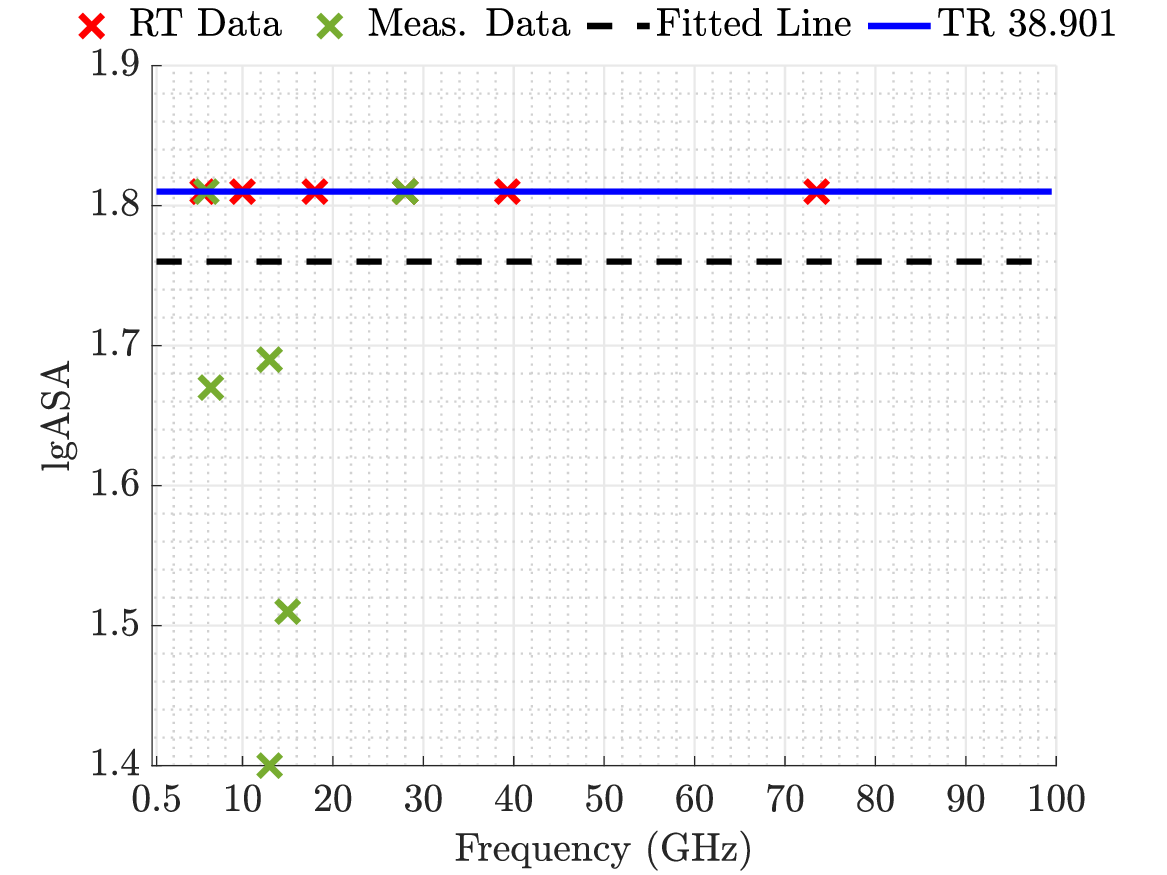}
        \caption{RT and Meas. data for mean of lgASA in the UMa LOS scenario over 0.5-100 GHz from 3GPP Rel-14 and Rel-19, with a WM fitted line (1.76 \cite{r1-2502415}) and the 3GPP TR 38.901 model (1.81 \cite{tr38901v18}).}
        \label{fig:uma_los_mean_asa_wls}
    \end{subfigure}
    
    \vspace{1em}

    \begin{subfigure}[b]{0.48\textwidth}
        \centering
        \includegraphics[width=\linewidth]{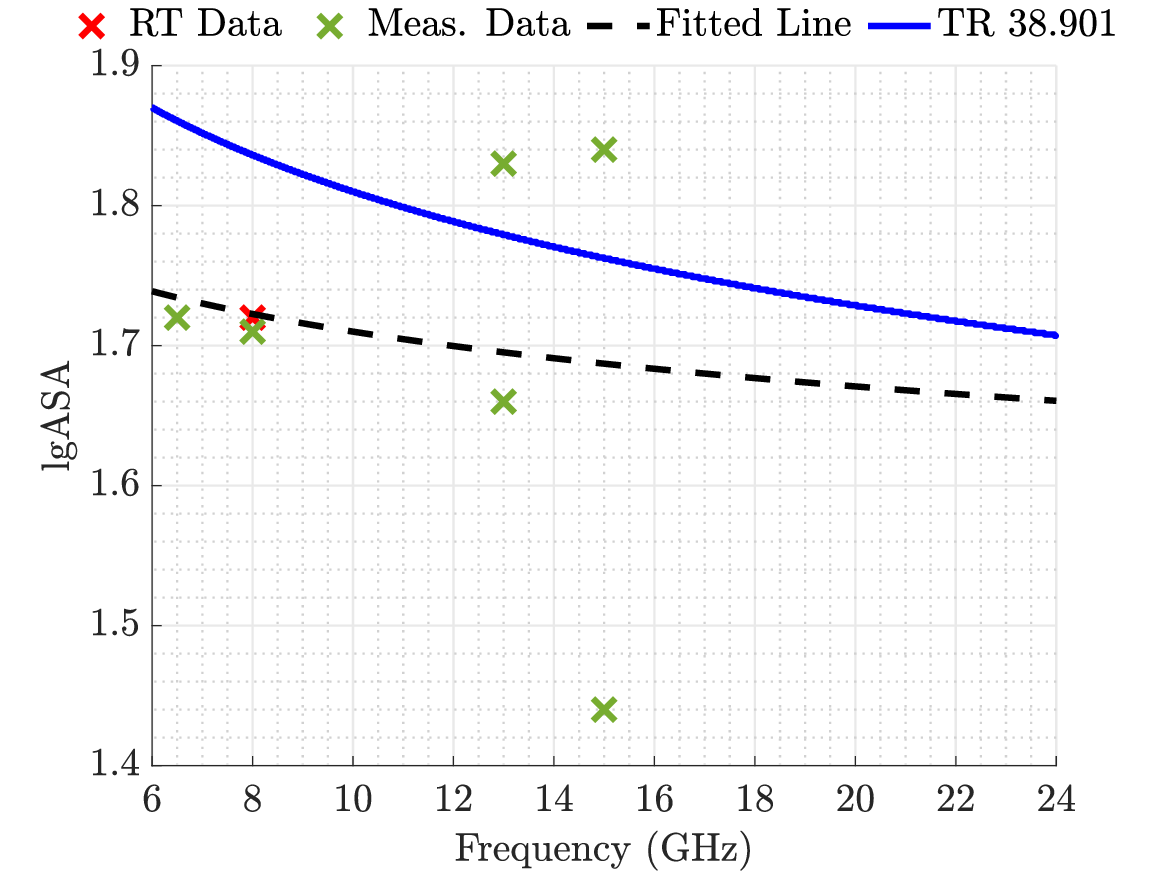}
        \caption{RT and Meas. data for mean of lgASA in the UMa NLOS scenario over 6–24 GHz from 3GPP Rel-19, with an OLS fitted line (-0.13 log$_{10}$(f) + 1.84~\cite{r1-2502415}) and the 3GPP TR 38.901 model (-0.27 log$_{10}$(f) + 2.08~\cite{tr38901v18}).}
        \label{fig:uma_nlos_mean_asa_rel19}
    \end{subfigure}
    \hfill
    \begin{subfigure}[b]{0.48\textwidth}
        \centering
        \includegraphics[width=\linewidth]{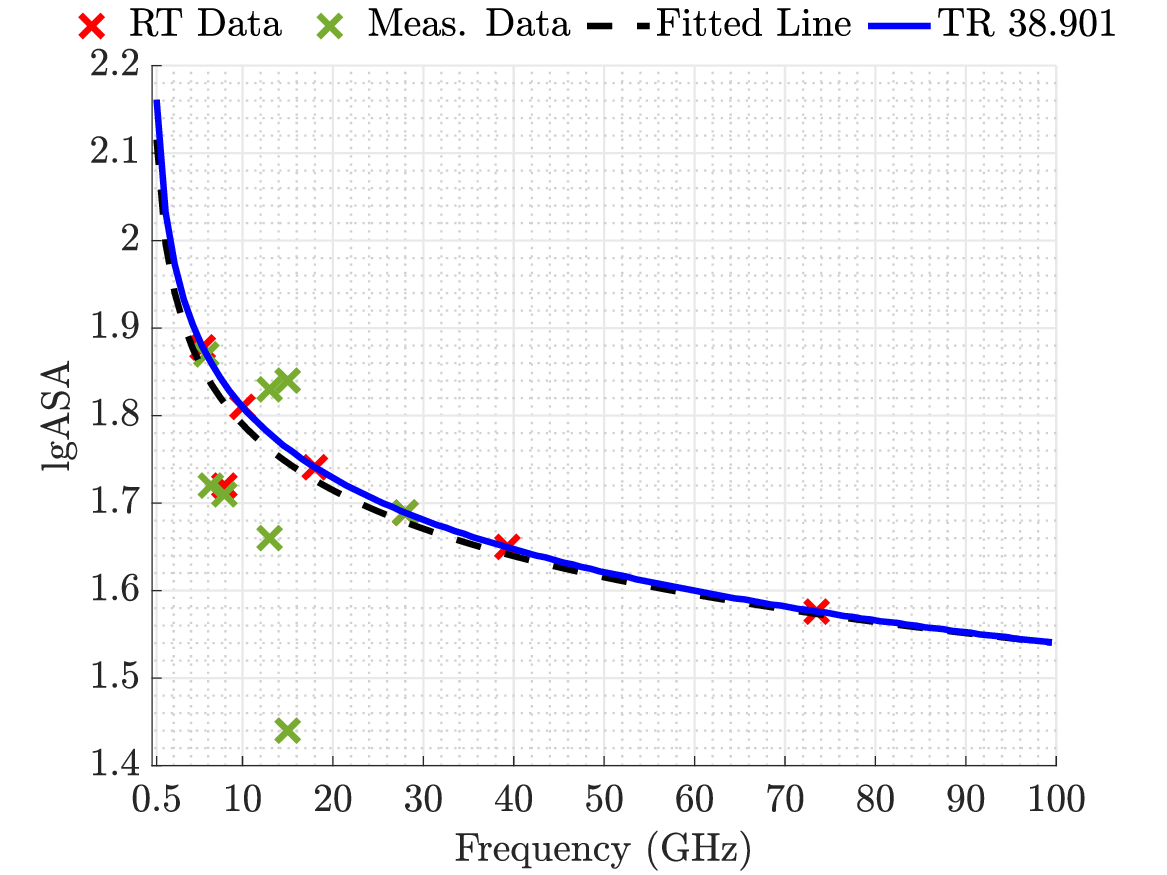}
        \caption{RT and Meas. data for mean of lgASA in the UMa NLOS scenario over 0.5-100 GHz from 3GPP Rel-14 and Rel-19, with a WLS fitted line (-0.25 log$_{10}$(f) + 2.04~\cite{r1-2502415}) and the 3GPP TR 38.901 model (-0.27 log$_{10}$(f) + 2.08~\cite{tr38901v18}).}
        \label{fig:uma_nlos_mean_asa_wls}
    \end{subfigure}

    \caption{Curve fitting of RT and Meas. data for the mean of lgASA in UMa LOS and NLOS channel conditions using  ordinary least squares (OLS), weighted mean (WM) and weighted least squares (WLS) methods. (a) and (c) use data from 3GPP Rel-19 only (6–24 GHz), while (b) and (d) use combined data from Rel-14 and Rel-19 (0.5–100 GHz). All subfigures include a comparison with the existing 3GPP TR 38.901 model.}
    \label{fig:uma_mean_asa}
\end{figure*}

%%%%%%%%% UMa ASA STD %%%%%%%%%%%%%
\begin{figure*}[h]
    \centering
    \begin{subfigure}[b]{0.48\textwidth}
        \centering
        \includegraphics[width=\linewidth]{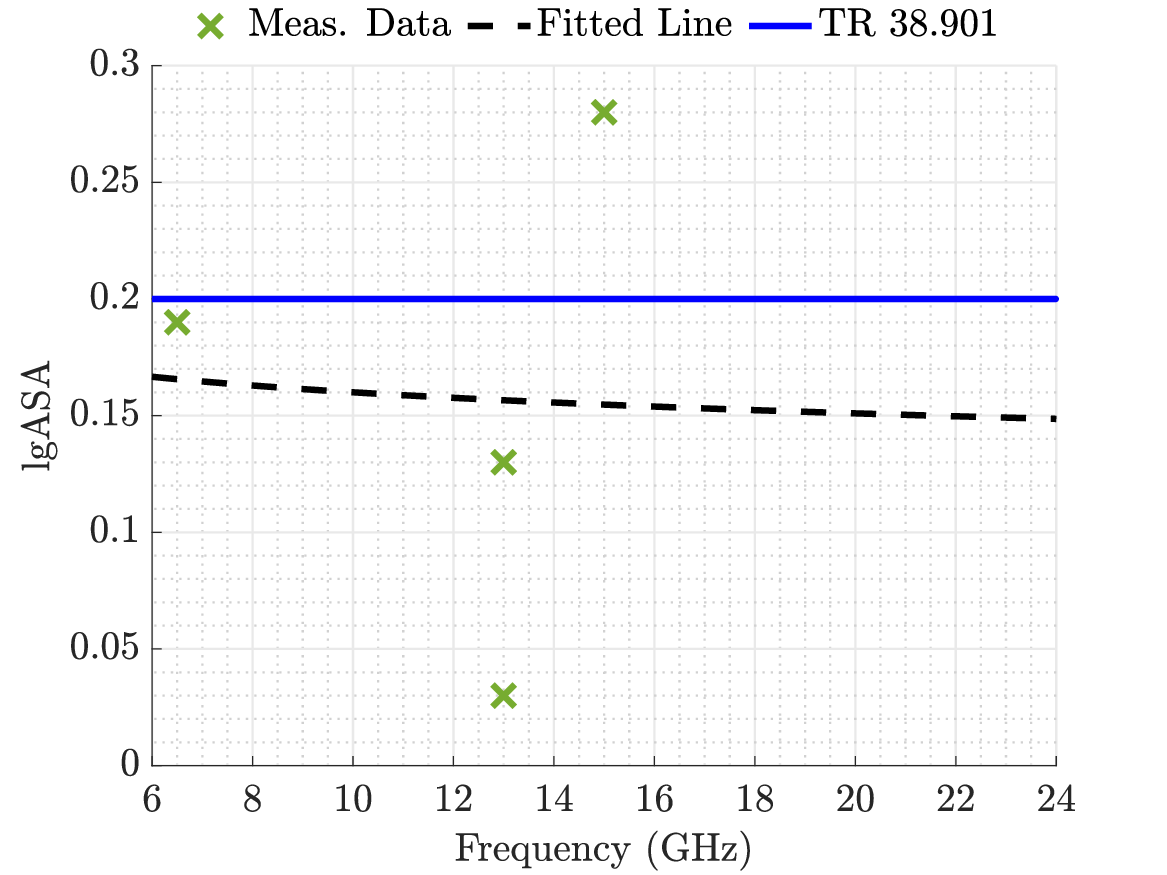}
        \caption{Meas. data for standard deviation of lgASA in the UMa LOS scenario over 6–24 GHz from 3GPP Rel-19, with an OLS fitted line (-0.03 log$_{10}$(f) + 0.19~\cite{r1-2502415}) and the 3GPP TR 38.901 model (0.20~\cite{tr38901v18}).}
        \label{fig:uma_los_std_asa_rel19}
    \end{subfigure}
    \hfill
    \begin{subfigure}[b]{0.48\textwidth}
        \centering
        \includegraphics[width=\linewidth]{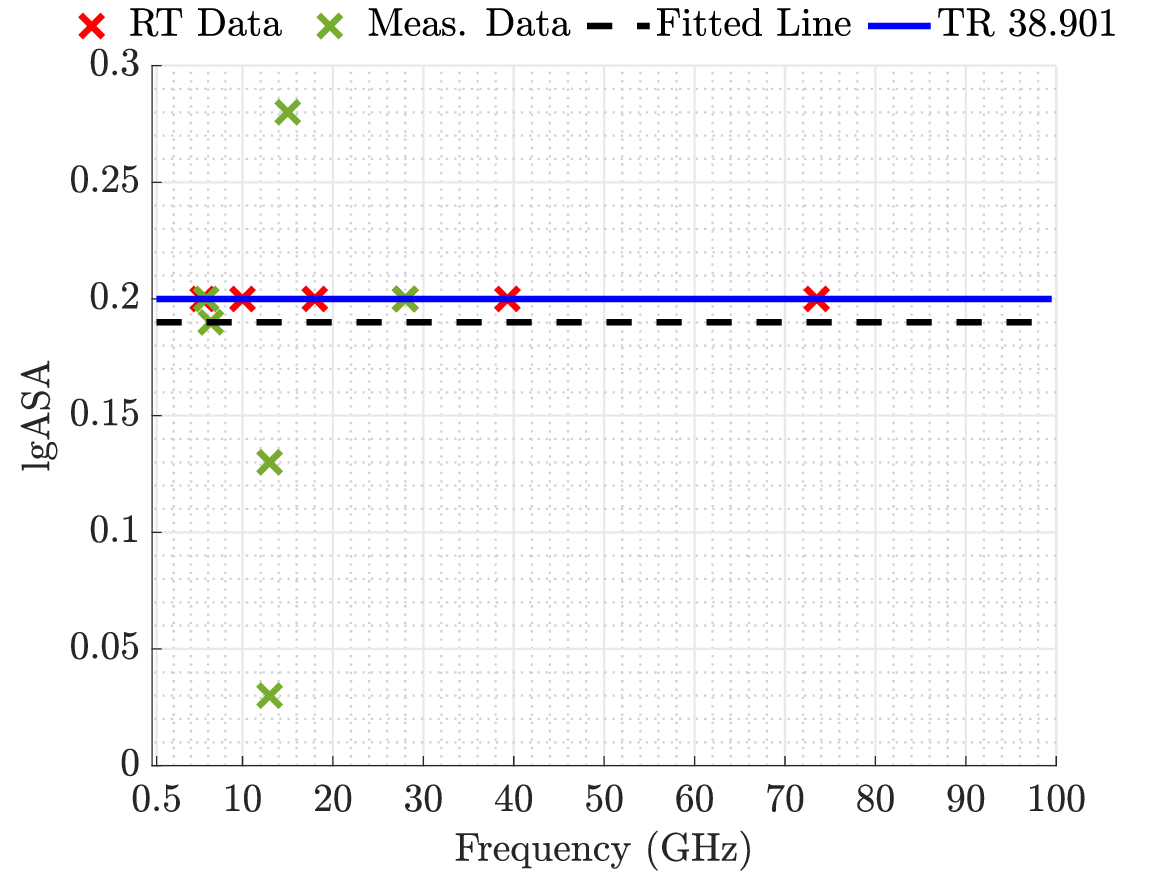}
        \caption{RT and Meas. data for standard deviation of lgASA in the UMa LOS scenario over 0.5-100 GHz from 3GPP Rel-14 and Rel-19, with a WM fitted line (0.19~\cite{r1-2502415}) and the 3GPP TR 38.901 model (0.20~\cite{tr38901v18}).}
        \label{fig:uma_los_std_asa_wls}
    \end{subfigure}
    
    \vspace{1em}

    \begin{subfigure}[b]{0.48\textwidth}
        \centering
        \includegraphics[width=\linewidth]{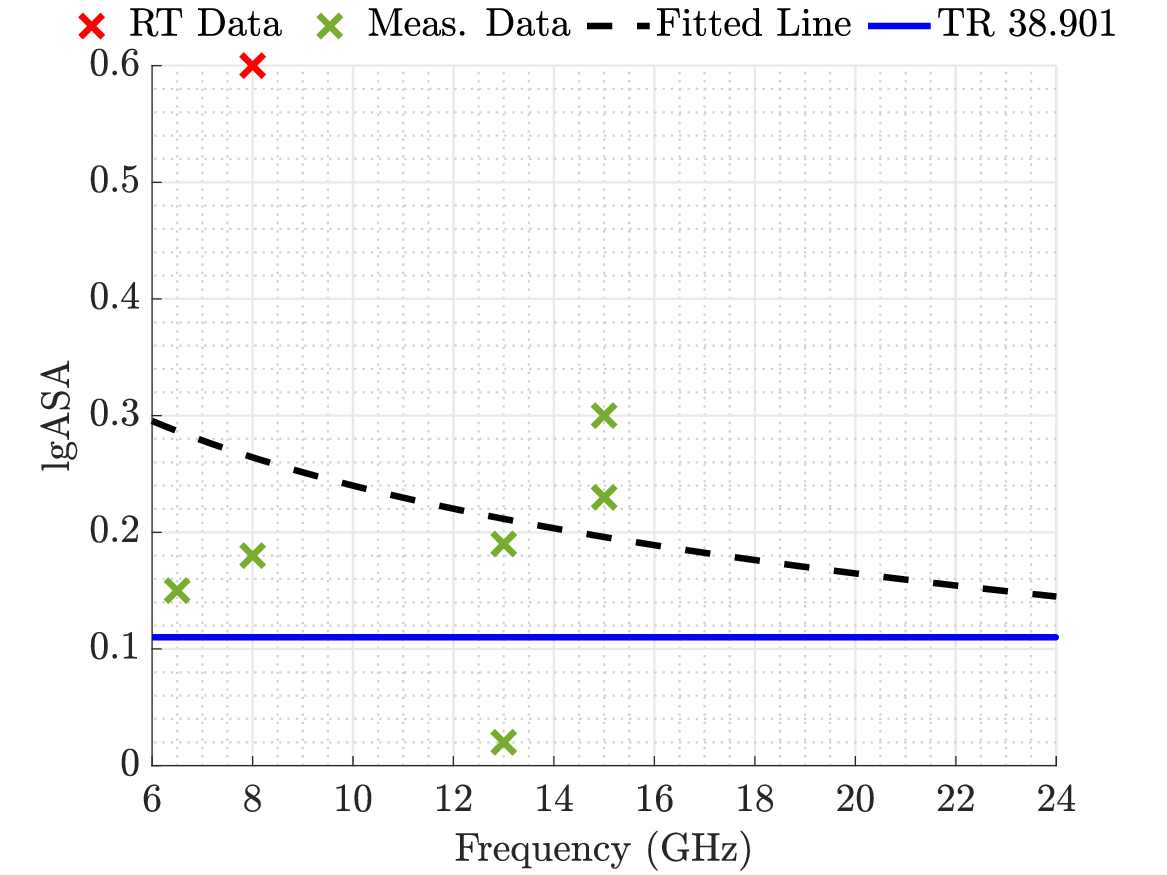}
        \caption{RT and Meas. data for standard deviation of lgASA in the UMa NLOS scenario over 6–24 GHz from 3GPP Rel-19, with an OLS fitted line (-0.25 log$_{10}$(f) + 0.49~\cite{r1-2502415}) and the 3GPP TR 38.901 model (0.11~\cite{tr38901v18}).}
        \label{fig:uma_nlos_std_asa_rel19}
    \end{subfigure}
    \hfill
    \begin{subfigure}[b]{0.48\textwidth}
        \centering
        \includegraphics[width=\linewidth]{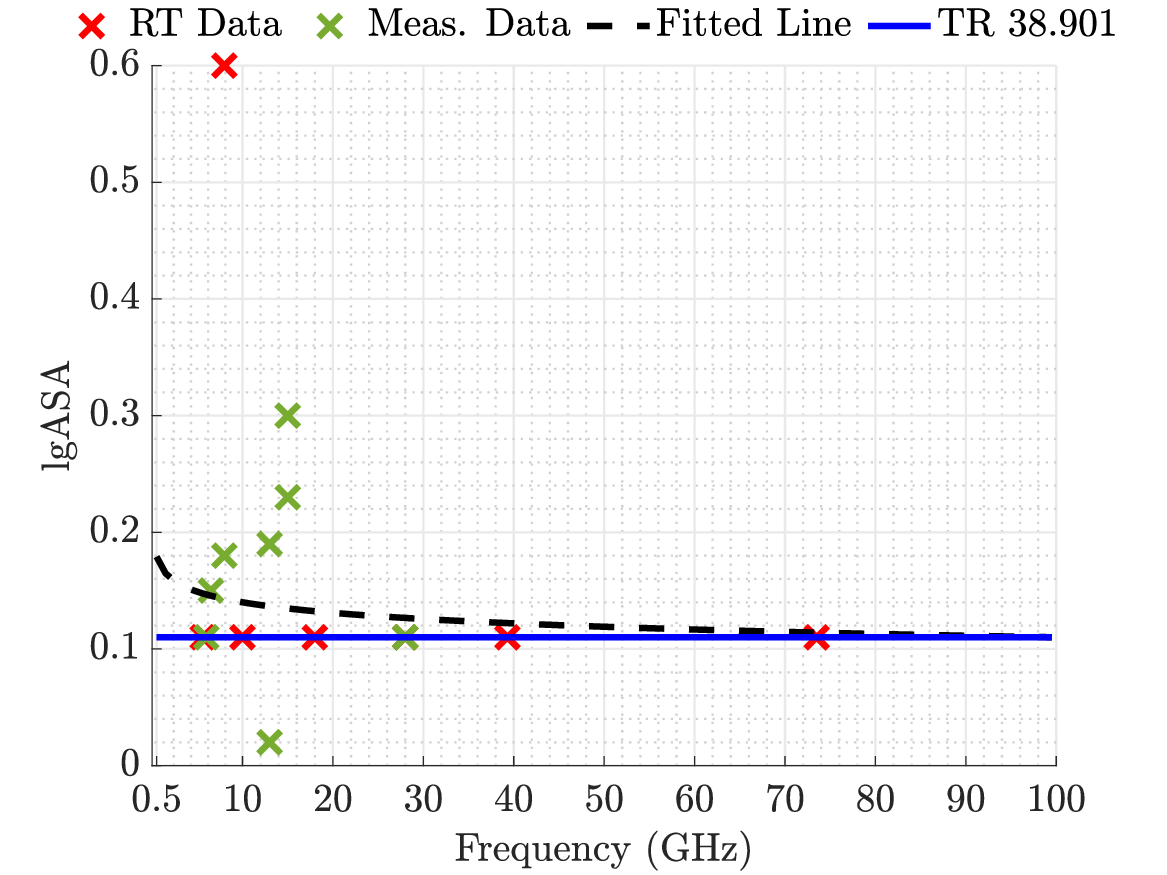}
        \caption{RT and Meas. data for standard deviation of lgASA in the UMa NLOS scenario over 0.5-100 GHz from 3GPP Rel-14 and Rel-19, with a WLS fitted line (-0.03 log$_{10}$(f) + 0.17~\cite{r1-2502415}) and the 3GPP TR 38.901 model (0.11~\cite{tr38901v18}).}
        \label{fig:uma_nlos_std_asa_wls}
    \end{subfigure}

    \caption{Curve fitting of RT and Meas. data for the standard deviation of lgASA in UMa LOS and NLOS channel conditions using ordinary least squares (OLS), weighted mean (WM) and weighted least squares (WLS) methods. (a) and (c) use data from 3GPP Rel-19 only (6–24 GHz), while (b) and (d) use combined data from Rel-14 and Rel-19 (0.5–100 GHz). All subfigures include a comparison with the existing 3GPP TR 38.901 model.}
    \label{fig:uma_std_asa}
\end{figure*}

\clearpage
\subsection{UMa ZSA}
\begin{figure*}[h]
    \centering
    \begin{subfigure}[b]{0.48\textwidth}
        \centering
        \includegraphics[width=\linewidth]{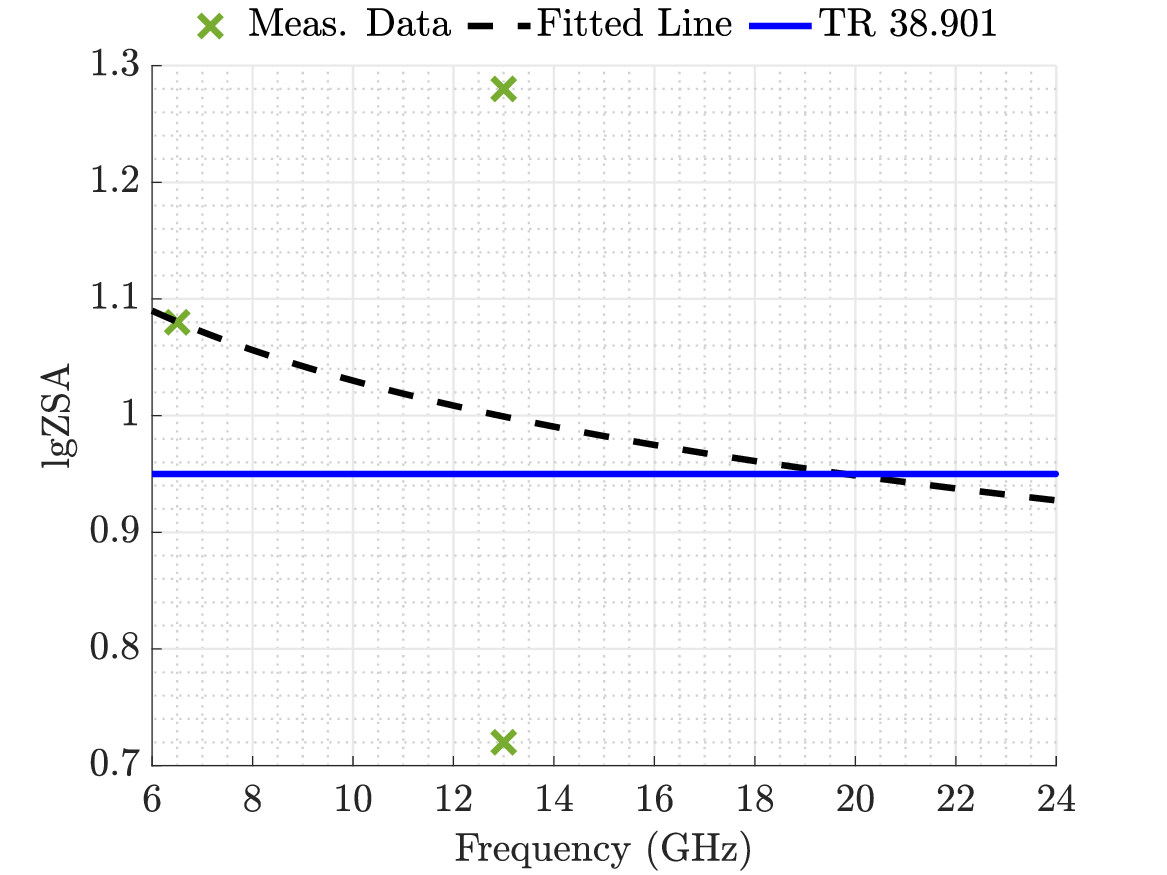}
        \caption{Meas. data for mean of lgZSA in the UMa LOS scenario over 6–24 GHz from 3GPP Rel-19, with an OLS fitted line (-0.27 log$_{10}$(f) + 1.30 ~\cite{r1-2502415}) and the 3GPP TR 38.901 model (0.95~\cite{tr38901v18}).}
        \label{fig:uma_los_mean_zsa_rel19}
    \end{subfigure}
    \hfill
    \begin{subfigure}[b]{0.48\textwidth}
        \centering
        \includegraphics[width=\linewidth]{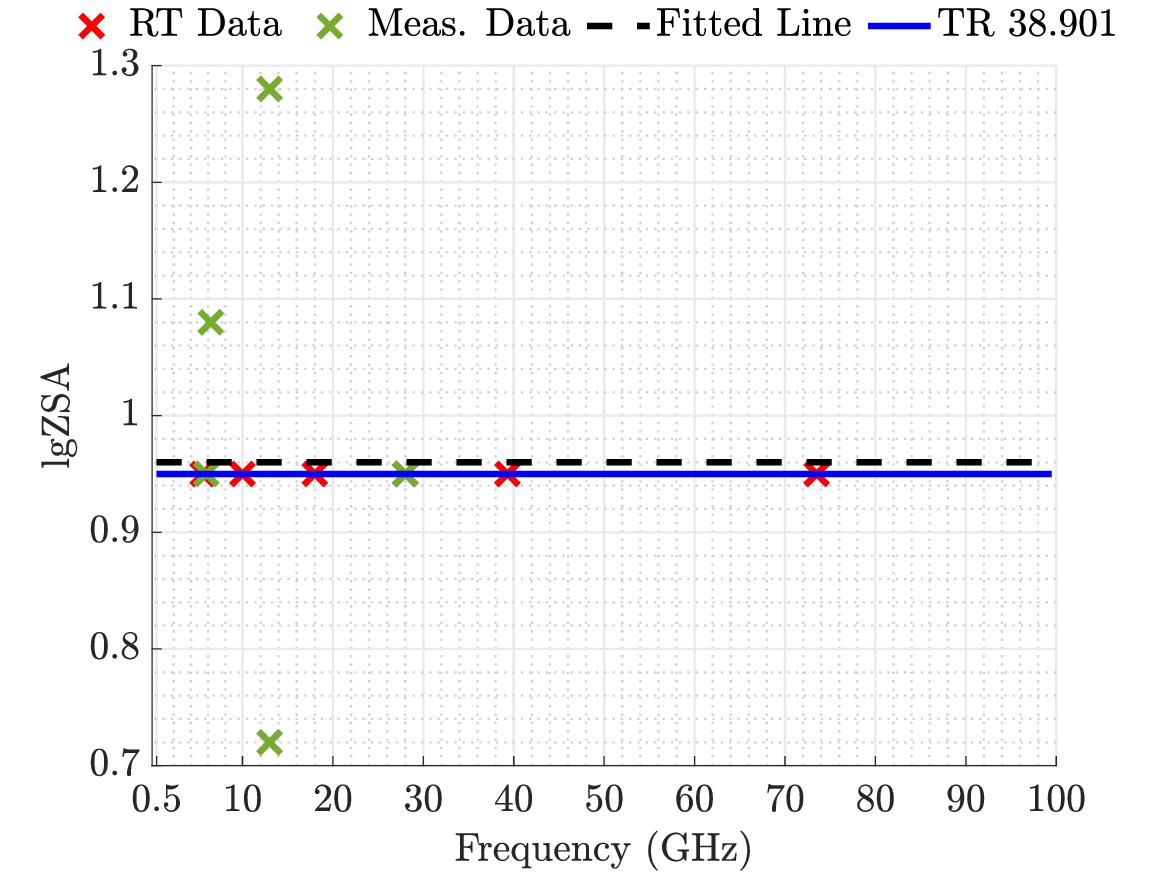}
        \caption{RT and Meas. data for mean of lgZSA in the UMa LOS scenario over 0.5-100 GHz from 3GPP Rel-14 and Rel-19, with a WM fitted line (0.96~\cite{r1-2502415}) and the 3GPP TR 38.901 model (0.95~\cite{tr38901v18}).}
        \label{fig:uma_los_mean_zsa_wls}
    \end{subfigure}
    
    \vspace{1em}

    \begin{subfigure}[b]{0.48\textwidth}
        \centering
        \includegraphics[width=\linewidth]{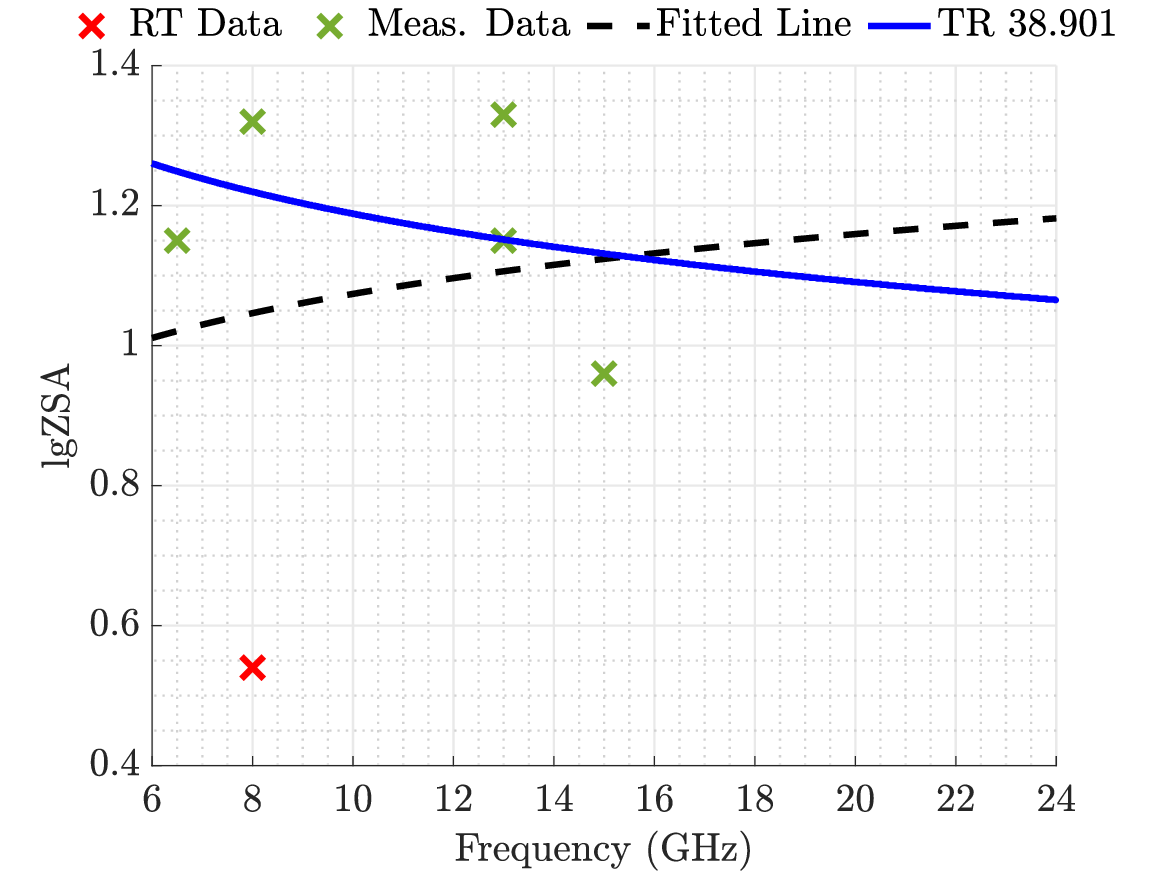}
        \caption{RT and Meas. data for mean of lgZSA in the UMa NLOS scenario over 6–24 GHz from 3GPP Rel-19, with an OLS fitted line (0.2839 log$_{10}$(f) + 0.79 ~\cite{r1-2502415}) and the 3GPP TR 38.901 model (-0.3236 log$_{10}$(f) + 1.512~\cite{tr38901v18}).}
        \label{fig:uma_nlos_mean_zsa_rel19}
    \end{subfigure}
    \hfill
    \begin{subfigure}[b]{0.48\textwidth}
        \centering
        \includegraphics[width=\linewidth]{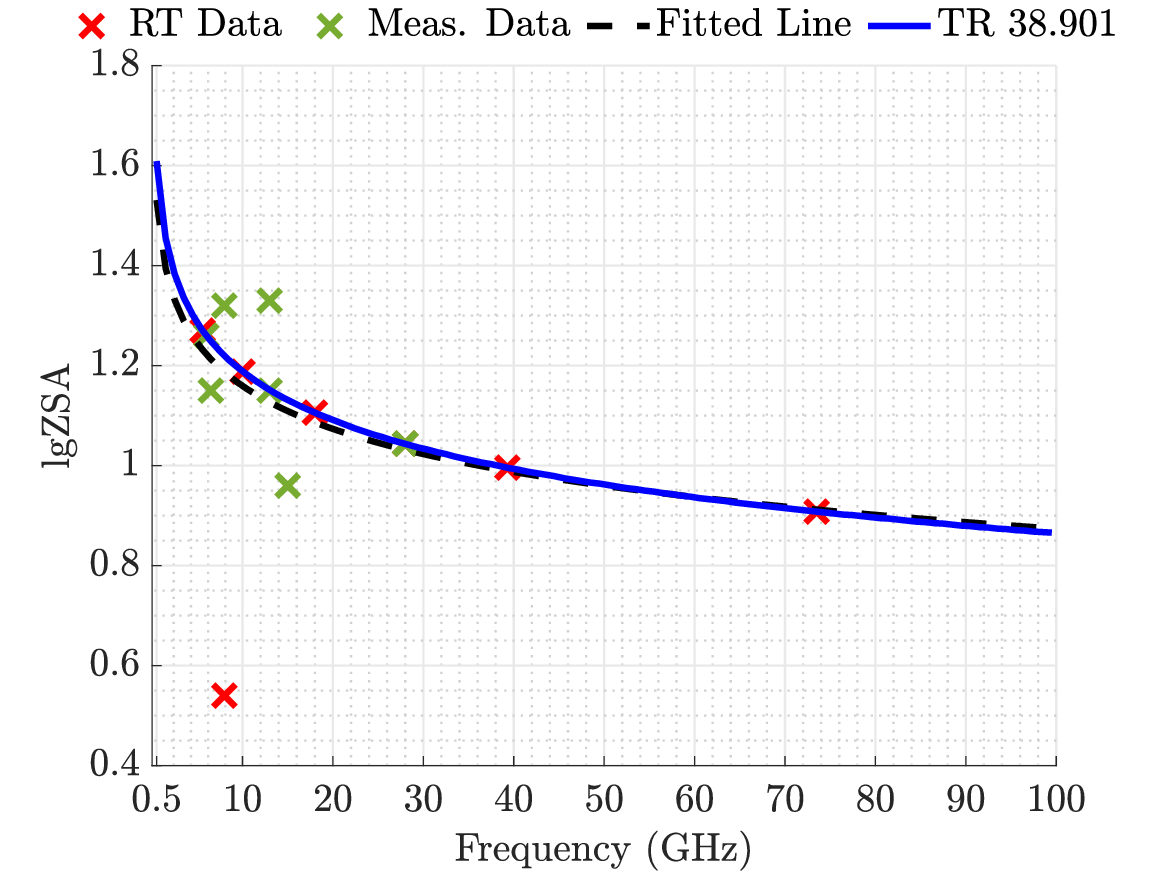}
        \caption{RT and Meas. data for mean of lgZSA in the UMa NLOS scenario over 0.5-100 GHz from 3GPP Rel-14 and Rel-19, with a WLS fitted line (-0.2856 log$_{10}$(f) + 1.445~\cite{r1-2502415}) and the 3GPP TR 38.901 model (-0.3236 log$_{10}$(f) + 1.512~\cite{tr38901v18}).}
        \label{fig:uma_nlos_mean_zsa_wls}
    \end{subfigure}

    \caption{Curve fitting of RT and Meas. data for the mean of lgZSA in UMa LOS and NLOS channel conditions using ordinary least squares (OLS), weighted mean (WM) and weighted least squares (WLS) methods. (a) and (c) use data from 3GPP Rel-19 only (6–24 GHz), while (b) and (d) use combined data from Rel-14 and Rel-19 (0.5–100 GHz). All subfigures include a comparison with the existing 3GPP TR 38.901 model.}
    \label{fig:uma_mean_zsa}
\end{figure*}

%%%%%%%%% UMa ZSA STD %%%%%%%%%%%%%
\begin{figure*}[h]
    \centering
    \begin{subfigure}[b]{0.48\textwidth}
        \centering
        \includegraphics[width=\linewidth]{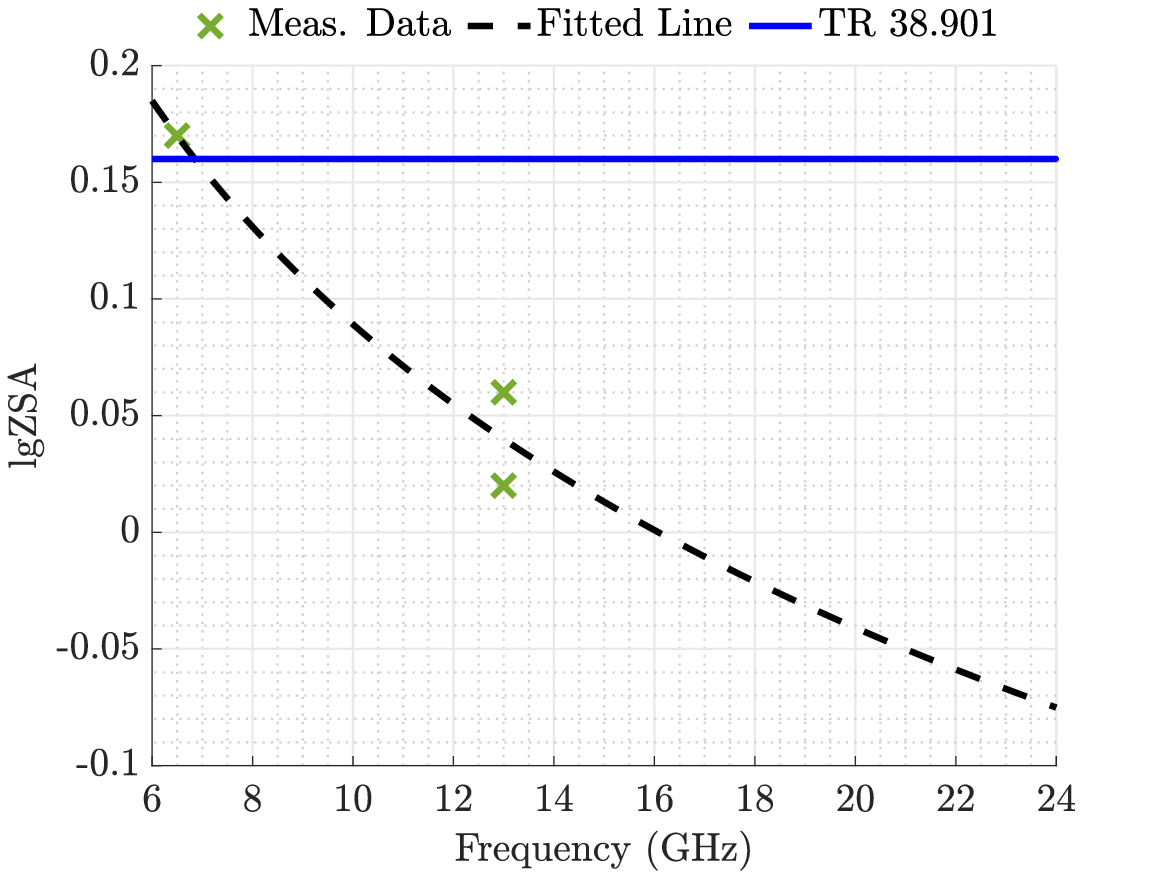}
        \caption{Meas. data for standard deviation of lgZSA in the UMa LOS scenario over 6–24 GHz from 3GPP Rel-19, with an OLS fitted line (-0.4319 log$_{10}$(f) + 0.521~\cite{r1-2502415}) and the 3GPP TR 38.901 model (0.16~\cite{tr38901v18}).}
        \label{fig:uma_los_std_zsa_rel19}
    \end{subfigure}
    \hfill
    \begin{subfigure}[b]{0.48\textwidth}
        \centering
        \includegraphics[width=\linewidth]{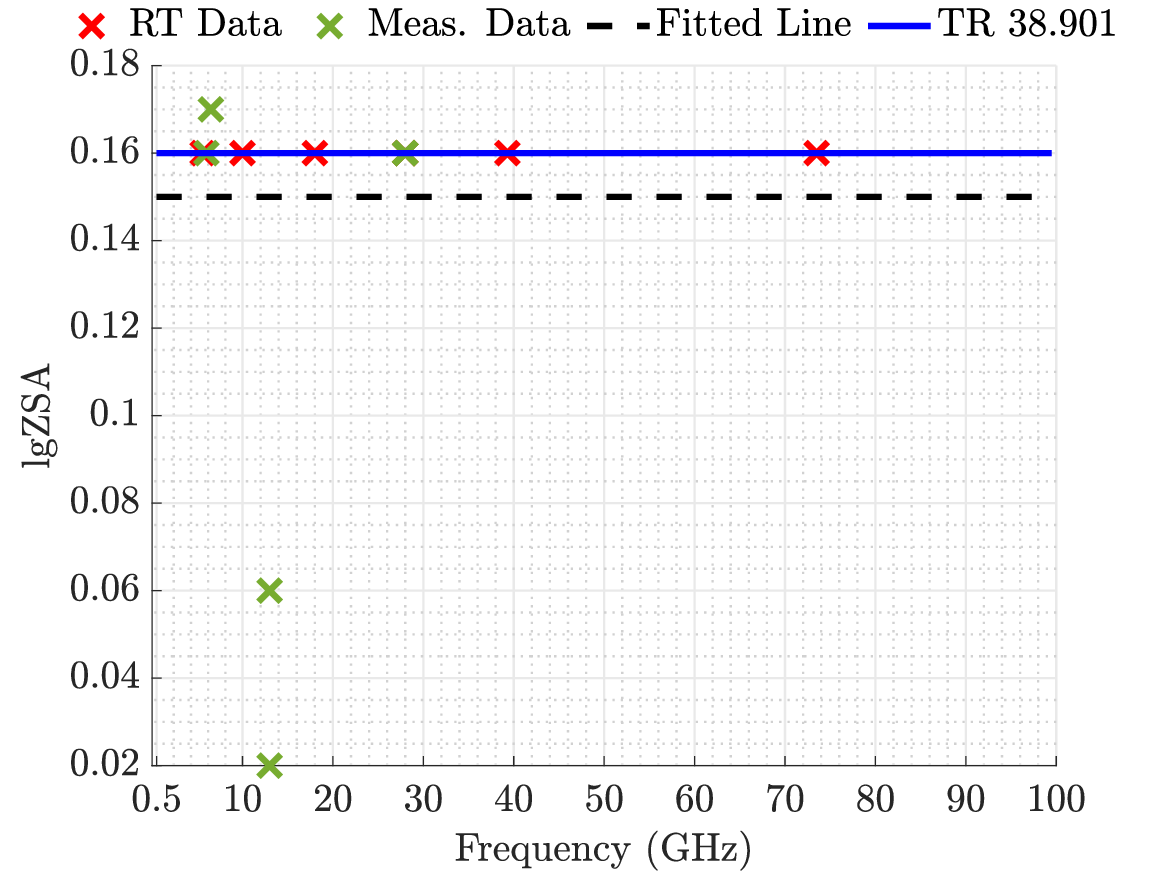}
        \caption{RT and Meas. data for standard deviation of lgZSA in the UMa LOS scenario over 0.5-100 GHz from 3GPP Rel-14 and Rel-19, with a WM fitted line (0.15~\cite{r1-2502415}) and the 3GPP TR 38.901 model (0.16~\cite{tr38901v18}).}
        \label{fig:uma_los_std_zsa_wls}
    \end{subfigure}
    
    \vspace{1em}

    \begin{subfigure}[b]{0.48\textwidth}
        \centering
        \includegraphics[width=\linewidth]{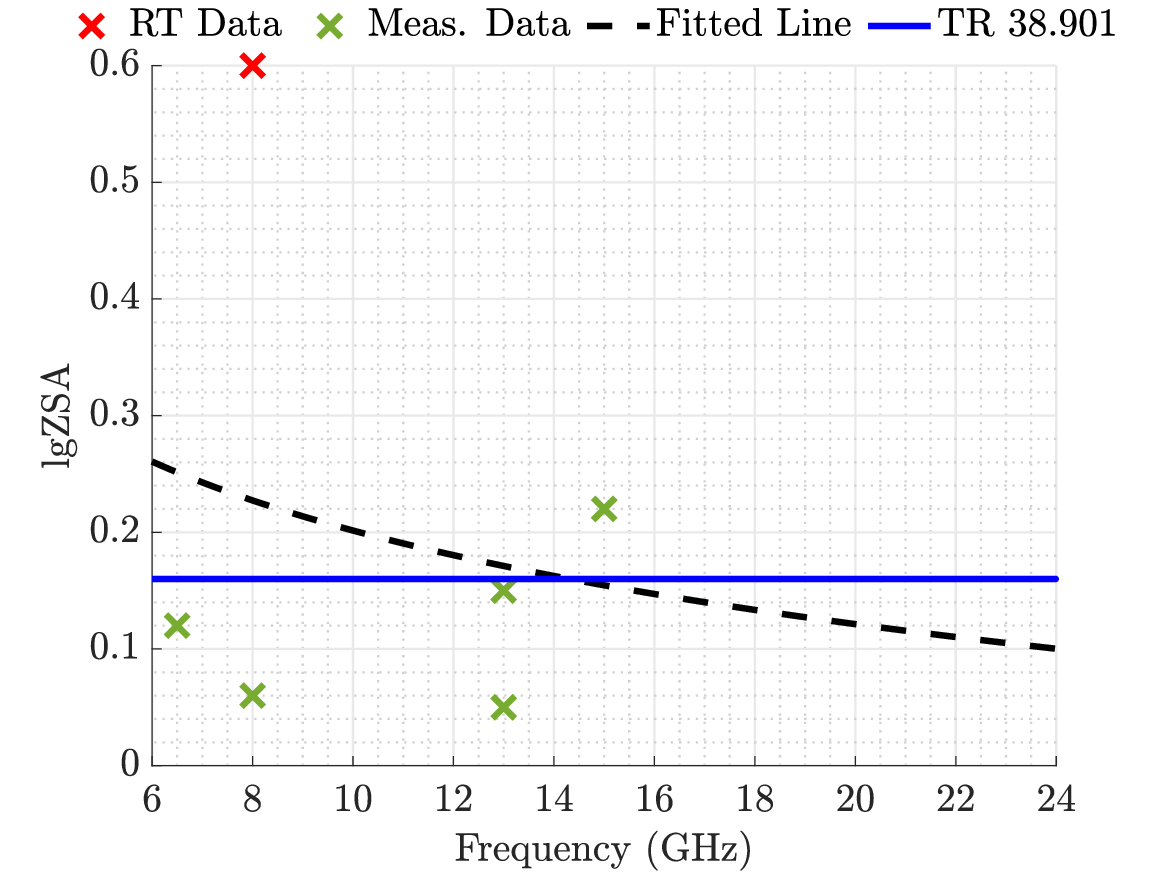}
        \caption{RT and Meas. data for standard deviation of lgZSA in the UMa NLOS scenario over 6–24 GHz from 3GPP Rel-19, with an OLS fitted line (-0.2665 log$_{10}$(f) + 0.468~\cite{r1-2502415}) and the 3GPP TR 38.901 model (0.16~\cite{tr38901v18}).}
        \label{fig:uma_nlos_std_zsa_rel19}
    \end{subfigure}
    \hfill
    \begin{subfigure}[b]{0.48\textwidth}
        \centering
        \includegraphics[width=\linewidth]{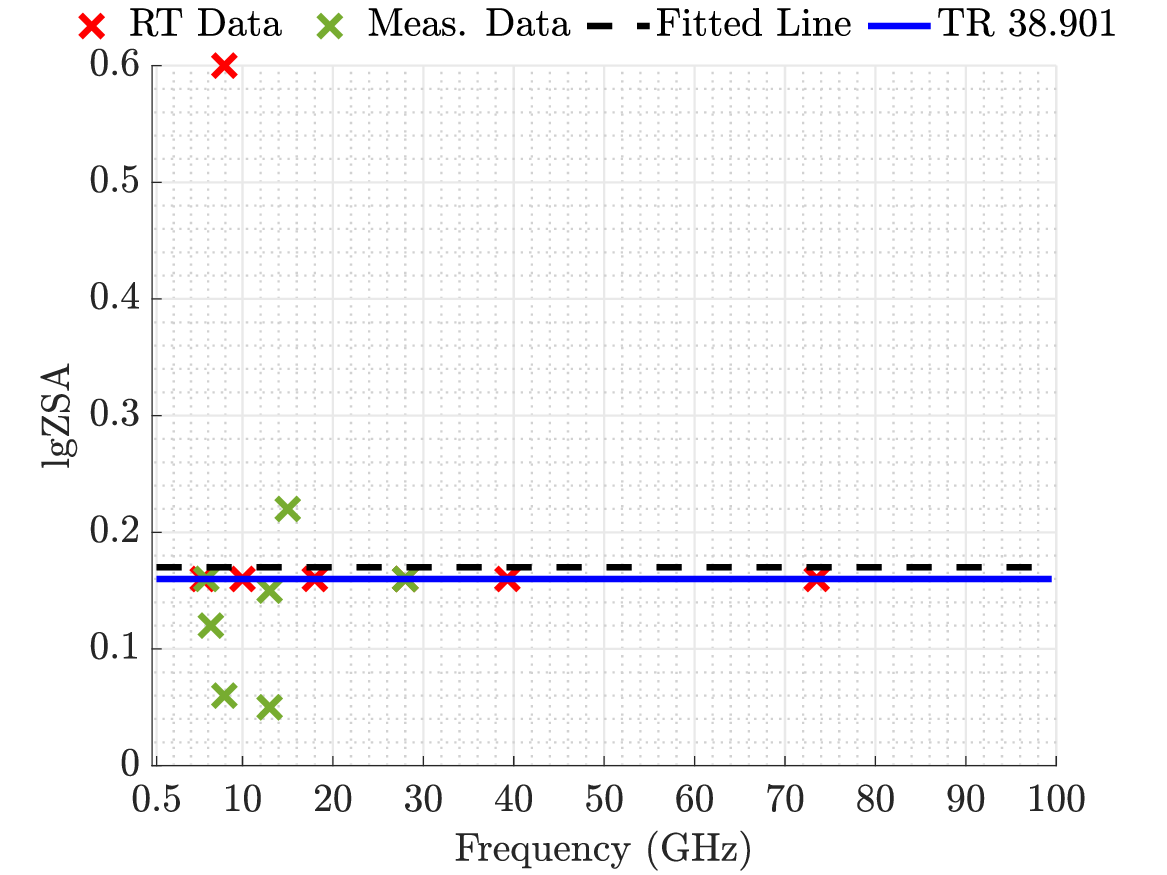}
        \caption{RT and Meas. data for standard deviation of lgZSA in the UMa NLOS scenario over 0.5-100 GHz from 3GPP Rel-14 and Rel-19, with a WM fitted line (0.17~\cite{r1-2502415}) and the 3GPP TR 38.901 model (0.16~\cite{tr38901v18}).}
        \label{fig:uma_nlos_std_zsa_wls}
    \end{subfigure}

    \caption{Curve fitting of RT and Meas. data for the standard deviation of lgZSA in UMa LOS and NLOS channel conditions using ordinary least squares (OLS) and weighted mean (WM) methods. (a) and (c) use data from 3GPP Rel-19 only (6–24 GHz), while (b) and (d) use combined data from Rel-14 and Rel-19 (0.5–100 GHz). All subfigures include a comparison with the existing 3GPP TR 38.901 model.}
    \label{fig:uma_std_zsa}
\end{figure*}

\clearpage
\subsection{SMa DS}
\begin{figure*}[h]
    \centering
    \begin{subfigure}[b]{0.48\textwidth}
        \centering
        \includegraphics[width=\linewidth]{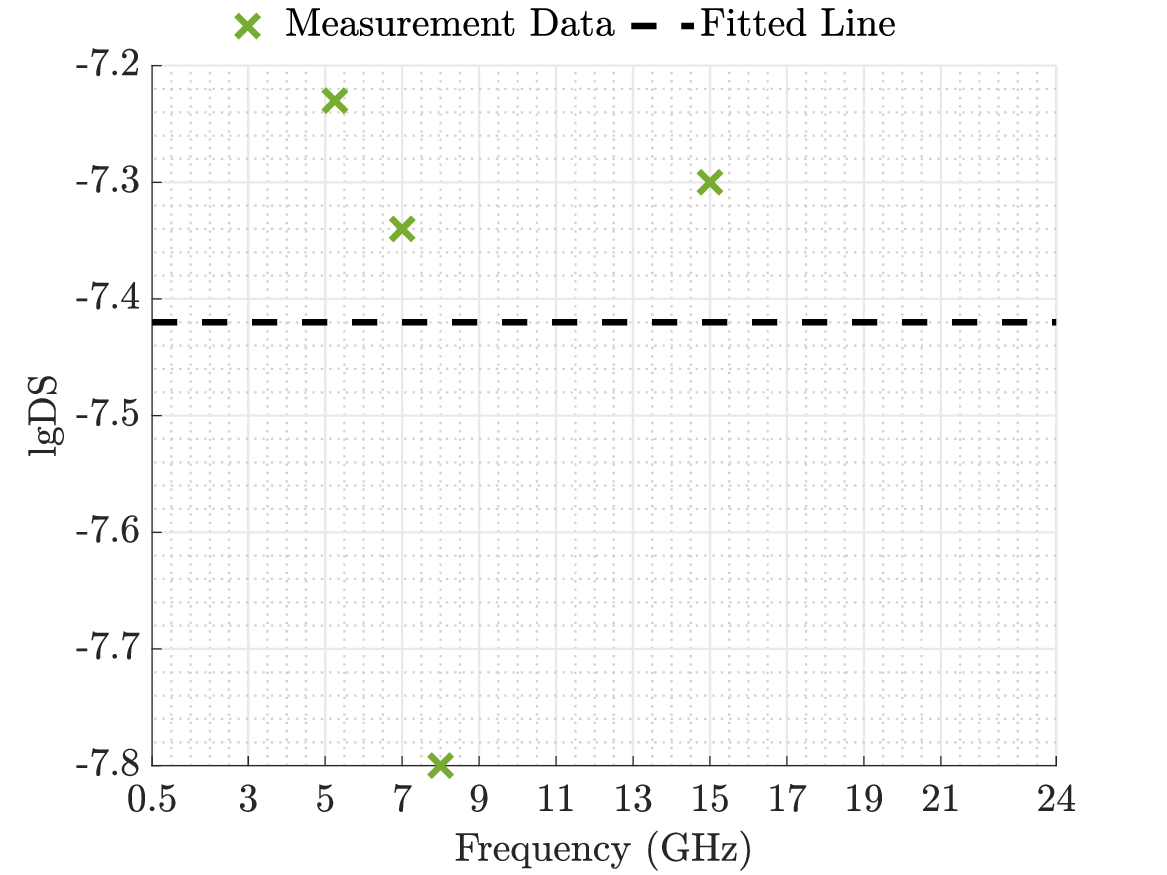}
        \caption{Meas. data for mean of lgDS in the SMa LOS scenario over 0.5-24 GHz, with an AM fitted line (-7.42~\cite{r1-2502415}).}
        \label{fig:sma_los_mean_ds_avg}
    \end{subfigure}
    \hfill
    \begin{subfigure}[b]{0.48\textwidth}
        \centering
        \includegraphics[width=\linewidth]{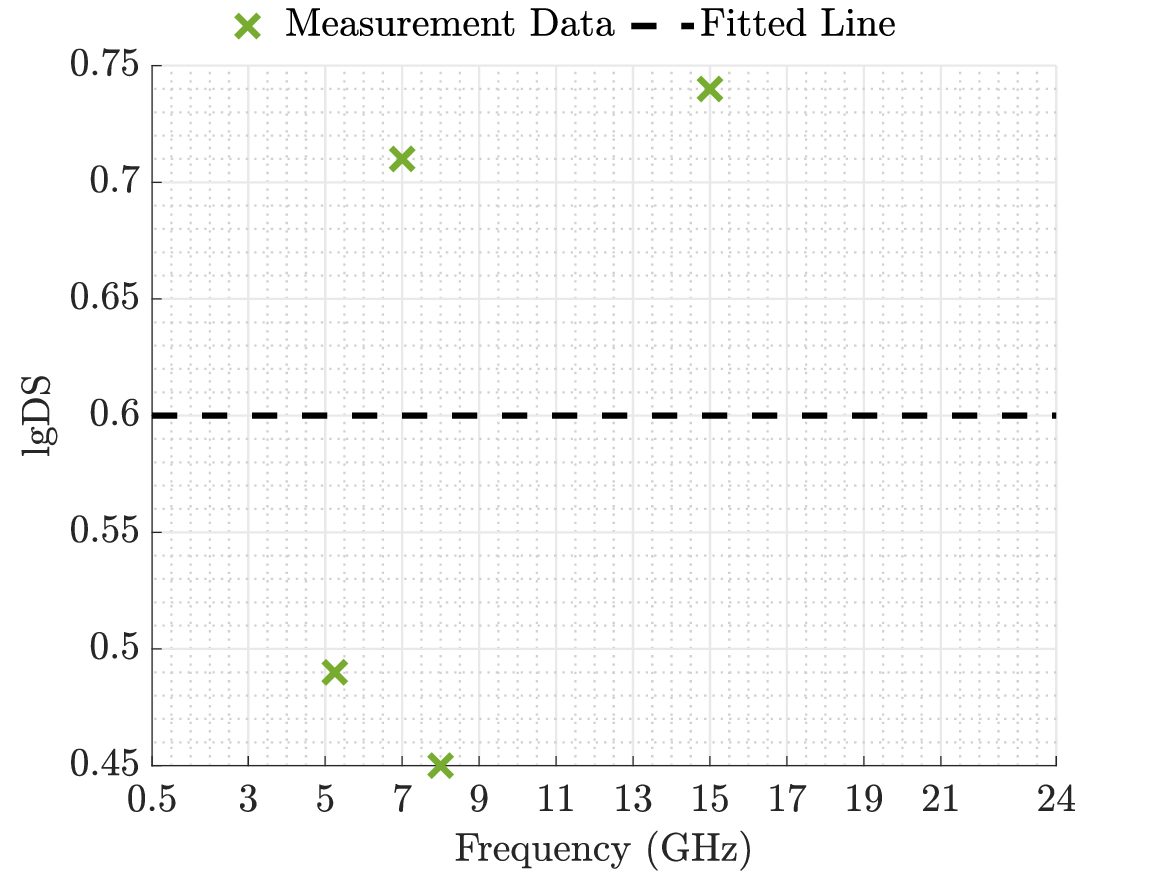}
        \caption{Meas. data for standard deviation of lgDS in the SMa LOS scenario over 0.5-24 GHz, with an AM fitted line (0.6~\cite{r1-2502415}).}
        \label{fig:sma_los_std_ds_avg}
    \end{subfigure}
    
    \vspace{1em}

    \begin{subfigure}[b]{0.48\textwidth}
        \centering
        \includegraphics[width=\linewidth]{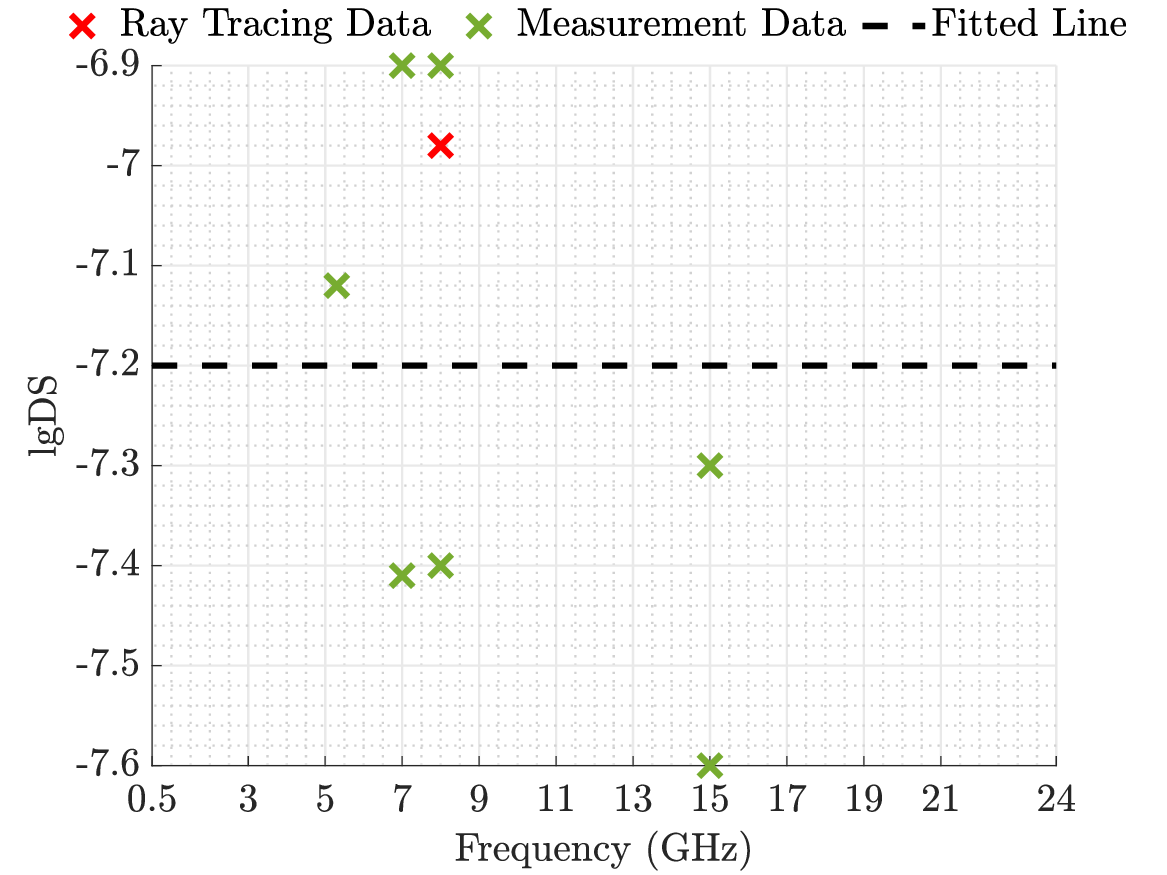}
        \caption{RT and Meas. data for mean of lgDS in the SMa NLOS scenario over 0.5-24 GHz, with an AM fitted line (-7.20~\cite{r1-2502415}).}
        \label{fig:sma_nlos_mean_ds_avg}
    \end{subfigure}
    \hfill
    \begin{subfigure}[b]{0.48\textwidth}
        \centering
        \includegraphics[width=\linewidth]{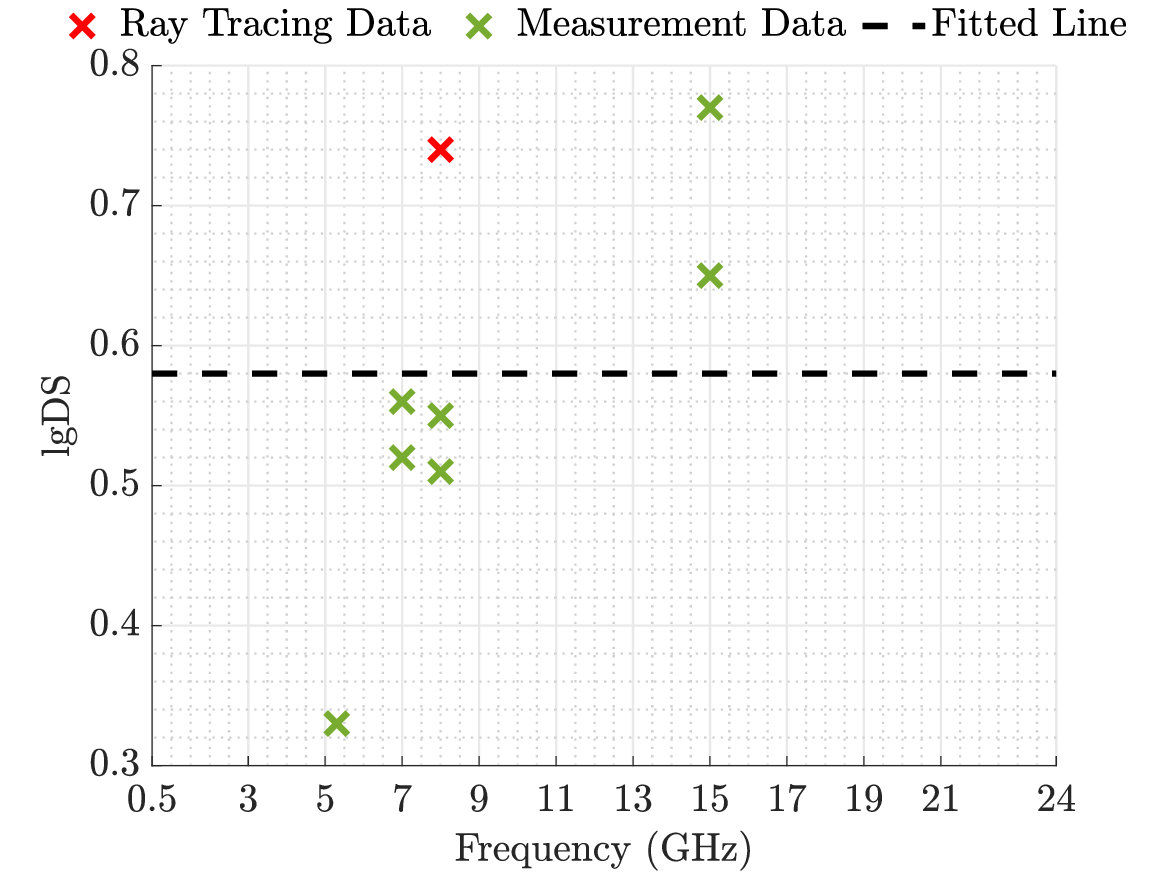}
        \caption{RT and Meas. data for standard deviation of lgDS in the SMa NLOS scenario over 0.5-24 GHz, with an AM fitted line (0.58~\cite{r1-2502415}).}
        \label{fig:sma_nlos_std_ds_avg}
    \end{subfigure}

    \caption{Curve fitting of RT and Meas. data for the mean and standard deviation of lgDS in SMa LOS and NLOS channel conditions using arithmetic mean (AM).}
    \label{fig:sma_mean_std_ds}
\end{figure*}

\clearpage
\subsection{SMa ASD}
\begin{figure*}[h]
    \centering
    \begin{subfigure}[b]{0.48\textwidth}
        \centering
        \includegraphics[width=\linewidth]{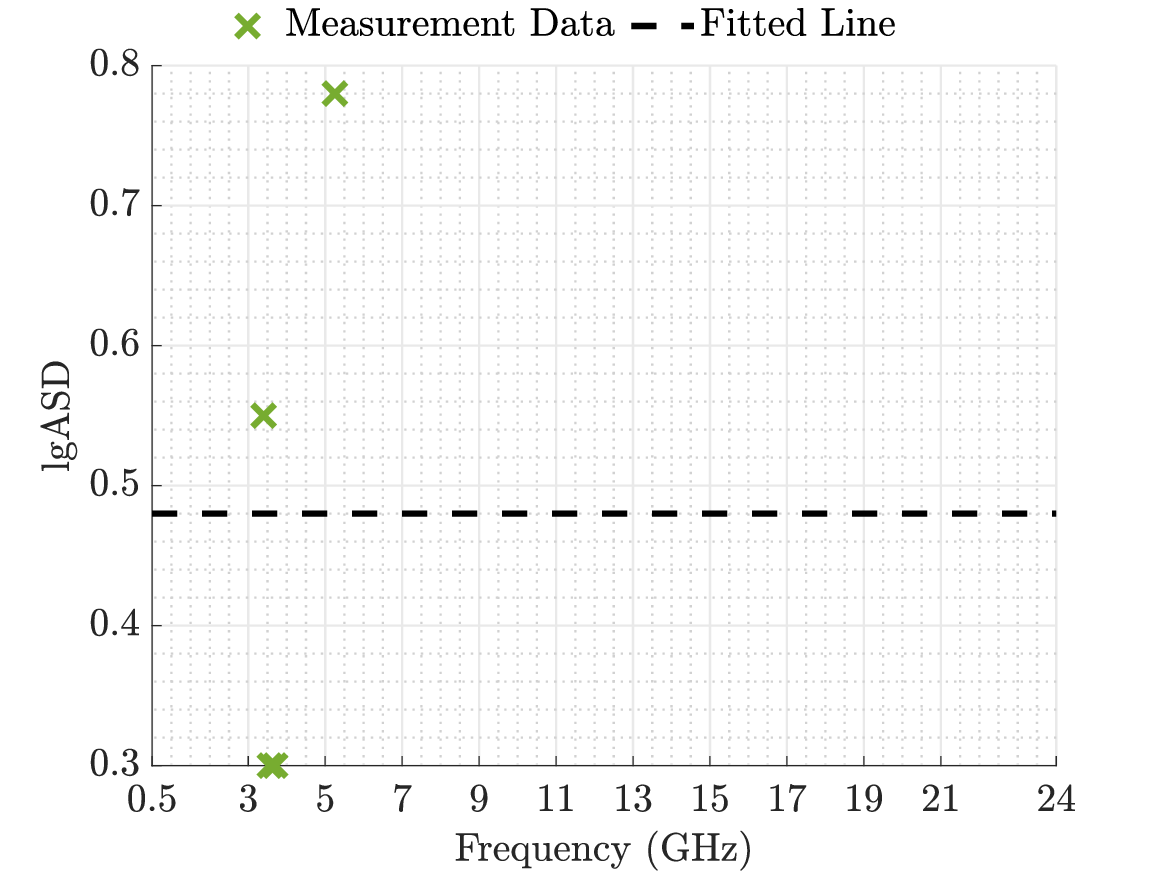}
        \caption{Meas. data for mean of lgASD in the SMa LOS scenario over 0.5-24 GHz, with an AM fitted line (0.48~\cite{r1-2502415}).}
        \label{fig:sma_los_mean_asd_avg}
    \end{subfigure}
    \hfill
    \begin{subfigure}[b]{0.48\textwidth}
        \centering
        \includegraphics[width=\linewidth]{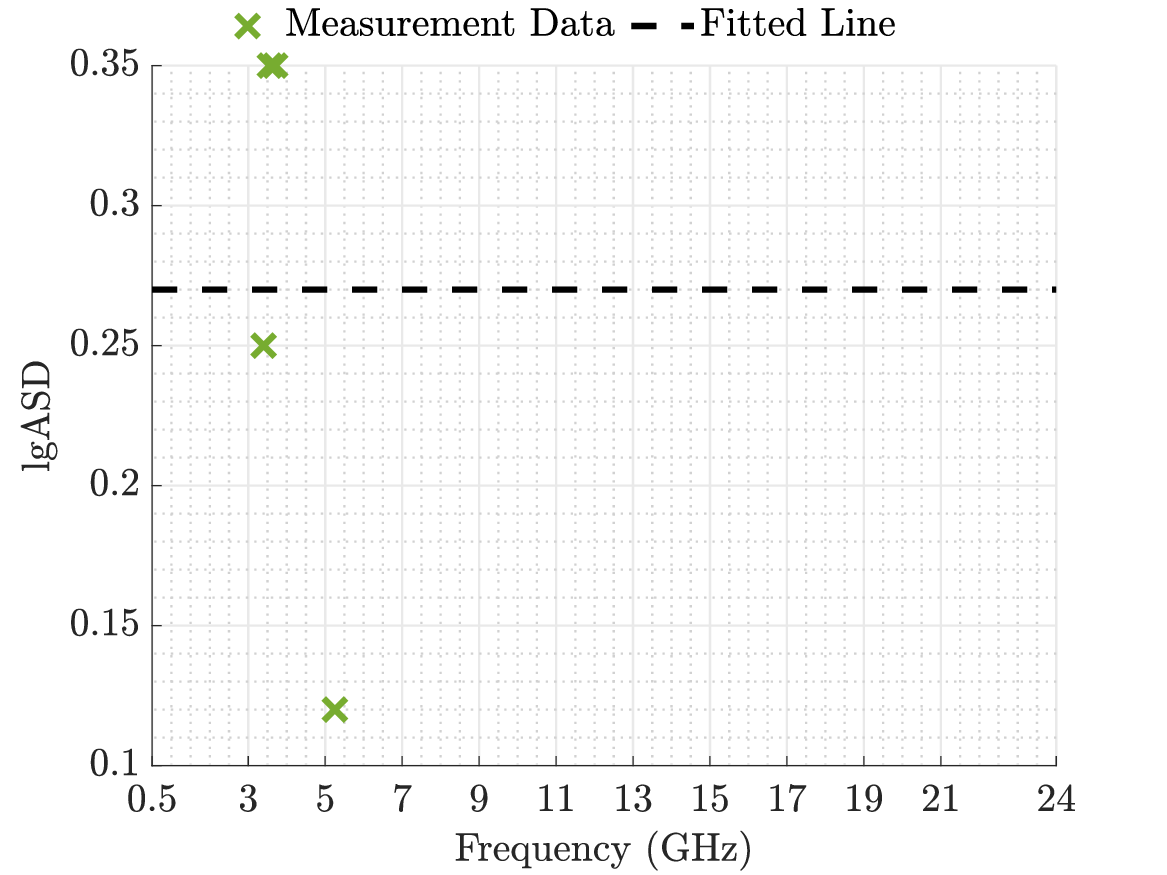}
        \caption{Meas. data for standard deviation of lgASD in the SMa LOS scenario over 0.5-24 GHz, with an AM fitted line (0.27~\cite{r1-2502415}).}
        \label{fig:sma_los_std_asd_avg}
    \end{subfigure}
    
    \vspace{1em}

    \begin{subfigure}[b]{0.48\textwidth}
        \centering
        \includegraphics[width=\linewidth]{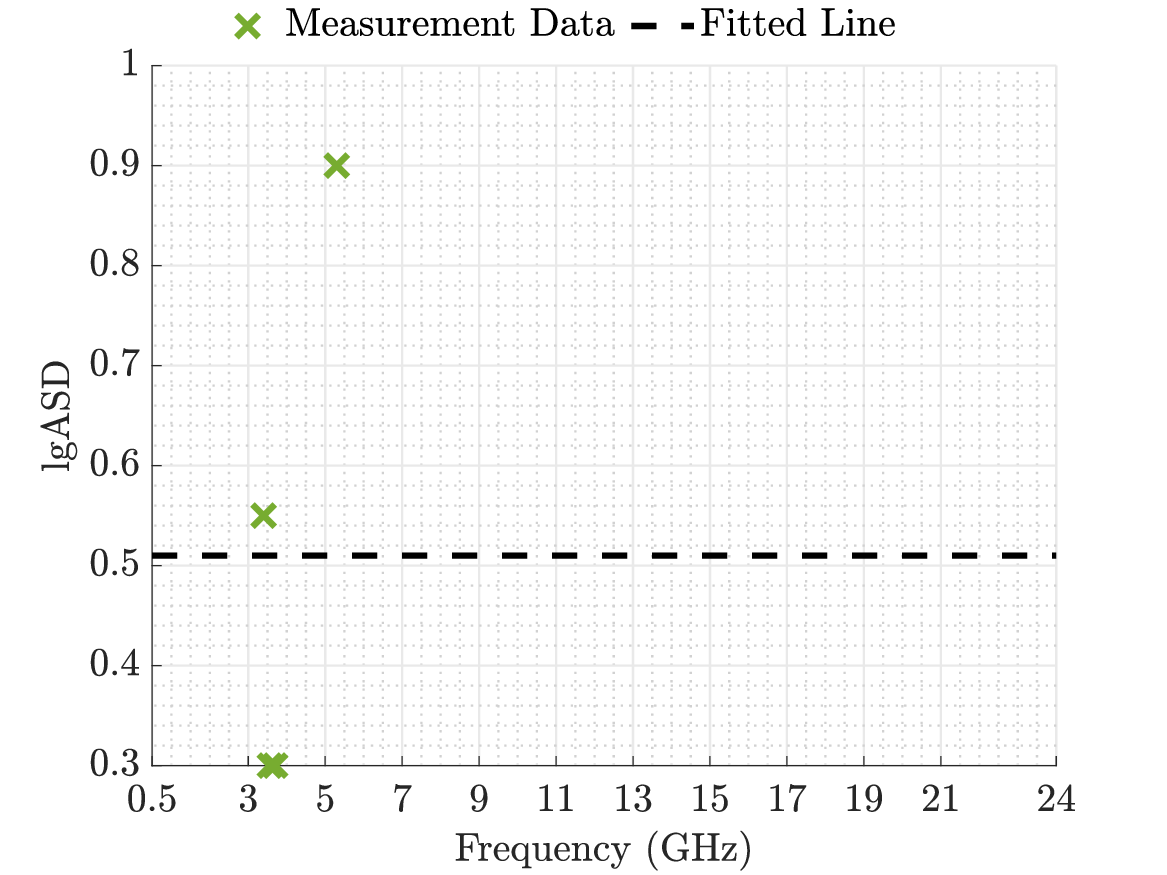}
        \caption{Meas. data for mean of lgASD in the SMa NLOS scenario over 0.5-24 GHz, with an AM fitted line (0.51~\cite{r1-2502415}).}
        \label{fig:sma_nlos_mean_asd_avg}
    \end{subfigure}
    \hfill
    \begin{subfigure}[b]{0.48\textwidth}
        \centering
        \includegraphics[width=\linewidth]{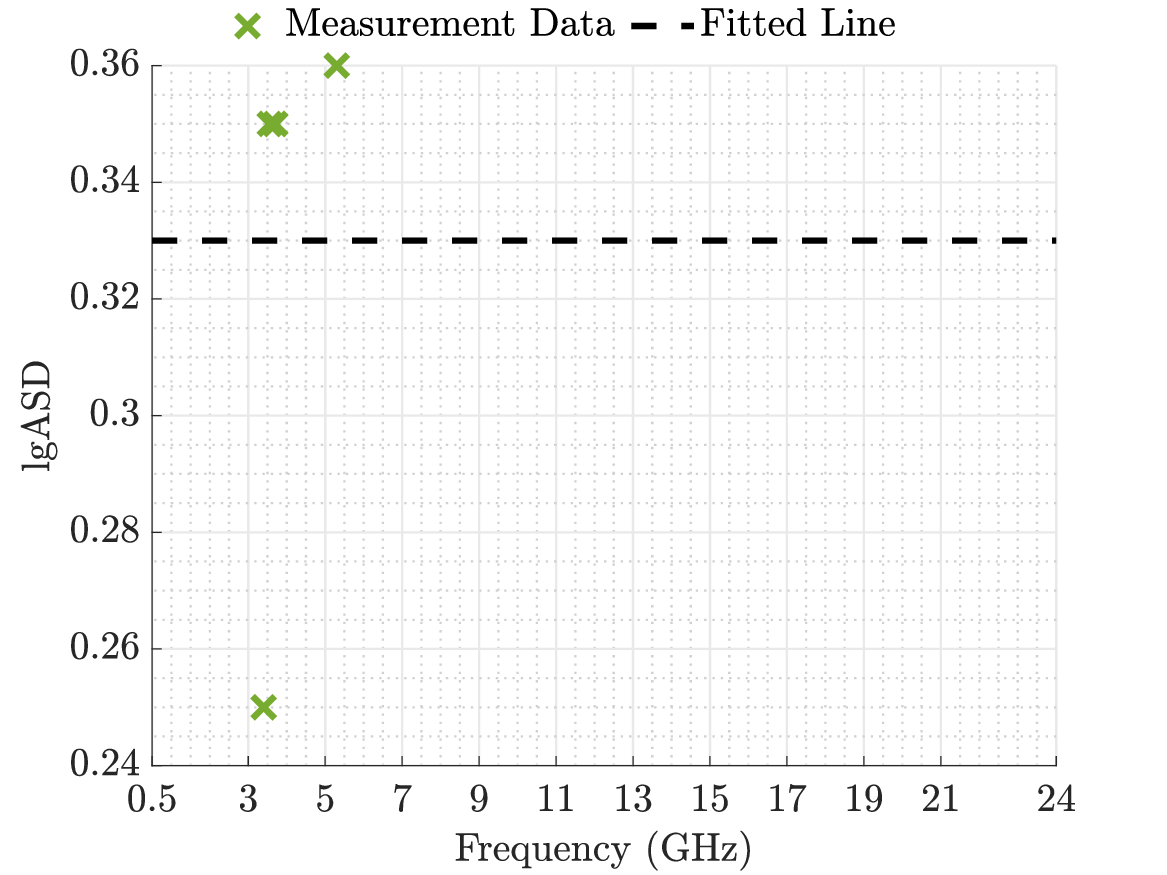}
        \caption{Meas. data for standard deviation of lgASD in the SMa NLOS scenario over 0.5-24 GHz, with an AM fitted line (0.33~\cite{r1-2502415}).}
        \label{fig:sma_nlos_std_asd_avg}
    \end{subfigure}

    \caption{Curve fitting of RT and Meas. data for the mean and standard deviation of lgASD in SMa LOS and NLOS channel conditions using arithmetic mean (AM).}
    \label{fig:sma_mean_std_asd}
\end{figure*}

\clearpage
\subsection{SMa ASA}
\begin{figure*}[h]
    \centering
    \begin{subfigure}[b]{0.48\textwidth}
        \centering
        \includegraphics[width=\linewidth]{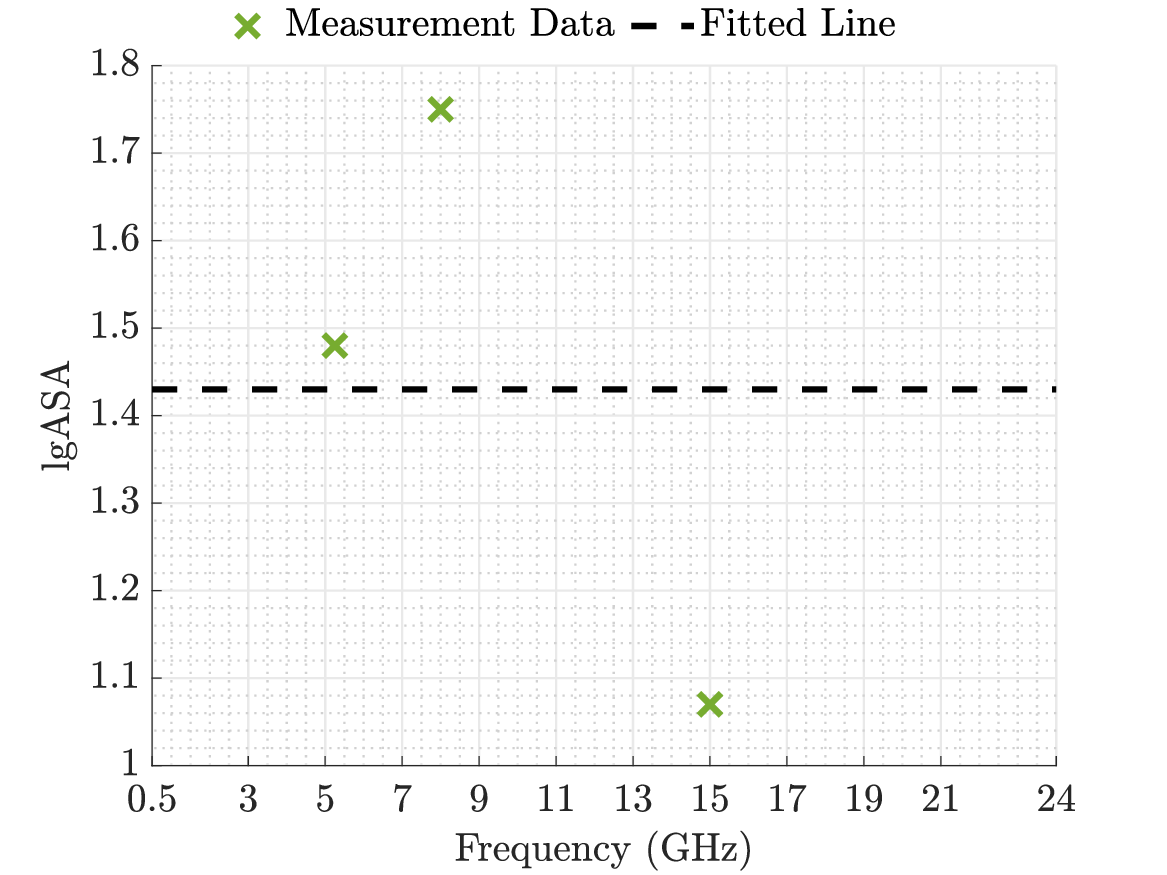}
        \caption{Meas. data for mean of lgASA in the SMa LOS scenario over 0.5-24 GHz, with an AM fitted line (1.43~\cite{r1-2502415}).}
        \label{fig:sma_los_mean_asa_avg}
    \end{subfigure}
    \hfill
    \begin{subfigure}[b]{0.48\textwidth}
        \centering
        \includegraphics[width=\linewidth]{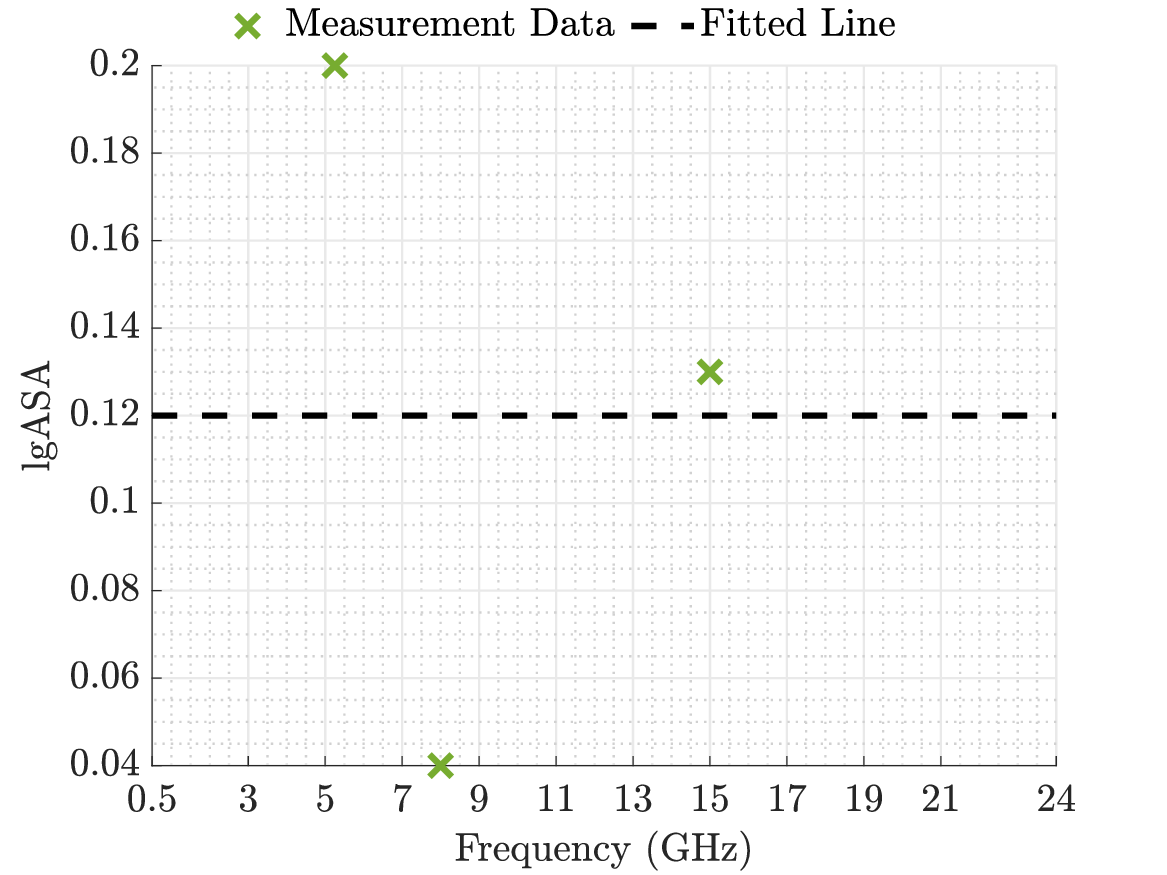}
        \caption{Meas. data for standard deviation of lgASA in the SMa LOS scenario over 0.5-24 GHz, with an AM fitted line (0.12~\cite{r1-2502415}).}
        \label{fig:sma_los_std_asa_avg}
    \end{subfigure}
    
    \vspace{1em}

    \begin{subfigure}[b]{0.48\textwidth}
        \centering
        \includegraphics[width=\linewidth]{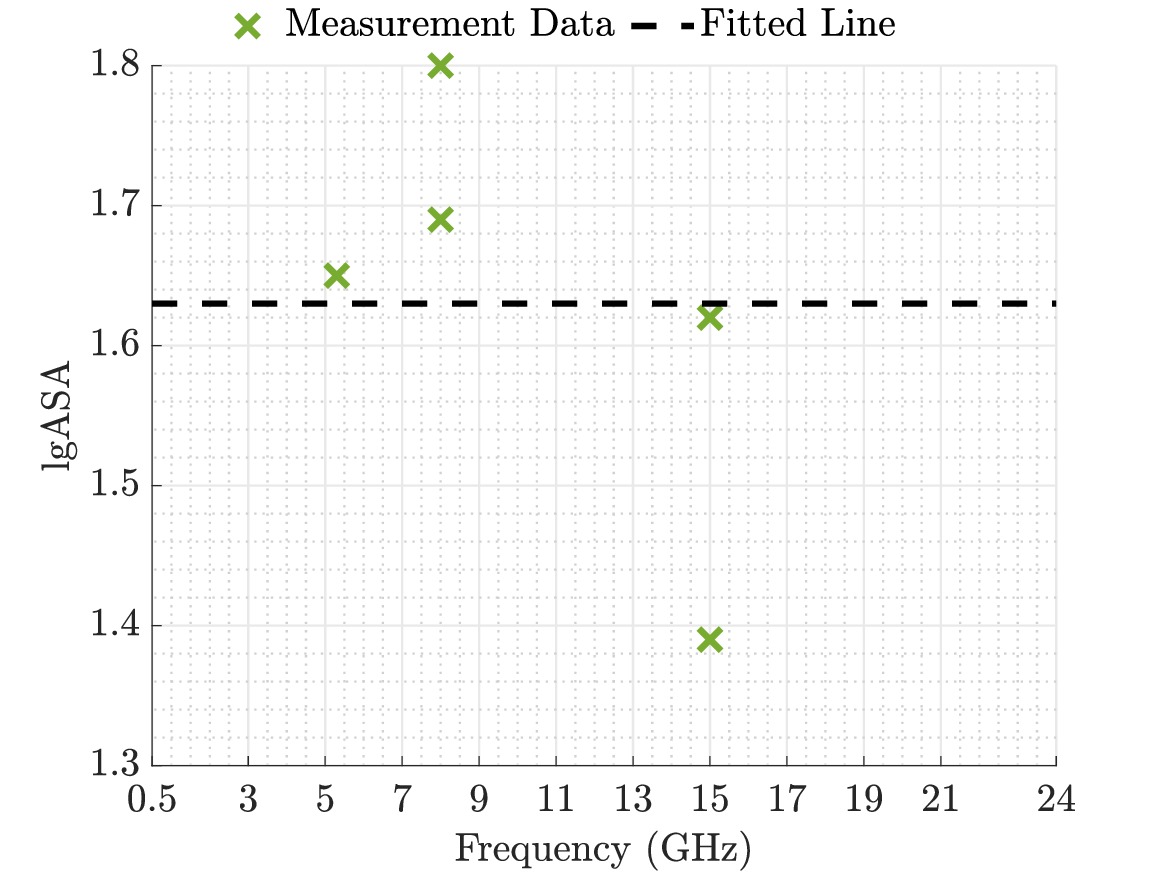}
        \caption{Meas. data for mean of lgASA in the SMa NLOS scenario over 0.5-24 GHz, with an AM fitted line (1.63~\cite{r1-2502415}).}
        \label{fig:sma_nlos_mean_asa_avg}
    \end{subfigure}
    \hfill
    \begin{subfigure}[b]{0.48\textwidth}
        \centering
        \includegraphics[width=\linewidth]{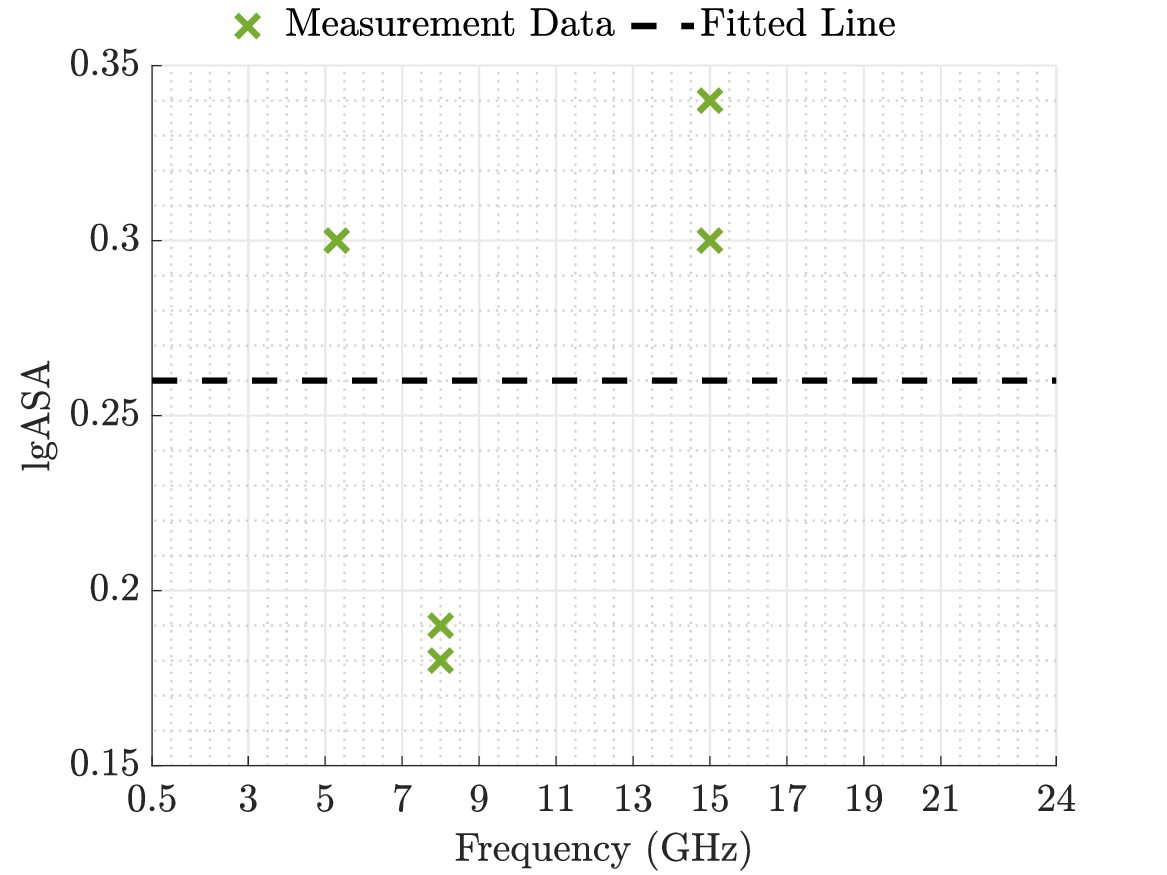}
        \caption{Meas. data for standard deviation of lgASA in the SMa NLOS scenario over 0.5-24 GHz, with an AM fitted line (0.26~\cite{r1-2502415}).}
        \label{fig:sma_nlos_std_asa_avg}
    \end{subfigure}

    \caption{Curve fitting of RT and Meas. data for the mean and standard deviation of lgASA in SMa LOS and NLOS channel conditions using arithmetic mean (AM).}
    \label{fig:uma_mean_std_asa}
\end{figure*}

\clearpage
\subsection{SMa ZSA}
\begin{figure*}[h]
    \centering
    \begin{subfigure}[b]{0.48\textwidth}
        \centering
        \includegraphics[width=\linewidth]{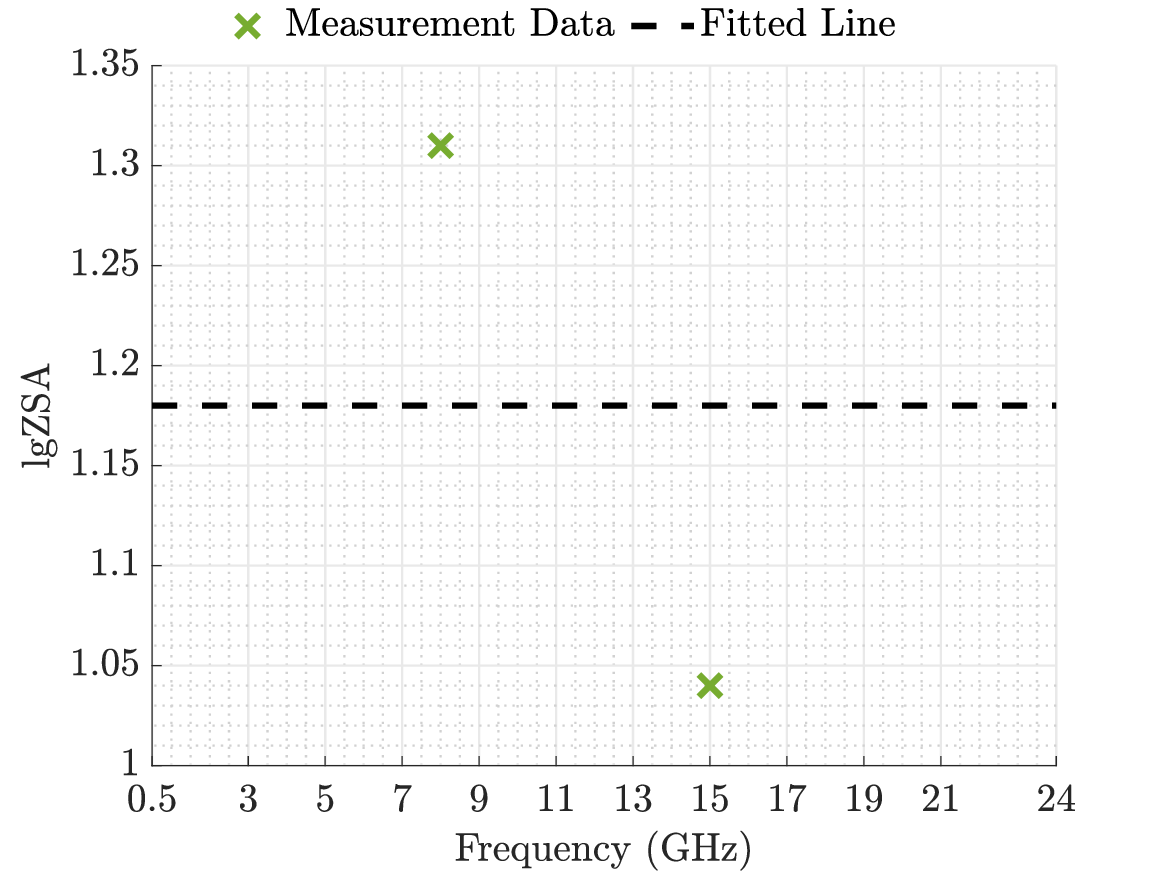}
        \caption{Meas. data for mean of lgZSA in the SMa LOS scenario over 0.5-24 GHz, with an AM fitted line (1.18~\cite{r1-2502415}).}
        \label{fig:sma_los_mean_zsa_avg}
    \end{subfigure}
    \hfill
    \begin{subfigure}[b]{0.48\textwidth}
        \centering
        \includegraphics[width=\linewidth]{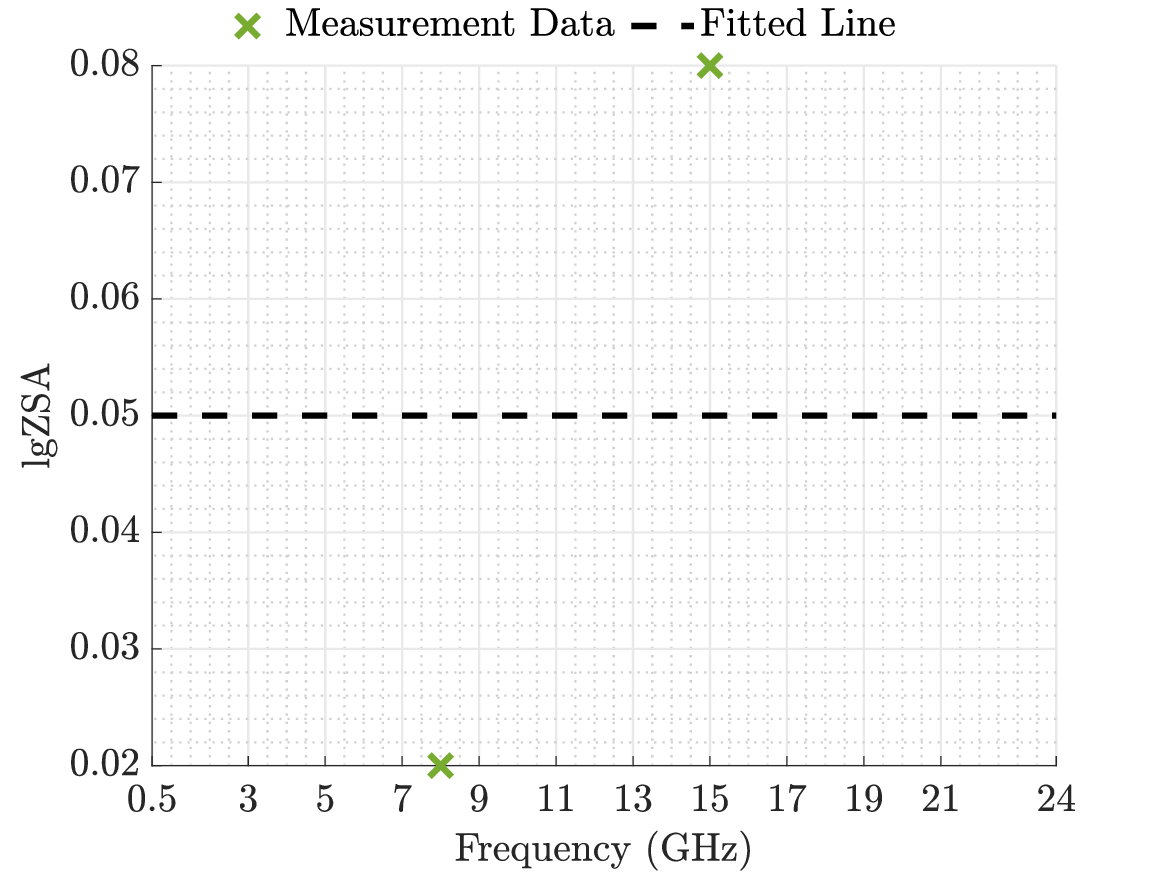}
        \caption{Meas. data for standard deviation of lgZSA in the SMa LOS scenario over 0.5-24 GHz, with an AM fitted line (0.05~\cite{r1-2502415}).}
        \label{fig:sma_los_std_zsa_avg}
    \end{subfigure}
    
    \vspace{1em}

    \begin{subfigure}[b]{0.48\textwidth}
        \centering
        \includegraphics[width=\linewidth]{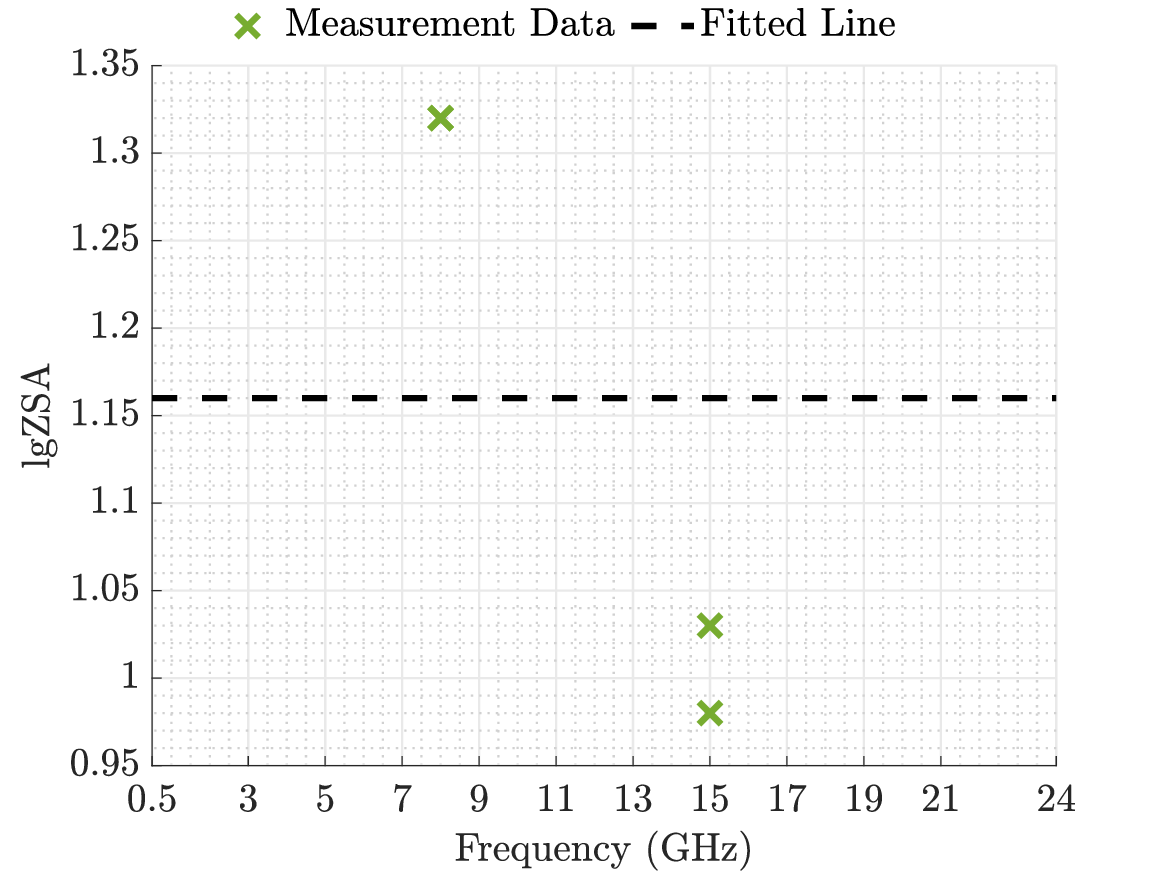}
        \caption{Meas. data for mean of lgZSA in the SMa NLOS scenario over 0.5-24 GHz, with an AM fitted line (1.16~\cite{r1-2502415}).}
        \label{fig:sma_nlos_mean_zsa_avg}
    \end{subfigure}
    \hfill
    \begin{subfigure}[b]{0.48\textwidth}
        \centering
        \includegraphics[width=\linewidth]{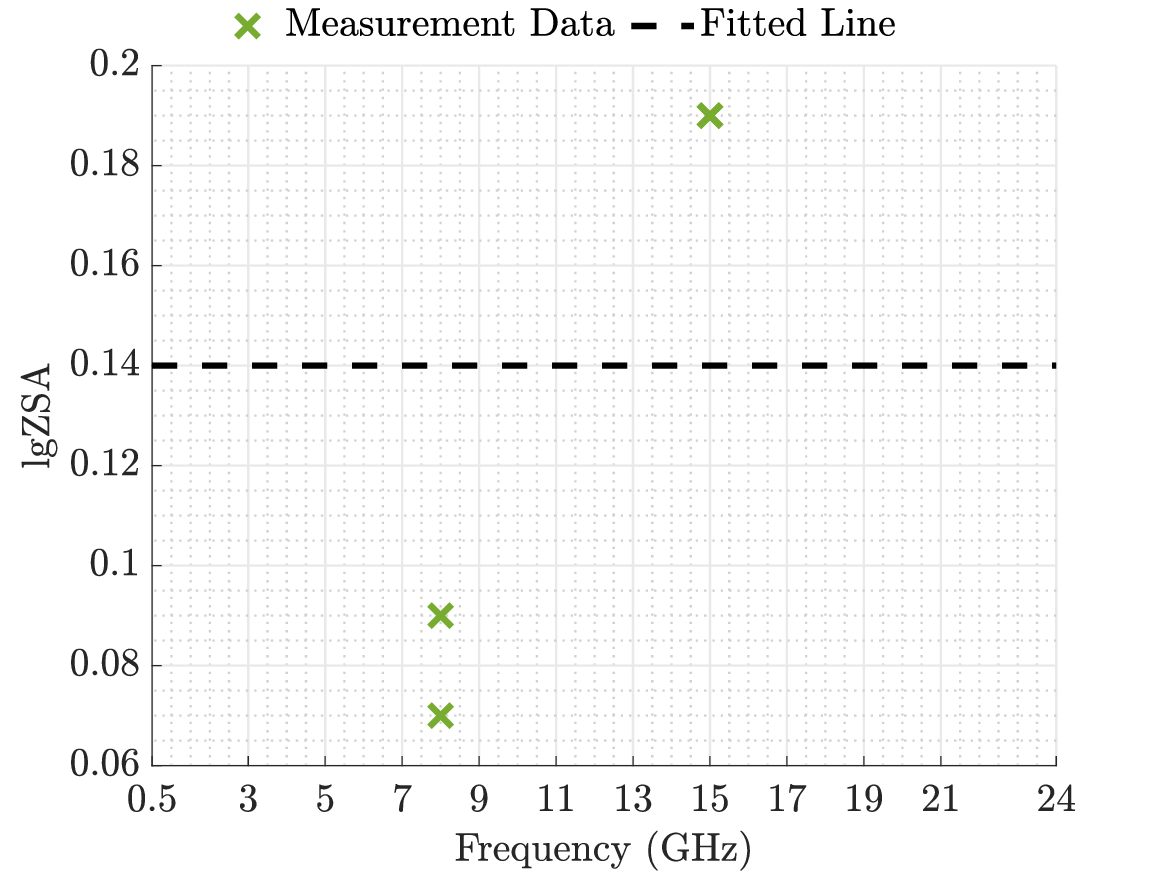}
        \caption{Meas. data for standard deviation of lgZSA in the SMa NLOS scenario over 0.5-24 GHz, with an AM fitted line (0.14~\cite{r1-2502415}).}
        \label{fig:sma_nlos_std_zsa_avg}
    \end{subfigure}

    \caption{Curve fitting of RT and Meas. data for the mean and standard deviation of lgZSA in SMa LOS and NLOS channel conditions using arithmetic mean (AM).}
    \label{fig:uma_mean_std_zsa}
\end{figure*}

\clearpage
\subsection{Material Penetration Loss Model for Plywood}
\begin{figure}[h!]
    \centering
    \includegraphics[width=\linewidth]{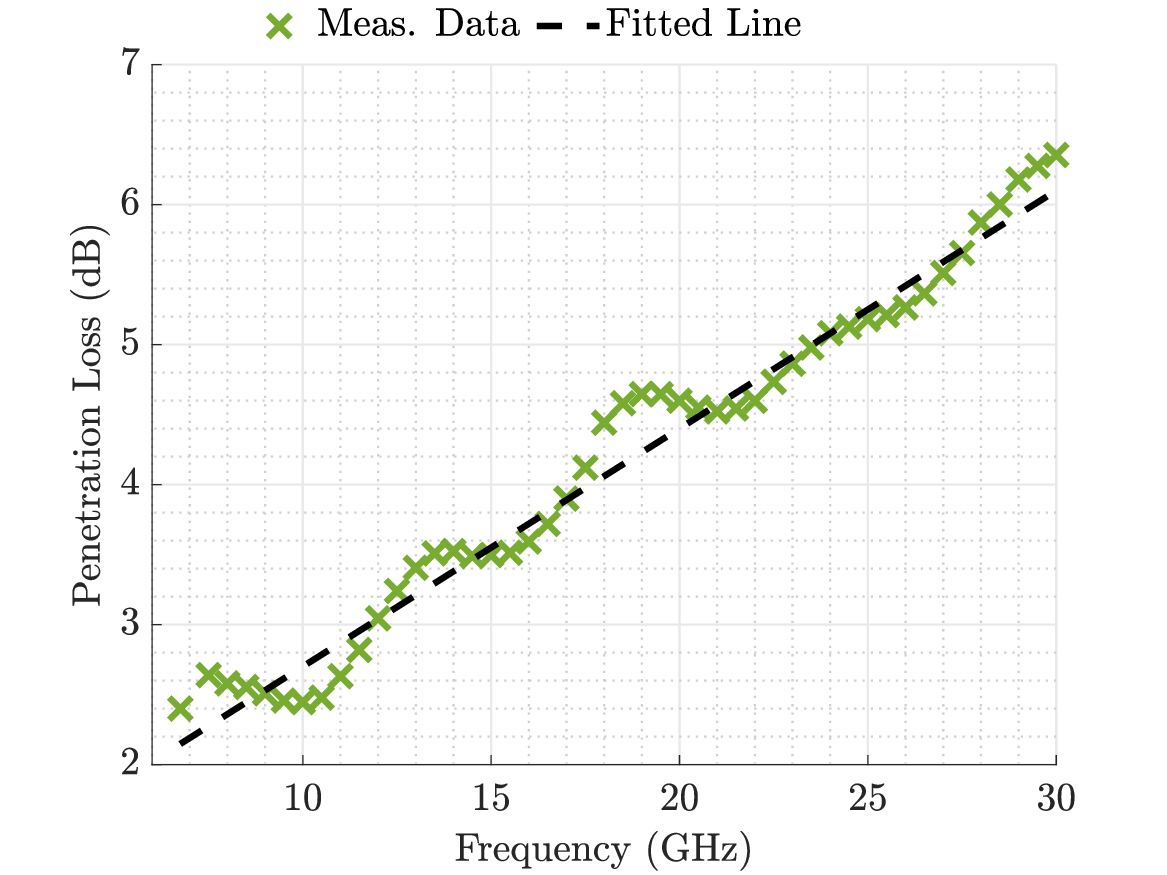}
    \caption{Meas. data for penetration loss of plywood over 0.5-30 GHz, with an OLS fitted line (1.03 + 0.17 f ~\cite{r1-2502415}). This empirical model is extrapolated to estimate plywood penetration loss over the extended frequency range up to 100 GHz.}
    \label{fig:plywood_pen_loss}
\end{figure}

%% file: conclusion.tex
This document summarizes the measurement campaigns, datasets, and curve-fitting results considered by 3GPP in the Rel-19 study on validate using measurements the channel model of 3GPP TR 38.901 at least for 7-24 GHz. The datasets and methodologies used to refine the various channel-modeling parameters documented herein may serve as a reference for future 3GPP channel-modeling studies aimed at extending the channel model to other frequency bands. In addition, the research community is encouraged to use this document to access the data considered by 3GPP for channel modeling and to understand the methodologies employed in deriving channel model parameters within the standardization process. Furthermore, researchers are encouraged to conduct measurement campaigns across different frequency bands, scenarios, and diverse environments to provide additional input to future 3GPP standardization efforts, as data from a wide range of geographic regions can support the development of more robust and generalized standardized stochastic channel models.